\documentclass[traditabstract]{aa} 

\usepackage{makecell}
\usepackage{graphicx}
\usepackage{natbib}
\usepackage{subfigure}
\usepackage{lipsum} 
\usepackage{txfonts}
\usepackage{hyperref}
\usepackage{multirow}
\usepackage{amsmath}
\begin{document}

   \title{An optical perspective on early-stage AGN with extreme radio flares}

   \author{L. Crepaldi\inst{1,2,3,4}\thanks{luca.crepaldi.1@phd.unipd.it}
          \and   
          M. Berton\inst{2}
          \and
          B. Dalla Barba\inst{2,5,6}
          \and
          G. La Mura\inst{3,7}
          \and
          E. J\"arvel\"a \inst{8,9}\thanks{Dodge Family Prize Fellow at The University of Oklahoma}
          \and
          A. Vietri\inst{1,2}
          \and
          S. Ciroi\inst{1}
          }

   \institute{
    $^1$ Dipartimento di Fisica e Astronomia "G. Galilei", University of Padova, Vicolo dell'osservatorio 3, 35122, Padova, Italy;\\
    $^2$ European Southern Observatory, Alonso de C\'ordova 3107, Casilla 19, 19001, Santiago, Chile;\\
    $^3$ INAF - Osservatorio Astronomico di Cagliari, Via della Scienza 5, 09047, Selargius, Italy; \\
    $^4$ Departamento de Física y Astronomía, Universidad de La Serena, Av. Cisternas 1200 N, La Serena, Chile; \\
    $^5$ Università degli studi dell’Insubria, Via Valleggio 11, 22100, Como, Italy; \\
    $^6$ INAF - Osservatorio Astronomico di Brera, Via E. Bianchi 46, 23807, Merate, Italy; \\    
    $^7$ Laboratory of Instrumentation and Experimental Particle Physics, Av. Prof. Gama Pinto, 2 - 1649-003, Lisboa, Portugal; \\
    $^8$ Homer L.\,Dodge Department of Physics and Astronomy, The University of Oklahoma, 440 West Brooks Street, Norman, 73019, OK, USA; \\
    $^9$ Department of Physics and Astronomy, Texas Tech University, Box 41051, Lubbock, 79409-1051, TX, USA.\\
    }

   \date{Received ...; accepted ...}

 
  \abstract
   {In the last decade of Active Galactic Nuclei (AGN) monitoring programs, the Metsähovi Radio Observatory detected multiple times seven powerful flaring narrow-line Seyfert 1 (NLS1) galaxies at 37~GHz. Several hypotheses have been proposed, but the understanding of this unique phenomenon is still far. To look at the case from a different point of view, we performed an emission line analysis of the optical spectra, with the aim of identifying similarities among the sources, that can be in turn possibly tied with the radio behavior. Our data were obtained with the Gran Telescopio Canarias. The results we obtained show that six out of seven sources have typical properties for the NLS1 class, and one of them is an intermediate Seyfert galaxy. We found on average black hole masses above the median value for the class (> 10$^7$ M$_{\odot}$), and a strong Fe~II emission, which could be a proxy for an intense ongoing accretion activity. Although interesting, the characteristics we found are not unusual for these kind of AGN: the optical spectra of our sources do not related with their unique radio properties. Therefore, further multi-wavelength studies will be necessary to narrow the field of hypotheses for this peculiar phenomenon.}


   \maketitle
%
\newcommand{\kms}{km s$^{-1}$}
\newcommand{\ergs}{erg s$^{-1}$}
\newcommand{\chired}{$\chi^2_\nu$}
\newcommand{\hb}{H$\beta$}

\section{Introduction}
Since its classification \citep{Osterbrock85}, the class of active galactic nuclei (AGN) known as narrow-line Seyfert 1 (NLS1) galaxies has been a source of new discoveries. NLS1s have by definition a full-width at half maximum (FWHM) of H$\beta$ lower than 2000 \kms. Such limit, even though it is frequently seen on this class, is not strict and it is more a historical convention than a real threshold \citep{Marziani18a}. Two more criteria, however, clearly distinguish NLS1s: a flux ratio [O~III]/H$\beta<3$ and often visible prominent Fe~II multiplets \citep{Cracco16, Marziani18b, Marziani21}. Moreover, they are characterized by low-to-intermediate-mass supermassive black holes (10$^6$-10$^8$ M$_\odot$), which induce a low rotational velocity of the line-emitting clouds and in turn narrow permitted emission lines \citep{Zhou06, Cracco16, Chen18}. Observations suggest an accretion close to or above the Eddington limit \citep{Boroson92}. As other sources with similar black hole masses, NLS1s belong to the so-called population A of the quasar main sequence \citep{Sulentic15}. Their black hole masses bring to think that they are still in an early phase of their life cycle and that they will eventually evolve, possibly through several accretion episodes, into broad-line Seyfert 1 galaxies \citep{Mathur00}. Despite the belief that only massive elliptical galaxies are able to launch powerful relativistic jets \citep{2000ApJ...543L.111L}, some NLS1s are found to be jetted sources \citep{Abdo09a, Abdo09c, Foschini11}, that, when face-on, produce $\gamma$-ray emission \citep{Foschini15}. 
To date $\sim$70 jetted NLS1s have been confirmed through $\gamma$-ray detections \citep{Romano18, Paliya19} or radio imaging \citep{2015richards1, 2016lister1, 2018berton1, 2020chen1, 2022chen1}, and several dozen new candidates have been identified \citep{2021foschini1, 2022foschini1}. In particular, jetted NLS1s may be an early evolutionary stage of flat-spectrum radio quasars (FSRQs) and the parent population of peaked sources, as low-luminosity compact sources \citep{Berton16c, Foschini17, Vietri24}. The jets in NLS1s are less powerful than those in FSRQs, likely due to the non-linear scaling relations between the jet power and the black hole mass, and the magnetic flux \citep{Heinz03, Foschini15, Chamani21}. These objects, indeed, are less luminous e.g. in radio than FSRQs \citep{Berton16c}, and they also lack the diffuse radio emission \citep{Berton18a}. Hence, because of the lack of relativistic beaming, some jetted NLS1s observed at large angles may actually appear as radio-quiet\footnote{Radio-quiet sources have a radio-loudness parameter, defined as the flux ratio F$_{5~GHz}$/F$_{B-band}$ \citep{Kellermann89}, lower than 10. On the other hand radio-loud sources have a radio loudness parameter bigger than 10. The use of radio loudness to identify jets should be, however, avoided, in favor of the more physical classification into jetted and non-jetted sources \citep{Padovani17, Jarvela17, Berton21c}.}.

To increase the number of jetted NLS1s, in 2012 the Metsähovi Radio Observatory (MRO) started a monitoring campaign at 37~GHz \citep{Lahteenmaki17}. Two of the samples were selected according to completely different criteria but based on properties that could correlate with their activity in radio. The first includes NLS1s residing in very dense Mpc-scale environments, such as superclusters \citep{2017jarvela1}, and for the second sample sources favorable to a 37~GHz detection were selected, according to their spectral energy distributions (SED). Among the two samples, eight sources were detected. In particular, seven of them were detected multiple times, showing flares at Jy-level flux densities \citep{Lahteenmaki18} with timescales of the order of days \citep{Jarvela23}. Such a huge bump in the amplitude was not expected since all of these sources had either weak or no known radio emission before the MRO detections. For one of the sources, \cite{Romano23} observed an X-ray brightening soon after an MRO detection. Moreover, the same source was identified as a new $\gamma$-ray emitter. From the literature, only two sources were detected at 1.4~GHz, at mJy levels, in the Faint Images of the Radio Sky at Twenty-Centimeters survey (FIRST, \citealp{1995becker1, 2015helfand1}), and the National Radio Astronomy Observatory Very Large Array Sky Survey (NRAO NVSS, \citealp{1998condon1}).

To untangle the situation, several follow-up radio observations have been carried out. In 2019 the sample was observed with the the Karl G. Jansky Very Large Array (JVLA) in A configuration at 1.6, 5.2, and 9.0~GHz \citep{Berton20b, Jarvela21}. Again in 2022, it was observed with the JVLA in A configuration at 10, 15, 22, 33, and 45~GHz and with the Very Long Baseline Array at 15~GHz \citep{Jarvela23}. In the meantime, single-dish monitoring has been ongoing. Besides the MRO monitoring, the sources are part of the Owens Valley Radio Observatory 40~m radio telescope AGN monitoring program at 15~GHz \citep{Jarvela23}. All the data acquired showed steep spectra up to 45~GHz, or no detection at all, especially at high radio frequencies. Although previous cases of such a high-frequency excess are known in the literature \citep{Antonucci88}, the variability observed, combined with the Jy-level flux densities, are rarely, if ever, seen in AGN. Because the observations are not simultaneous, the different beam sizes have to be considered, especially comparing single-dish with interferometric observations, with resolved-out structures in the latter. Although a minor difference can be present, resolved-out emission cannot explain such a huge difference in the flux density \citep{Jarvela23}. Moreover, contamination by close sources has been ruled out \citep{Lahteenmaki18,Jarvela23}. Even though these recent data helped in ruling out some hypotheses, they did not provide a unique and clear answer.

Some viable interpretations for this phenomenon were extensively discussed in \cite{Jarvela23}. One of the possibilities is that these NLS1s could harbor small-scale (likely a few parsecs) relativistic beamed jets, whose low-frequency radio emission is synchrotron self-absorbed or free-free absorbed by ionized gas. The origin of the ionization is unclear, but it could either be due to the abundant star formation often observed in NLS1s \citep{Sani10}, or by collision between the jet and the ISM surrounding the AGN, which produces via shocks a cocoon of ionized gas surrounding the jet head \citep{Bicknell97}. A detailed study of the radio spectral index maps of these sources revealed that the latter is actually the most likely mechanism to account for this unique behavior \citep{Jarvela21}. Jets-in-jets magnetic reconnection or magnetic reconnection in the black hole magnetosphere are not ruled out either. The latter is particularly interesting since it does not require the presence of relativistic jets.

The implications of this discovery are far-reaching because such large-amplitude variability, coupled with the very short timescales, has never before been observed for this class of sources. This study focuses on the optical spectra of the seven flaring NLS1s, comparing each other and understanding if and how they differ from the spectra of the general NLS1s population. Moreover, we calculated in several ways some of the most important physical parameters for NLS1s such as the broad-line region (BLR) radius, the black hole mass, the Eddington ratio, and the emission lines properties. This paper is organized as follows. In Sec.~\ref{sec:sample} we introduce the sample, in Sec.~\ref{sec:data_reduction} we outline the observations performed and the data reduction techniques, in Sec.~\ref{sec:spectral_analysis} we describe the emission lines analysis and how we calculated the physical parameters, in Sec.~\ref{sec:results} we present our results, in Sec.~\ref{sec:discussion} we discuss these results, and in Sec.~\ref{sec:conclusions} we provide a summary and the conclusions of this work. Throughout this work, we adopt a standard $\rm\Lambda CDM$ cosmology, with a Hubble constant $H_0 = 70$~km s$^{-1}$ Mpc$^{-1}$, and $\Omega_\Lambda = 0.73$ \citep{Komatsu11}.

\section{Sample}
\label{sec:sample}
As described before, the sample is composed of seven NLS1s detected in a flaring state at 37~GHz by the MRO (Tab.~\ref{tab:sample}). The sources span a redshift range from 0.0769 to 0.4511. Three out of seven are from the very dense Mpc-scale environment sample and the remaining four are from the SED-based sample. The radio-loudness parameters (RL), has been calculated using the integrated radio fluxes at 5.2~GHz reported in \cite{Berton20b}, and the optical fluxes in the B band. The latter are measured from the optical spectra, using the Johnson-B filter response curve which is centered at $\lambda_0$=4500\r{A} with a width of $\Delta \lambda$=1050\r{A}. All the sources are radio-quiet (Tab.~\ref{tab:sample}), even though as we already described they present strong flares which could suggest the presence of relativistic jets. Six sources are hosted in disk-like host galaxies, identified trough a photometric decomposition of their near-infrared images \citep{Jarvela18,Olguiniglesias20,Varglund22}, and in one case the host galaxy morphology is unknown. For a complete overview on the most recent radio properties of the sample, see \cite{Jarvela23}.

\section{Observations and data reduction}
\label{sec:data_reduction}
The observations were performed with the long-slit spectrometer Optical System for Imaging and low-Intermediate-Resolution Integrated Spectroscopy (OSIRIS) mounted on the Gran Telescopio Canarias (GTC) as part of the programme GTC26-22B (P.I. E. J\"arvel\"a). The observations were performed using a 0.6 arcsec wide slit, and two different grisms, R1000B and R1000R, to widen the spectral range from 3630\r{A} to 10'000\r{A}. The two grisms have a spectral resolution of 1018 and 1122, respectively, and are thus sufficient to sample the main emission lines in the spectra properly. For all the observations, we carried out the standard data reduction using the software Image Reduction and Analysis Facility (IRAF), applying bias and flat-field corrections, and wavelength and flux calibrations. For each source we combined all the single spectra, to maximize the signal-to-noise ratio (S/N), and the final spectra of the two grisms, to obtain the final extended spectrum. However, we were not able to combine the spectra from the two grisms for all sources as in some cases one of the two spectra was too noisy or with bad features, decreasing the quality of the combined spectrum. In those cases, we decided to keep only the best spectrum for the analysis, acquired with either R1000B or R1000R grism.

We corrected the spectra for galactic absorption using the Cardelli, Clayton and Mathis extinction law with $R_\nu=3.1$ \citep{Cardelli89}, which relies on the total extinction parameter, \textit{A(V)}. Such parameter is proportional to the column density of neutral hydrogen atoms, $N_H$, by a factor $5.3\times10^{-22}$. We derived $N_H$ by means of H~I profiles templates\footnote{https://www.astro.uni-bonn.de/hisurvey/AllSky\_profiles/index.php} \citep{2005A&A...440..775K}. The total extinction parameters are shown in Tab.~\ref{tab:spectra}.

We tried to apply the host galaxy subtraction by using a technique based on the principal component analysis \citep{Lamura07, Chen18}. However, the shape of the continuum is affected by the combination of the two grisms, and by the contamination of the second spectral order in the R1000R grism. This prevented us from obtaining reliable models of the host spectra. Furthermore, the spectral range covered by only one of the two grisms is also insufficient to perform accurate host modeling. Therefore, we decided to proceed without subtracting the host galaxy. It is worth noting that, in fact, the host contribution is relatively small in sources with z$>$0.1 \citep{Letawe07}, and our analysis mostly focuses on emission lines, which are only marginally affected by the host contribution.

For the iron subtraction, we used the templates available on the Serbian Virtual Observatory\footnote{http://servo.aob.rs/FeII\_AGN/} \citep{2010ApJS..189...15K, 2012ApJS..202...10S}. It produces a dedicated iron template with several Fe~II lines between 4000\r{A} and 5500\r{A}, according to the source's spectral features. By means of this template, we computed the parameter \textit{R}4570=\textit{F}(Fe~II$\lambda$4570)/\textit{F}(H$\beta$) which is shown in Tab.~\ref{tab:spectra}.

\begin{table*}
    \caption{Properties of the sample.}
    \centering
    \begin{tabular}{l c c c c c c c} \\
\hline \hline
SDSS name & Alias & RA & Dec & $z$ & Sample & RL & Host\\
\tiny{(1)} & \tiny{(2)} & \tiny{(3)} & \tiny{(4)} & \tiny{(5)} & \tiny{(6)} & \tiny{(7)} & \tiny{(8)} \\
\hline
J102906.69+555625.2 & J1029 & 10 29 06.69 & +55 56 25.25 & 0.4511 & D-env & -- & -- \\
J122844.81+501751.2 & J1228 & 12 28 44.82 & +50 17 51.24 & 0.2627 & D-env & 2.4 & D$^{(3)}$ \\
J123220.11+495721.8 & J1232 & 12 32 20.12 & +49 57 21.82 & 0.2625 & D-env & 0.1 & D$^{(3)}$ \\
J150916.18+613716.7 & J1509 & 15 09 16.17 & +61 37 16.80 & 0.2012 & SED & -- & D$^{(3)}$ \\
J151020.06+554722.0 & J1510 & 15 10 20.05 & +55 47 22.11 & 0.1497 & SED & 0.5 &  D,b$^{(1)}$ \\
J152205.41+393441.3 & J1522 & 15 22 05.50 & +39 34 40.45 & 0.0769 & SED & 1.7 & D,b,PB$^{(1)}$,merger \\
J164100.10+345452.7 & J1641 & 16 41 00.10 & +34 54 52.67 & 0.1640 & SED & 8.0 & D$^{(2)}$ \\
\hline
    \end{tabular}
    \tablefoot{Columns: (1) SDSS source name; (2) short name; (3) right ascension [hh mm ss.s] (J2000); (4) declination [dd mm ss.s] (J2000); (5) spectroscopic redshift; (6) sample of origin of the source. D-env: very dense Mpc-scale environment sample, SED: SED sample; (7) radio loudness parameter \citep{Kellermann89}; (8) host galaxy morphology, composed by: D=disk, b=bar, PB=pseudo-bulge ($^{(1)}$\citealp{Jarvela18}, $^{(2)}$\citealp{Olguiniglesias20}, and $^{(3)}$\citealp{Varglund22}).}
    \label{tab:sample}
\end{table*}

\section{Spectral analysis}
\label{sec:spectral_analysis}
 
\subsection{Line profiles}
To extract the information from the spectra, we fit the line profiles of the main emission lines with several models, using our own \texttt{python3} code. The S/N of the spectra measured at 5100\r{A} spans between 12 and 65 (Tab.~\ref{tab:spectra}). We analyzed the most prominent emission lines, such as H$\beta$, [O~III]$\lambda\lambda$4959,5007, H$\alpha$+[N~II]$\lambda\lambda$6548,6583 and, when visible, the [S~II]$\lambda\lambda$6716,6731. We measured the spectroscopic redshift using all the forbidden lines present in the spectra, and we double-checked the results with the permitted lines. Even though low-ionization lines are usually less perturbed than high-ionization lines \citep{Komossa08}, and thus more suitable for redshift measurements, the redshifts we obtained are comparable to those already reported in the literature.

\begin{figure}
\begin{center}
\includegraphics[trim={0.5cm 0 0 0}, clip, width=1.1\columnwidth]{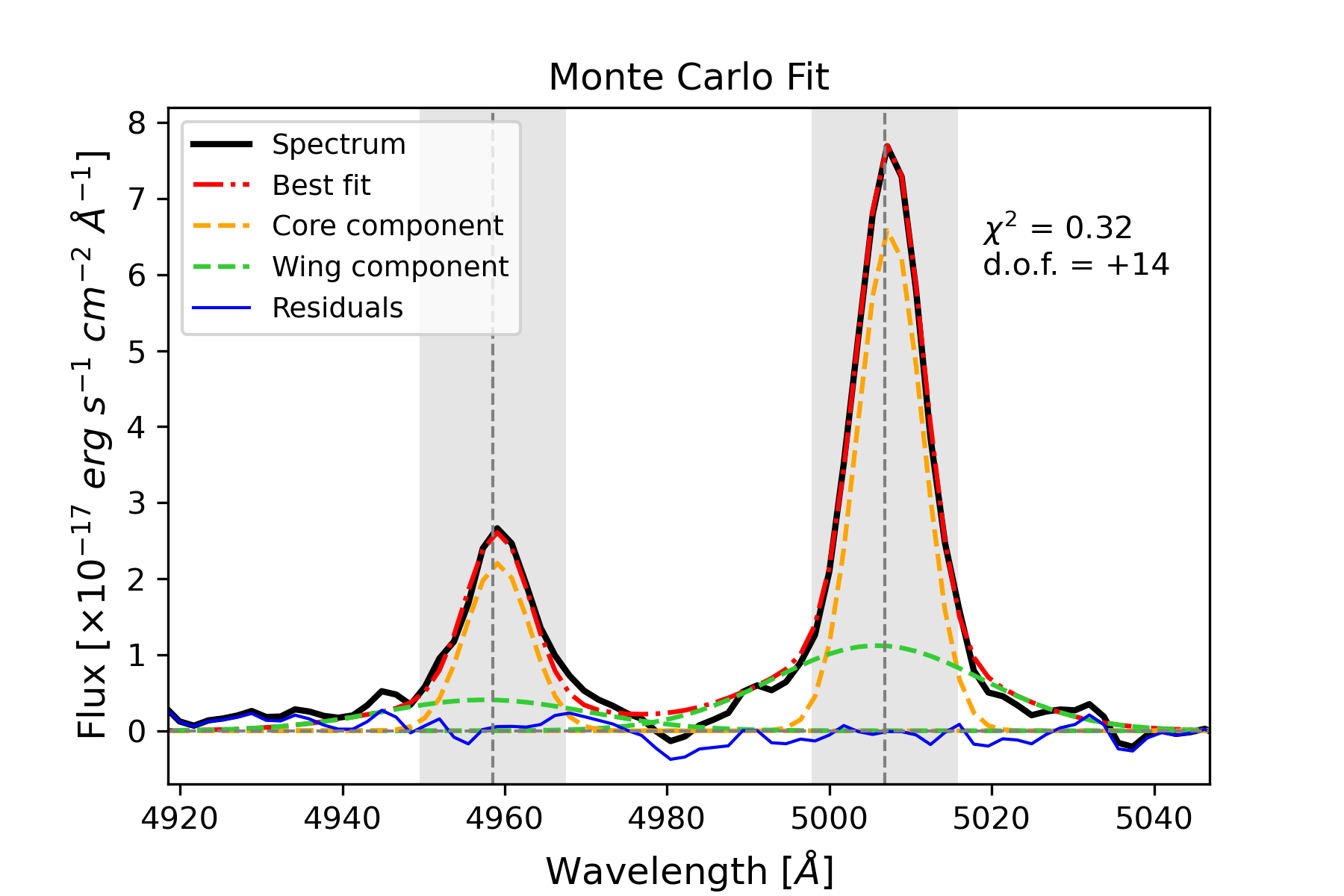}
\caption{[O~III]$\lambda\lambda$4959,5007 emission lines of J1641. The light gray shadows represent the ranges where the $\chi^2$ is assessed. 
\label{fig:J1641_GoF_o3}}
\end{center}
\end{figure}

The first lines we modeled were the [O~III]$\lambda\lambda$4959,5007. Usually in NLS1s the [O~III] lines show two main components. A core component, produced by the narrow-line region (NLR) gas which is at the same redshift as the host galaxy, and a wing component, associated with outflowing gas, on the blue side of the line core. Since the outflowing gas clouds can have different velocities, multiple components can be necessary to properly fit their emission profile. Therefore, we fit each [O~III] with two or three Gaussians, one for the core component and one or two for the wing component. To reduce the number of free parameters, we tied certain parameters of the [O~III]$\lambda$4959 with those of the [O~III]$\lambda$5007. In particular, we constrained the central position to the rest frame wavelength and the FWHM, corrected for instrumental resolution, of all the components. Moreover, the flux ratio between the components of the two lines was fixed to the theoretical value of 1/3 \citep{2007MNRAS.374.1181D}. Due to the spectra being quite noisy, a goodness-of-fit test, such as the $\chi^2$, needs particular attention to deliver reliable results, especially when multiple emission lines are fit simultaneously. For [O~III] lines we set an interval around the rest frame position of the two lines, in which we computed the $\chi^2$, avoiding the non-fitted region between the two lines that would inevitably increase the $\chi^2$ value. Such intervals have been calculated as a multiple of the FWHM of the two [O~III] lines, around the central wavelengths of both lines (Fig.~\ref{fig:J1641_GoF_o3}). Since the outflow emission, when present, is on the blue side of the lines, the bluer side of the interval is often larger compared to the redder side (see Tab.~\ref{tab:spectra}). Finally, exploiting the [O~III]$\lambda$5007 flux, we measured R5007=\textit{F}([O~III]$\lambda$5007)/\textit{F}(H$\beta$) (Tab.~\ref{tab:spectra}).

To fit \hb\ we tried several models for its line profile, with combinations of Gaussian and Lorentzian functions. Overall the best ones turned out to be the two Gaussians model (2G model), the three Gaussians model (3G model) and the Lorentzian+Gaussian model (LG model). One of the Gaussians in the 2G and 3G models and the only Gaussian in the LG model represent the narrow component of the H$\beta$. We constrained their central position to the H$\beta$ rest frame wavelength and their FWHM, corrected for instrumental resolution, with the FWHM of the [O~III]$\lambda$5007 core component, leaving the amplitude free to vary. We tied the FWHMs since the narrow component of the Balmer lines is supposed to come from the same region, i.e. NLR, and thus with the same velocity dispersion, as the core component of the forbidden lines. Besides the narrow component, the Gaussians left, one in the 2G model and two in the 3G model, represent the broad emission of H$\beta$. In the 3G model, we had to adopt two Gaussians instead of one, due to the impossibility of properly fitting the wings of the line profile with only one function. In the LG model instead, the broad emission is represented by a Lorentzian function, which is often a good representation of the BLR in NLS1s \citep{Cracco16}. For the fitting functions of the broad component of H$\beta$, we calculated the second-order moment, defined as

\begin{equation}
    \sigma^2 = \frac{\int \lambda^2 F_{H\beta_b}(\lambda) d\lambda}{\int F_{H\beta_b}(\lambda) d\lambda} - \left(\frac{\int \lambda F_{H\beta_b}(\lambda) d\lambda}{\int F_{H\beta_b}(\lambda) d\lambda} \right)^2 \; .
    \label{eq:sec-mom}
\end{equation}
This parameter has been calculated for models built using only Gaussian functions, since for a Lorentzian function it is, by definition, equal to infinite.

When visible we also modeled the [S~II]$\lambda\lambda$6716,6731. We fitted each [S~II] profile with a Gaussian function. To reduce the number of free parameters, we constrained the FWHM and the shift, with respect to the rest frame position, of the Gaussian for [S~II]$\lambda$6716 with those retrieved with the Gaussian for [S~II]$\lambda$6731, leaving the flux free to vary.

Finally, we fitted the complex line profile of H$\alpha$+[N~II]$\lambda\lambda$6548,6583, which, as stated before, are blended together. For the H$\alpha$ line profile, we applied the same model that was used for H$\beta$ (2G, 3G or LG model), with the same FWHM, corrected for instrumental resolution, and the respective shifts for all the components. For the [N~II]$\lambda\lambda$6548,6583 we adopted a 2G model, in which we constrained the position to the rest frame wavelength, and the FWHM using [S~II] or [O~III] width. This is particularly important since the [N~II] are completely blended with H$\alpha$, and therefore impossible to model without any priors. We preferred [S~II] over [O~III], when possible, as a reference for [N~II], because they are always observed with the same grism, and they are both low-ionization lines. We also fixed the flux ratio of the two Gaussians to the value 1/3.049 \citep{2023AdSpR..71.1219D}. To investigate the internal extinction, we exploited the Balmer decrement calculating the $\mathcal{R}$ ratio, defined as the ratio between the H$\alpha$ and H$\beta$ narrow component fluxes. Following \cite{Cardelli89}, and assuming a theoretical ratio of 2.86, the internal extinction can be expressed as

\begin{equation}
    A(V) = 7.215 \log \left( \frac{\mathcal{R}}{2.86} \right) \; .
    \label{eq:A(V)}
\end{equation}

Once the best-fit model has been selected for all the lines, based on the $\chi^2$ and on a visual inspection, we performed a Monte Carlo method, repeating the fit one thousand times. For each iteration, we added a Gaussian noise to the line profile, proportional to the standard deviation of the signal in the continuum between 5050\r{A} and 5150\r{A}. Using a high number of iterations, for each parameter we measured the mean value $\overline{X}$ and the standard error $err_{\overline{X}}$ \citep{Ahn&Fessler03} defined as

\begin{equation}
    \overline{X} \pm err_{\overline{X}} = \frac{1}{N} \sum_{i=1}^N X_i \pm \sqrt{\frac{\sigma^2_N}{N}} \; ,
\end{equation}
where $N$ is the number of Monte Carlo iterations, $X_i$ the \textit{i}-th value of the parameter, and $\sigma^2_N$ the variance of the parameter expressed as

\begin{equation}
    \sigma^2_N = \frac{\sum_{i=1}^N (X_i-\overline{X})^2 }{N-1} \; .
\end{equation}

\subsection{BLR radius}
\label{sec:BLR_radius}

We can reasonably assume that the BLR is in a photoionization equilibrium state. The photoionization degree depends on the intensity of the photoionizing radiation, which is proportional to the intensity of all the radiation produced, for AGN, by the accretion flow. Likewise, for the equilibrium assumption, this translates into the intensity of the emission lines produced in the BLR. Furthermore, there is an empirical relation which binds the BLR radius with the luminosity of the accretion disk, such as the 5100\r{A} continuum, or with the luminosity of emission lines that come from the BLR, such as H$\beta$. Exploiting H$\beta$ emission, \cite{Greene10} derived a relation for the BLR radius, expressed as

\begin{equation}
    \log \left( \frac{R_{\rm{BLR}}}{\mathrm{l.d.}} \right) = (1.85 \pm 0.05) + (0.53 \pm 0.04) \log \left( \frac{L(\rm{H}\beta)}{10^{43}~\mathrm{erg~s^{-1}}} \right) \; ,
    \label{eq:Rblr_Hbeta}
\end{equation}
where $L$(H$\beta)$ is the integrated luminosity of the line profile, and the BLR radius is expressed in light days. 
A similar relation has been found by \cite{Bentz13} using the continuum emission at 5100\r{A}. As mentioned before, the continuum has a small contribution coming from the host galaxy, but we expect it to be inside the flux calibration error and therefore negligible. 
\cite{Bentz13} derived the coefficients of such relation, relying on reverberation mapping data, which is expressed as

\begin{equation}
    \log \left( \frac{R_{\rm{BLR}}}{\mathrm{l.d.}} \right) = (1.53 \pm 0.03) + (0.53 \pm 0.03) \log \left( \frac{\lambda L_{\lambda}(5100\textrm{\r{A}})}{10^{44}~\mathrm{erg~s^{-1}}} \right) \; ,
    \label{eq:Rblr_cont}
\end{equation}
with the BLR radius expressed in light days.
Recently, an implementation of Eq.~\ref{eq:Rblr_cont}, for AGN with high Eddington ratios, such as NLS1s, has been derived \citep{Du19, Paliya24}. Indeed, \cite{Du19} found that the $R_{\rm{H\beta}}$-$L_{5100}$ relationship \citep{Kaspi00,Bentz09a,Bentz13} diverges from the more precise results obtained through the most recent reverberation mapping data. By means of the sample analyzed, they found a strong dependence on Fe~II$\lambda$4570 emission, modifying Eq.~\ref{eq:Rblr_cont} as

\begin{equation}
\begin{aligned}
    \log \left( \frac{R_{\rm{BLR}}}{\mathrm{l.d.}} \right) = & ~(1.65 \pm 0.06) + (0.45 \pm 0.03) \log \left( \frac{\lambda L_{\lambda}(5100\textrm{\r{A}})}{10^{44}~\mathrm{erg~s^{-1}}} \right) \\
    & - (0.35 \pm 0.08)~R4570 \; ,
    \label{eq:Rblr_Fe}
\end{aligned}
\end{equation}
where \textit{R}4570 is the ratio described in Sec.~\ref{sec:data_reduction}. Therefore we calculated the BLR radius through Eq.~\ref{eq:Rblr_Fe} whenever the \textit{R}4570 was available. Only for one source the Fe~II$\lambda$4570 emission was not measurable. In this case we calculated the BLR radius using Eq.~\ref{eq:Rblr_Hbeta} and \ref{eq:Rblr_cont}. For all the formulas described we performed an error propagation for the uncertainties estimation.
It is important to remember that all the cited equations for the BLR radius calculation were derived empirically, from real data. Therefore, they are never free from biases.

\subsection{Black hole mass}

Assuming virialized gas orbiting the black hole, the virial theorem can be used to express the black hole mass as

\begin{equation}
    M_{\rm{BH}} = f\frac{R_{\rm{BLR}} v^2}{G} \; ,
    \label{eq:M_BH}
\end{equation}
where $R_{\rm{BLR}}$ is the radius of the BLR, $\nu$ is the rotational velocity of the gas, \textit{G} is the gravitational constant, and $f$ is the so-called scaling factor. In Eq.~\ref{eq:M_BH}, the two main unknowns are the BLR radius and the velocity. These quantities can be derived in many different ways. In the case of the BLR radius, we already described in the previous section the three types of relations that can be used starting from different observables, namely H$\beta$ luminosity (Eq.~\ref{eq:Rblr_Hbeta}), 5100\r{A} continuum luminosity (Eq.~\ref{eq:Rblr_cont}), and 5100\r{A} continuum luminosity and \textit{R}4570 (Eq.~\ref{eq:Rblr_Fe}). Two independent parameters can also be used as a proxy for the rotational velocity \textit{v}. The first is the FWHM of the H$\beta$ broad component. We derived it using the functions in the line fitting process, which modeled only the broad emission of H$\beta$. In particular, the cited functions were a Gaussian for the 2G model, two Gaussians for the 3G model, and a Lorentzian for the LG model. Alternatively, we adopted the second-order moment (Eq.~\ref{eq:sec-mom}) as a proxy for \textit{v}. As mentioned before, we did not calculate such a parameter for the LG model, as it is equal to infinite for the Lorentzian function. Several studies emphasized how the second-order moment is more reliable than the FWHM(H$\beta_b$) as a proxy for the rotational velocity, being less affected by inclination effects and the BLR geometry \citep{Peterson04, Peterson11, Peterson18}. Nevertheless, we decided to use both methods, reducing the probability of systematics and biases that each single assumption can have.

The parameter \textit{v} represents the rotational velocity of the gas in the BLR, while the FWHM(H$\beta_b$) is instead an observable of the velocity dispersion. Such difference is included in Eq.~\ref{eq:M_BH} by the $f$ factor. The black hole mass formula defined before is a theoretical relation for a completely virialized gas in a perfect Keplerian motion, a case that basically never occurs in AGN. The main sources of uncertainty are the BLR geometry and its inclination. It is clear that the measurable velocities for a flat BLR, seen face-on or edge-on, are different. In the former, the rotational velocity can even be close to zero, leading to an underestimation of the black hole mass \citep{Decarli08}. However, although not negligible, in NLS1s the inclination effect might be not so significant \citep{Vietri18, Berton20a}, since a significant vertical structure in the BLR can be present \citep{Kollatschny11, Kollatschny13a}. On the other hand, a spherical BLR geometry is even harder to handle with just a relatively simple formula. The $f$ factor accounts for the differences between the theoretical formula and the actual black hole mass, by correcting the rotational velocity observables. The most recent knowledge shows that a Keplerian motion of the BLR clouds is present \citep{Peterson99, Gravity18}, with possibly additional components such as turbulent vertical motions originating in a disk wind \citep{Gaskell09, Kollatschny13a}. Moreover, it has been found that $f$ is inversely proportional to the FWHM(H$\beta$) \citep{Shen14}. \cite{Mandal21} estimated a typical range for the $f$ factor between 0.8 and 5, mostly derived from a comparison with reverberation mapping observations.
For this work, we decided to use the $f$ factors estimated by \cite{Collin06} which are $f_{\sigma}$=3.93 and $f_{FWHM}$=2.12, for the rotational velocity obtained by the second-order moment and the FWHM(H$\beta_b$), respectively.

Summarizing, when the H$\beta$ profile was modeled by either the 2G or the 3G model, we got two different black hole mass estimates, using Eq.~\ref{eq:Rblr_Fe} for the BLR radius, and considering the ways in which we calculated the \textit{v} parameter. When we applied the LG model instead, we got the two black hole mass estimates using only the FWHM(H$\beta_b$) as a proxy for the rotational velocity, and Eq.~\ref{eq:Rblr_Hbeta} and \ref{eq:Rblr_cont} for the BLR radius. In all cases to retrieve the associated error for each black hole mass, we applied a standard error propagation. 

\subsection{Eddington ratio}
\label{sec:Edd_ratio}

For AGN, the Eddington ratio is defined as

\begin{equation}
    \epsilon = \frac{L_{\rm{bol}}}{L_{\rm{Edd}}} = \frac{L_{\rm{bol}}}{1.3\times10^{38}M_{\rm{BH}}/M_{\odot}} \; ,
    \label{eq:Edd}
\end{equation}
where $L_{bol}$ is the bolometric luminosity, and $L_{Edd}$ is the Eddington luminosity. Objects with high $\epsilon$, even higher than 1, usually show strong Fe~II multiplets and narrow H$\beta$, while low Eddington objects show broader H$\beta$ and a weak Fe~II emission. Such differences also translate to a different position in the quasar main sequence \citep{Marziani18b}. For NLS1s a typical Eddington ratio is between 0.1 and 1 \citep{Boroson92, Williams02, Williams04, Grupe10, Xu12}, but sometimes even super-Eddington accretion has been observed \citep{Chen18, Tortosa22}. The bolometric luminosity can be derived by exploiting simple relations with observables, such as the 5100\r{A} continuum luminosity formula \citep{Netzer19} expressed as

\begin{equation}
    L_{\rm{bol}} = k_{bol}\times \lambda L_\lambda(5100\textrm{\r{A}}) \; ,
    \label{eq:L_bol}
\end{equation}
where $k_{bol}$ is the bolometric correction factor defined as

\begin{equation}
    k_{bol} = 40 \left[ \frac{\lambda L_\lambda(5100\textrm{\r{A}})}{10^{42}~\mathrm{erg~s^{-1}}} \right]^{-0.2} \; .
    \label{eq:k_bol}
\end{equation}
The main uncertainty of $k_{bol}$ comes from the inclination of the AGN accretion disk with respect to the line of sight. \cite{Netzer19} estimated that for type-1 AGN the bolometric correction factor decreases by a factor of $\sim$ 1.4 on average, and by a factor of $\sim$ 2.5 for face-on accretion disks. Since the inclination of the analyzed sources is unknown, and consequently to account for all the possible inclinations, we decided to take $\frac{k_{bol}}{2}$ as bolometric correction factor, and $\Delta k_{bol}=\frac{k_{bol}}{2}-\frac{k_{bol}}{2.5}$ as its uncertainty.
In general, Eq.~\ref{eq:L_bol} and \ref{eq:k_bol} are particularly affected by the jet presence and can be partially affected by the host galaxy contribution, since both these components can contribute to the continuum luminosity. However, in our case, the presence of the jet is still debated, and the host component is only marginal, therefore we decided to keep such relation to estimate $\epsilon$. Alternatively, we also used an approach less affected by non-nuclear contributions. 
We retrieved the 5100\r{A} continuum luminosity for Eq.~\ref{eq:L_bol} and \ref{eq:k_bol} using the $L_\lambda(5100\textrm{\r{A}})$-\textit{L}(H$\beta_b$) relation \citep{Ilic17,DallaBonta20} defined as

\begin{equation}
\begin{aligned}
    \log (\lambda L_\lambda(5100\textrm{\r{A}}) ) = & (43.396\pm0.018) + (1.003\pm0.022) \times \\
    & [\log(L(\rm{H}\beta_b)) - 41.746] \; .
    \label{eq:L5100_der}
\end{aligned}
\end{equation}
This approach allowed us to indirectly calculate the continuum luminosity, and consequently the bolometric luminosity, exploiting the H$\beta$ properties. Its flux is directly proportional to the ionizing continuum of the AGN, which is free from jet and host contamination. For the Eddington ratio calculation we averaged the black hole mass of each source. Also in this case we applied a proper error propagation for the error calculation through all the described steps.

\begin{table*}
    \caption{Observational parameters derived from the optical spectra.}
    \centering
    \begin{tabular}{l c c c c c c c c c c c} \\
\hline \hline
Source & S/N & [O~III] $\chi^2_{range}$ & FWHM(H$\beta$) & FWHM(H$\beta_b$) & $\sigma$ & \textit{F}(H$\beta$) & \textit{F}(H$\beta_b$) & \textit{F}$_{cont}$ & $F_{SDSS}$ & \textit{R}4570 & A(V) \\
\tiny{(1)} & \tiny{(2)} & \tiny{(3)} & \tiny{(4)} & \tiny{(5)} & \tiny{(6)} & \tiny{(7)} & \tiny{(8)} & \tiny{(9)} & \tiny{(10)} & \tiny{(11)} & \tiny{(12)} \\
 \hline
J1029 & 12 & -2/+1 & 2132 & 2320 & 2856 & 1.39 & 1.38 & 2.74 & 4.85 & 1.34 & -- \\
J1228 & 65 & -3/+2 & 971  & 1360 & 1807 & 1.97 & 1.88 & 13.96 & 14.66 & 2.81 & 3.590 \\
J1232 & 33 & -2/+1 & 1518 & 1700 & 1631 & 8.14 & 8.01 & 17.67 & 15.95 & 1.58 & 1.826\\
J1509 & 26 & -2/+1 & 1744 & 1861 & 2193 & 2.20 & 2.18 & 5.77 & 8.19 & 1.06 & 3.067 \\
J1510 & 14 & -3/+2 & 1031 & 1193 & 2307 & 1.99 & 1.87 & 8.09 & 12.97 & 2.37 & 3.764 \\
J1522 & 42 & -2/+1 & 886  & 1181 & --   & 1.34 & 1.13 & 17.48 & 33.87 & --   & 3.285  \\
J1641 & 25 & -1/+1 & 1140 & 4050 & 1767 & 0.90 & 0.79 & 10.48 & 21.23 & 2.58 & -- \\
\hline
    \end{tabular}
    \tablefoot{Columns: (1) Source name; (2) S/N measured in the 5100\r{A} continuum;
    (3) values multiplied to the FWHMs of [O~III]$\lambda\lambda$4959,5007 for the $\chi^2$ calculation; (4) FWHM of the H$\beta$ profile [\kms]; (5) FWHM of the H$\beta$ broad component [\kms]; (6) square root of the second-order moment as in Eq.~\ref{eq:sec-mom} [\kms]; (7) H$\beta$ flux [×10$^{-15}$ \ergs cm$^{-2}$]; (8) H$\beta$ broad flux [×10$^{-15}$ \ergs cm$^{-2}$]; (9) flux density at 5100\r{A} [×10$^{-17}$ \ergs cm$^{-2}$ \r{A}$^{-1}$]; (10) flux density at 5100\r{A} on SDSS spectra [×10$^{-17}$ \ergs cm$^{-2}$ \r{A}$^{-1}$]; (11) flux ratio between Fe~II$\lambda$4570 multiplets and H$\beta$; (12) internal extinction as in Eq.~\ref{eq:A(V)}.}
    \label{tab:spectra}
\end{table*}

\section{Results}
\label{sec:results}

\subsection{SDSS J102906.69+555625.2}
J1029, with a redshift of 0.4511, is the farthest source of the sample. Due to its distance, the H$\alpha$ and [S~II] region is inside the portion of the spectrum contaminated by the sky emission. Therefore, in this case, we only managed to analyze H$\beta$ and [O~III]$\lambda\lambda$4959,5007 emission lines. Moreover, due to a bad feature on the bluer part of the spectrum, probably of instrumental origin, we kept only the R1000R grism spectrum. The fitting models we adopted were the 3G model for H$\beta$ and the four Gaussians model for the [O~III]$\lambda\lambda$4959,5007 since these last do not show any asymmetric shape. This source, with values of 2132 \kms\ and 2856 \kms, has the highest FWHM(H$\beta$) and $\sigma$ respectively. Its FWHM(H$\beta_b$) is 2320 \kms, which means that it is not formally an NLS1 but it still shares all the properties of Population A sources. In turn, averaging the results in Table~\ref{tab:results}, it has the highest black hole mass in the sample, equal to (6.63$\pm$1.88) ×10$^7$ M$_{\odot}$. Such characteristics, coupled with a moderate Eddington ratio of 0.07 on average (Table~\ref{tab:results}), could be traits of an NLS1 in an evolved stage, possibly transitioning into a classical broad-line AGN.

\subsection{SDSS J122844.81+501751.2}
\begin{figure}
\begin{center}
\includegraphics[trim={0.5cm 0 0 0}, clip, width=1.1\columnwidth]{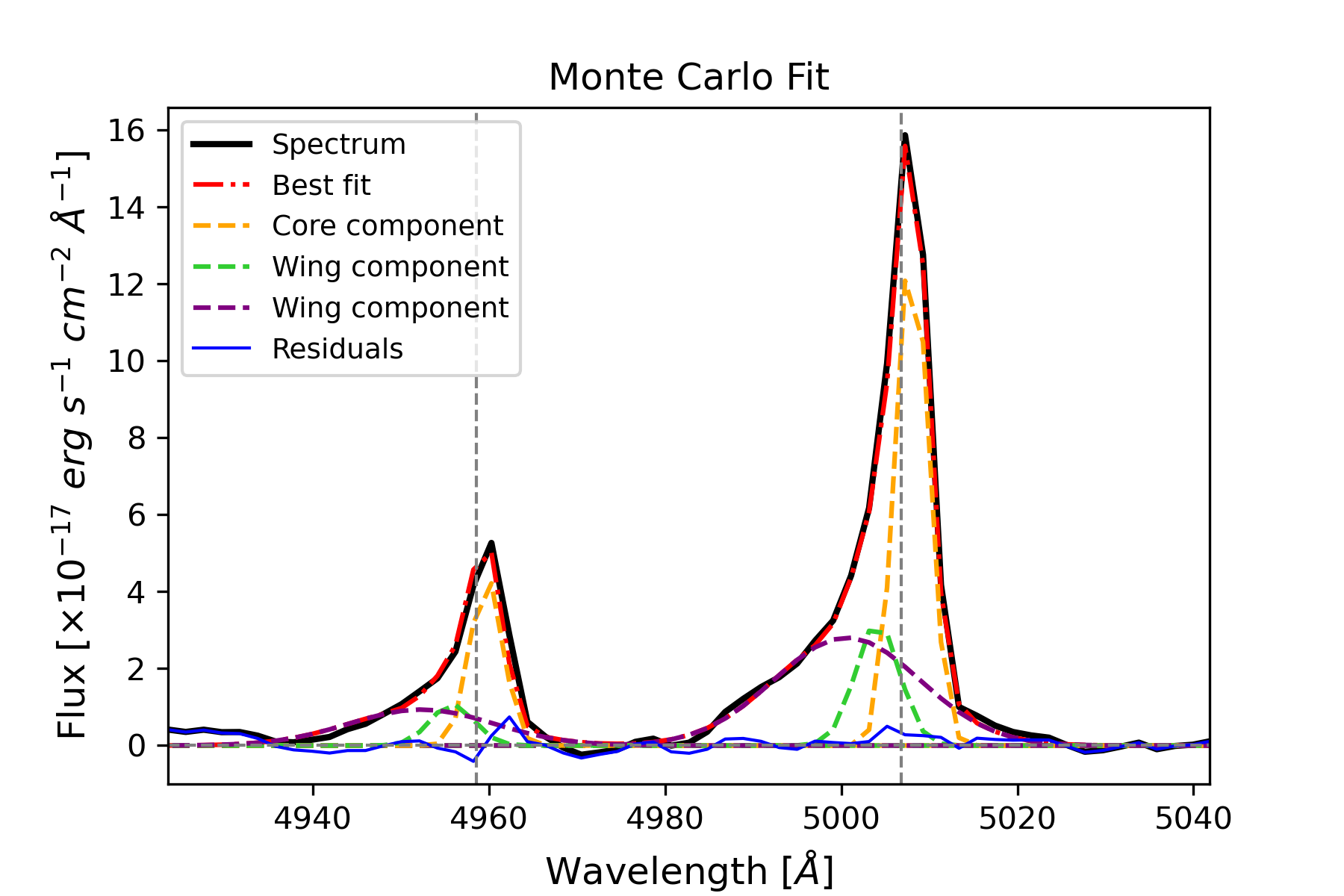}
\caption{[O~III]$\lambda\lambda$4959,5007 line profiles of J1228. 
\label{fig:J1228_o3}}
\end{center}
\end{figure}

Also in this case we kept only the R1000R grism spectrum, since the R1000B grism spectrum had a much worst S/N, making difficult, among all, the modeling of the continuum. For the emission lines the applied fitting models were the 3G model, for H$\beta$ and H$\alpha$, and the six Gaussians model, for the [O~III]$\lambda\lambda$4959,5007. Thanks to an S/N of 65, the highest of the sample, the asymmetric profiles on the two [O~III] emission lines are well visible (Fig.~\ref{fig:J1228_o3}). Therefore, six Gaussian functions were necessary to properly fit such profiles. In NLS1s, classified as high-accretion sources, the [O~III] blue wing is associated with powerful outflows that can be generated by the radiation pressure coming from the accretion disk or a jet \citep{Proga00, Greene05}. The wing components, usually broader than the core component, are likely produced in the innermost part of the NLR \citep{Berton16b}. In our case, the two Gaussians used to model the wing components are shifted of $-$163 and $-$371 \kms\ toward the blue side. The average mass of the black hole is (9.21$\pm$4.96) ×10$^6$ M$_{\odot}$, and the calculated Eddington ratios are 0.94$\pm$0.19 and 0.11$\pm$0.02. In agreement with these values (see Sec.~\ref{sec:R4570-Edd_ratio} for the discussion) the \textit{R}4570 is the highest in the sample (2.81), suggesting strong Fe~II multiplets emission.

\subsection{SDSS J123220.11+495721.8}
\label{sec:J1232}
\begin{figure}
\begin{center}
\includegraphics[trim={0.5cm 0 0 0}, clip, width=1.1\columnwidth]{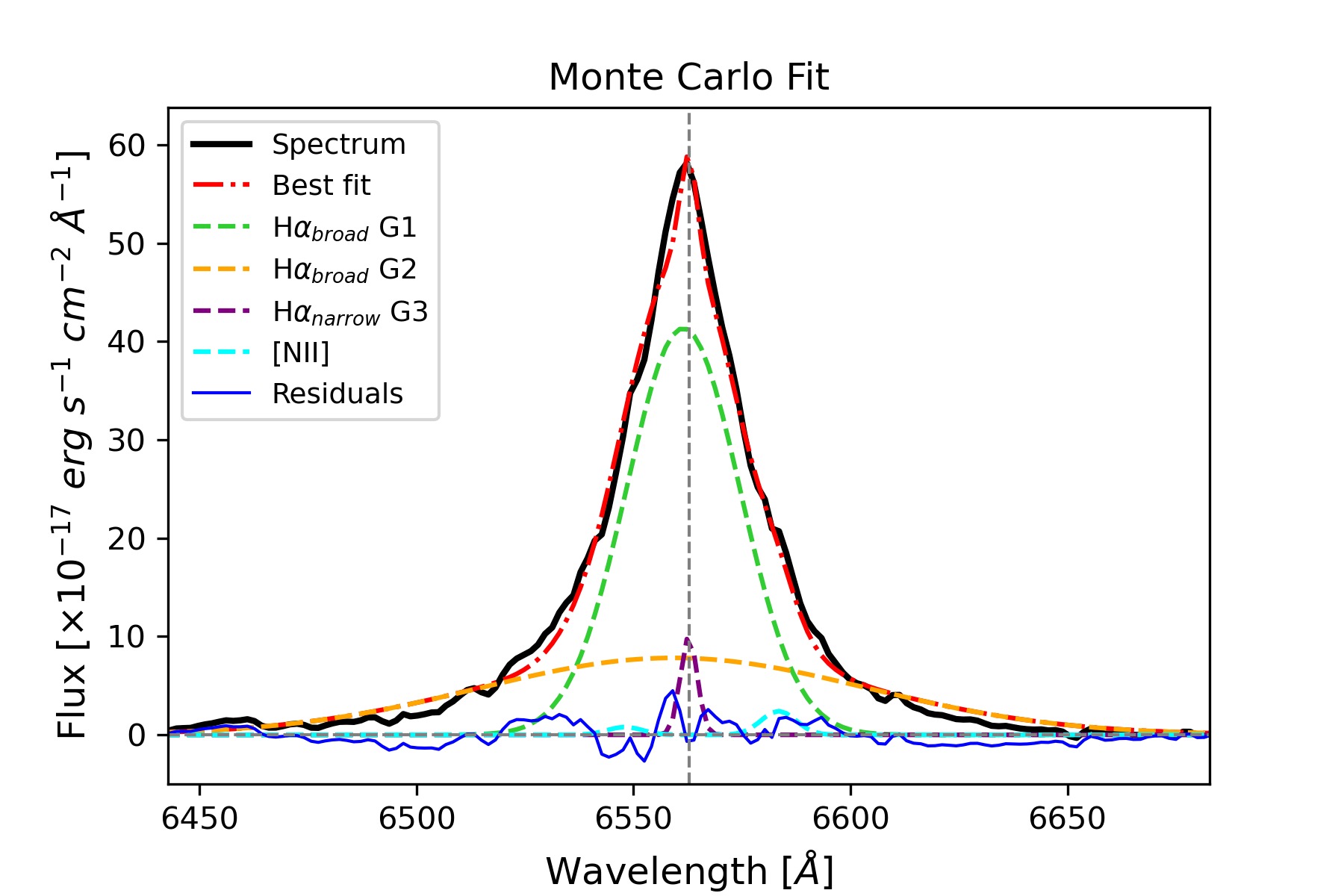}
\caption{H$\alpha$+[N II]$\lambda\lambda$6548,6583 lines profile of J1232. The cyan dashed line represents the [N~II] line profiles, which are much fainter than the H$\alpha$ emission. \label{fig:J1232_Halpha}}
\end{center}
\end{figure}

\begin{figure}
\begin{center}
\includegraphics[trim={0.5cm 0 0 0}, clip, width=1.1\columnwidth]{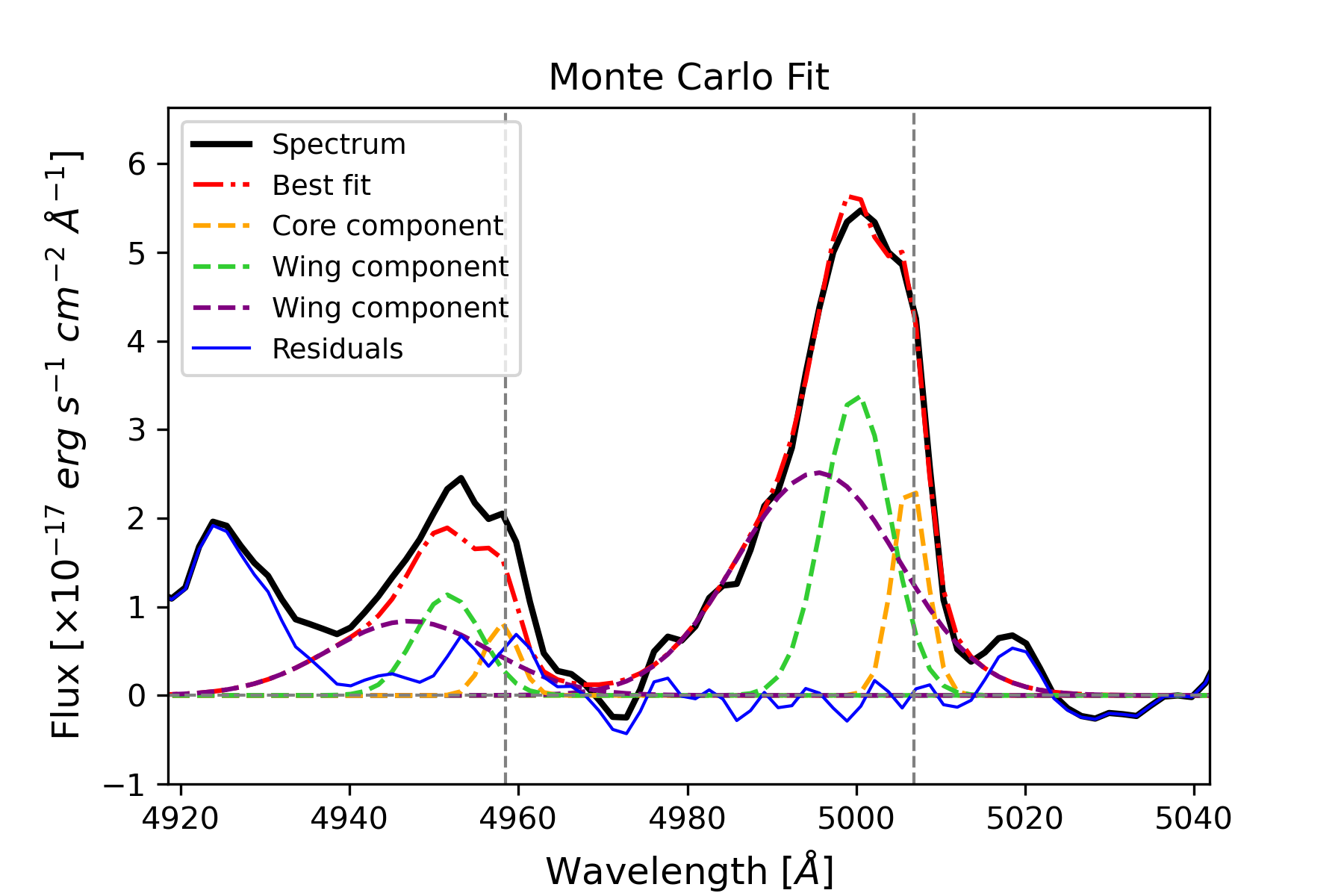}
\caption{[O~III]$\lambda\lambda$4959,5007 line profiles of J1232, which show wing components stronger than the core components. 
\label{fig:J1232_o3}}
\end{center}
\end{figure}

The emission line parameters for J1232 have been derived using the 3G fitting model for the H$\beta$ and H$\alpha$ lines, and the six Gaussians model for the [O~III]$\lambda\lambda$4959,5007. Along the whole spectrum, the permitted lines are much stronger than the forbidden lines (Fig.~\ref{fig:J1232_Halpha}). Moreover, the [O~III]$\lambda\lambda$4959,5007 profile shows core components fainter compared to the wing components (Fig.~\ref{fig:J1232_o3}). The two Gaussians used to model the blue side of the [O~III] lines are shifted about $-$411 \kms\ and $-$596 \kms. According to these, we can classify this source as a blue outlier \citep{Marziani03,Komossa08,Berton16b,Schmidt18}, following the criterion adopted by \cite{Zamanov02} ([O~III] wing component with a shift <$-$250 \kms). The origin of blue outliers could be due to a jet interacting with the NLR, as they correlate with the radio emission \citep{Berton21b}, but also due to winds produced by strong radiation pressure-driven outflows in a high-Eddington source \citep{Komossa08, Marziani16}. The latter is likely the case here since this is the source with the second highest Eddington ratio we found ($\sim$0.27).
Also in this case the \textit{R}4570=1.58 suggests strong Fe~II multiplets emission, especially considering the high flux of the \hb\ (8.14×10$^{-15}$ \ergs cm$^{-2}$), placing J1232 in the A4 population region of the quasar main sequence \citep{Sulentic15}. The mean black hole mass of (2.73$\pm$0.88) ×10$^7$ M$_{\odot}$ is well inside the range for NLS1s.

\subsection{SDSS J150916.18+613716.7}
For this source we used the 3G model for the fitting of the H$\beta$ and H$\alpha$ lines, and the six Gaussians model for the fitting of the [O~III]$\lambda\lambda$4959,5007 lines. Even thought the six Gaussians model was necessary to fit the [O~III] lines, they did not show strong asymmetric wings. The H$\beta$ line has a FWHM of 1744 \kms, inside the range for NLS1s. The mean black hole mass we obtained is (3.02$\pm$0.73) ×10$^7$ M$_{\odot}$ and the Eddington ratio is around 0.06. The \textit{R}4570 of 1.06 and the \hb\ flux of 2.20 ×10$^{-15}$ \ergs cm$^{-2}$, lowest and second highest values in the sample respectively, suggest faint Fe~II emission. It is worth noting that the faint iron emission in this case is only in comparison to the other sources in the sample, since all the sources for which we manage to measure the \textit{R}4570 belong to populations A3 and A4 of the quasar main sequence.

\subsection{SDSS J151020.06+554722.0}
With a black hole mass on average equal to (8.03$\pm$3.83) ×10$^6$ M$_{\odot}$, J1510 has the second least massive black hole in the sample, and an average Eddington ratio of 0.15. The applied fitting models to the spectrum were the 3G model for the H$\beta$ and H$\alpha$ lines, and the four Gaussians model for the [O~III]$\lambda\lambda$4959,5007. Even in this case, the [O~III] emission lines are quite symmetric, without strong evidence of outflows. A defect, possibly due to uncontrolled reflections inside the instrument, affected the blue side of the H$\beta$ profile. Therefore, the measured FWHM(H$\beta$) of 1031 \kms, despite already quite narrow, might have been slightly overestimated, causing an overestimation of the black hole mass. The second-order moment of the line, equal to 2307 \kms, is roughly twice the FWHM(H$\beta$), which can be explained by the overestimated width of the line.

\subsection{SDSS J152205.41+393441.3}
J1522 is the closest source, with a redshift of 0.0769. As for J1228, we used only the R1000R grism spectrum, because in comparison, the R1000B grism spectrum had an 8 to 10 times worse S/N. Here the [S~II] lines were too faint to be distinguished from the noise, therefore we used the FWHM of [O~III]$\lambda$5007 core component as a reference for the FWHM of [N~II] lines. The best fitting models turned out to be the four Gaussians model for the [O~III]$\lambda\lambda$4959,5007, and the LG model for the H$\beta$ and H$\alpha$ lines. Since we used a Lorentzian to fit the broad components of the emission lines, we did not calculate the second-order moment for this source. Using only the FWHM(H$\beta_b$) as a proxy for the velocity parameter, we got the lowest value for the black hole mass of the whole sample, on average equal to (3.14$\pm$0.54) ×10$^6$ M$_{\odot}$. Considering that J1522 has the second highest Eddington ratio derived from the measured continuum luminosity at 5100\r{A}, we could classify it as an NLS1 in an early evolutionary stage. However, the resulting lower $\epsilon$ using the derived continuum luminosity, as described in Sec.~\ref{sec:Edd_ratio}, suggests an overestimation of the former Eddington ratio likely due to contamination by the host galaxy. Such a hypothesis derives from the vicinity of the source, and from the presence of absorption lines in the spectrum, which can be produced only by the host galaxy. This contamination cannot be estimated since a proper host galaxy modeling was not possible, due to the unusable bluer spectrum.

\subsection{SDSS J164100.10+345452.7}
\label{sec:J1641}
\begin{figure}
\begin{center}
\includegraphics[trim={0cm 0 0 0}, clip, width=1.05\columnwidth]{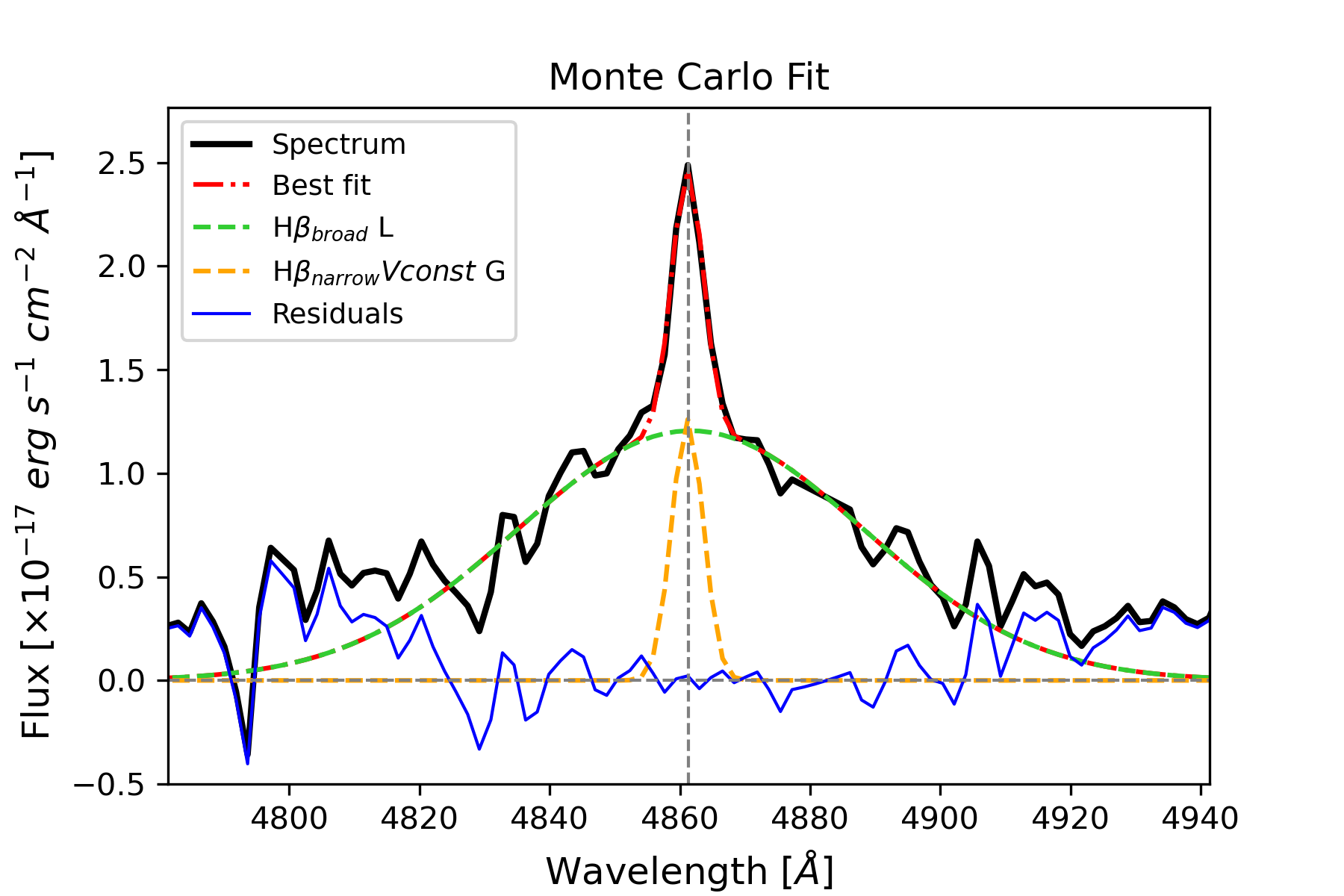}
\caption{H$\beta$ line profile of J1641. The broad and narrow components have a FWHM of 4050 \kms and 493 \kms, respectively. \label{fig:J1641_Hbeta}}
\end{center}
\end{figure}

J1641 shows peculiar line profiles compared to the rest of the sample. The broad components of the H$\beta$ line, with an FWHM of 4050 \kms\, is the broadest among the analyzed sources. This value, coupled with the FWHM of the narrow component of 493 \kms\, translates to a line profile (Fig.\ref{fig:J1641_Hbeta}) typical for intermediate Seyfert galaxies \citep{Osterbrock91,DallaBarba23}. To properly fit the H$\beta$ shape we used a 2G model. A four Gaussians model was used for the [O~III]$\lambda\lambda$4959,5007 lines. Two Gaussians for each [O~III] line were enough to achieve a good fit since no evidence of outflows was present. We did not perform a fitting of the H$\alpha$ profile since it was strongly contaminated by the sky emission, impossible to remove despite several attempts. According to the results, J1641 turned out to be one of the sources with the lowest Eddington ratio, equal to 0.17$\pm$0.04 and 0.01$\pm$0.01. This source was found to be the only $\gamma$-ray emitter of the sample, hinting at the presence of a relativistic beamed jet \citep{Lahteenmaki18}. Such a feature, coupled with the relatively low Eddington ratio and high FWHM(H$\beta$), might suggest an advanced evolutionary stage \citep{Foschini17}.

\begin{table*}
    \caption{Black hole mass and Eddington ratio results.}
    \centering
    \begin{tabular}{l c c c c c}\\
\hline \hline
Source & $R_{\rm{BLR}}$ & $\nu$ & $M_{\rm{BH}}$ & $\epsilon_m$ & $\epsilon_d$ \\
\tiny{(1)} & \tiny{(2)} & \tiny{(3)} & \tiny{(4)} & \tiny{(5)} & \tiny{(6)} \\
\hline
\multirow{2}{*}{J1029} & \multirow{2}{*}{15.60$\pm$4.43} & FWHM(H$\beta_b$) & (3.48$\pm$0.99) ×10$^7$ & \multirow{2}{*}{0.10$\pm$0.02} & \multirow{2}{*}{0.04$\pm$0.01}\\
      &  & $\sigma$(H$\beta$) & (9.77$\pm$2.77) ×10$^7$\\
\hline
\multirow{2}{*}{J1228} & \multirow{2}{*}{5.63$\pm$3.02} & FWHM(H$\beta_b$) & (4.32$\pm$2.31) ×10$^6$ & \multirow{2}{*}{0.94$\pm$0.19} & \multirow{2}{*}{0.11$\pm$0.02}\\
      &  & $\sigma$(H$\beta$) & (1.41$\pm$0.76) ×10$^7$\\
\hline
\multirow{2}{*}{J1232} & \multirow{2}{*}{16.84$\pm$5.44} & FWHM(H$\beta_b$) & (2.02$\pm$0.65) ×10$^7$& \multirow{2}{*}{0.38$\pm$0.08} & \multirow{2}{*}{0.16$\pm$0.03}\\
      &  & $\sigma$(H$\beta$) & (3.43$\pm$1.11) ×10$^7$\\
\hline
\multirow{2}{*}{J1509} & \multirow{2}{*}{11.77$\pm$2.85} & FWHM(H$\beta_b$) & (1.69$\pm$0.41) ×10$^7$ & \multirow{2}{*}{0.09$\pm$0.02} & \multirow{2}{*}{0.03$\pm$0.01}\\
      &  & $\sigma$(H$\beta$) & (4.34$\pm$1.05) ×10$^7$\\
\hline
\multirow{2}{*}{J1510} & \multirow{2}{*}{3.56$\pm$1.64} & FWHM(H$\beta_b$) & (2.10$\pm$0.96) ×10$^6$ & \multirow{2}{*}{0.25$\pm$0.05} & \multirow{2}{*}{0.05$\pm$0.01}\\
      &  & $\sigma$(H$\beta$) & (1.45$\pm$0.67) ×10$^7$\\
\hline
\multirow{2}{*}{J1522} & 11.53$\pm$1.12$^{(1)}$ & FWHM(H$\beta_b$) & (5.09$\pm$0.50) ×10$^6$ & \multirow{2}{*}{0.38$\pm$0.08} & \multirow{2}{*}{0.02$\pm$0.01}\\
      & 2.70$\pm$1.31$^{(2)}$ & FWHM(H$\beta_b$) & (1.19$\pm$0.58) ×10$^6$\\
\hline
\multirow{2}{*}{J1641} & \multirow{2}{*}{3.70$\pm$1.86} & FWHM(H$\beta_b$) & (2.51$\pm$1.26) ×10$^7$ & \multirow{2}{*}{0.17$\pm$0.04} & \multirow{2}{*}{0.01$\pm$0.01}\\
      &  & $\sigma$(H$\beta$) & (8.86$\pm$4.45) ×10$^6$\\
\hline
    \end{tabular}
    \tablefoot{Columns: (1) Source name; (2) BLR radius derived using Eq.~\ref{eq:Rblr_Fe}, $^{(1)}$ BLR radius derived using Eq.~\ref{eq:Rblr_cont}, $^{(2)}$ BLR radius derived using Eq.~\ref{eq:Rblr_Hbeta}; (3) proxy of the rotational velocity, either FWHM(H$\beta_{broad}$) or $\sigma($H$\beta$); (4) black hole mass [M$_{\odot}$]; (5) Eddington ratio calculated using the measured L$_{bol}$; (6) Eddington ratio calculated using the derived L$_{bol}$, exploiting Eq.~\ref{eq:L5100_der}.}
    \label{tab:results}
\end{table*}

\section{Discussion}
\label{sec:discussion}

In this paper, we analyzed the optical spectra of seven NLS1s with extreme radio features. We did so by performing a model fitting of the main emission lines. The goal was to investigate for similar characteristics among these sources by comparing their physical properties, such as black hole mass and Eddington ratio, and the emission lines profiles.

\subsection{Black hole mass}
From Table~\ref{tab:results}, the black hole mass calculated using the second-order moment turned out to be on average larger than that obtained with the FWHM(H$\beta_b$). It is due to the fact that the $\sigma$ we measured is systematically larger than the FWHM(\hb). This is not uncommon in NLS1, as $\sigma$ and FWHM(\hb) can be different from each other, yielding significantly different results \citep[e.g., see][]{Foschini15}. The physical meaning of such a discrepancy is not easy to address. The FWHM(H$\beta_b$) is, by definition, a directly measurable geometric property of the line, which is the reason why most studies tend to use this simple parameter \citep[e.g. see the large surveys by][]{Rakshit17a, Paliya24}. However, even a small difference in line width drives a large variability in the output, i.e. the black hole mass, because of its quadratic dependence shown in Eq.~\ref{eq:M_BH}. The way the line width is measured, for instance, the FWHM, has several limitations. This becomes particularly evident when an emission line presents a complex line profile. As described in Sec.~\ref{sec:results}, in all the cases but two the resulting broad profile in the H$\beta$ line is no longer represented by a single function, but it is instead a composition of two functions. However, the measured FWHM comes mostly only from one of the two, according to the amplitude of each component. It leads to overestimation of the mass when lines are broader and underestimation when lines are narrower \citep{Peterson18}. This means that two different H$\beta$ lines can yield the same black hole mass estimate, in all those cases where the main broad component is described by the same function. On the other hand, the second-order moment is more sensitive to the whole broad profile. This issue has already been widely discussed \citep{Peterson04, Peterson11}, and $\sigma$ is likely the most reliable proxy for the gas velocity.

In two cases, however, we modeled the H$\beta$ broad component using only one function, namely in J1522 and J1641. According to what we discussed until this point, in those two cases, the black hole masses derived with the FWHM(H$\beta_b$) and the second-order moment should be comparable. However, for J1522, the broad component in the H$\beta$ line was modeled using a Lorentzian profile, therefore a comparison between the two methods is literally impossible since the second-order moment of a Lorentzian function cannot be measured. However, since it shows all the typical traits of NLS1s, we believe that the estimated black hole mass calculated with the FWHM of the Lorentzian profile is reliable. J1641, instead, turned out to be an intermediate Seyfert galaxy \citep{Osterbrock76,Osterbrock77}, and not an NLS1. The same classification can be retrieved by looking at the Sloan Digital Sky Survey (SDSS) spectrum. The main characteristic that distinguishes the two classes is the ratio between the narrow and the broad component, which is much larger in the intermediate Seyfert galaxies.
The intermediate Seyfert galaxies are further subdivided according to the prominence of the broad component of the emission lines compared to the narrow component \citep{DallaBarba23}. The origin of these objects is widely discussed, spanning from inclination effects to intrinsic processes \citep{Barquin-Gonzalez24}. In the inclination hypothesis, the optical features visible in the spectra of intermediate Seyfert galaxies are due to the partial obscuration of the BLR \citep{Malkan98,Guainazzi05}.
In this case, there would be an underestimation of the real width of the broad components in the emission lines. As a consequence, it can affect the calculation of the black hole mass and the Eddington ratio. In particular, an underestimation of the line width yields an underestimation of the black hole mass, and consequently an overestimation of the Eddington ratio. In our case, this means that J1641 may have a more massive black hole than the estimated. Considering that J1641 was found to be a $\gamma$-ray source, a more massive black hole would strengthen our hypothesis about its advanced stage of evolution.

The results we found in this work are systematically larger than previously estimated \citep{Jarvela15}. This is most likely due to the fact that past calculations adopted a totally different approach, that led to lower mass values. To further investigate the reliability of the results we obtained with Eq.~\ref{eq:M_BH}, we additionally measured the black hole masses using a recent scaling law derived by \cite{Shen24}. They derived a relation comparing the masses of single epoch spectra with the ones obtained with the reverberation mapping technique, exploiting the $L_\lambda(5100\textrm{\r{A}})$ and the FWHM(H$\beta$). The black hole masses we measured using such relation are listed in Tab.~\ref{tab:app_results}. Comparing the results obtained with this scaling law and the virial theorem, we found very similar values, confirming the reliability of the main approach we used. \cite{Shen24} derived similar scaling laws also for Mg~II and C~IV emission lines. Nevertheless, these two relations cannot be used in our case, since both Mg~II and C~IV are not visible in the spectra analyzed because of spectral coverage.

On average, the black hole masses derived for the whole sample are well within the typical range of NLS1s population \citep{Peterson11}. Basically all of them, though, lie above the median value for the class (M$_{BH}$ = 1×10$^7$, \citealp{Cracco16}), except for J1522. This is in agreement with the typical values in jetted NLS1s. Indeed, jetted NLS1s tend to have larger black hole masses than what is usually observed in non-jetted NLS1s \citep{Foschini15, Berton15a}. Nevertheless, only based on the results obtained in this study, we cannot prove the presence of relativistic jets. It is worth noting, however, that the mass difference observed between jetted and non-jetted sources may be due to an observational bias. Since the power of relativistic jets scales nonlinearly with the black hole mass \citep{Heinz03, Foschini14}, more massive black holes have brighter relativistic jets, which in turn are easier to detect. Therefore, we may still be missing the population of jetted NLS1s of lower mass. J1522 could fall exactly into this population. The Lorentzian profile for the broad H$\beta$ suggests a source in an early stage of evolution \citep{Berton20a}, indicating that the relatively low black hole mass does not derive from an underestimation.

\subsection{Eddington ratio}
Regardless of the method used, the Eddington ratio of our sources is also consistent with typical NLS1 values, that is above 1\% and up to super-Eddington \citep{Boroson92, Sulentic00}. In particular, the results show $\epsilon$ closer to the lower limit of the range instead of high values of accretion. This, coupled with the black hole masses we retrieved, suggests sources in a middle-advanced stage of evolution. However, an exception is present, which is J1522. Indeed, in this case, as described before, the low black hole mass and the Lorentzian profile used for the broad H$\beta$ component are characteristics usually visible in unevolved sources.

\begin{figure}
\begin{center}
\includegraphics[trim={0.5cm 0 0 0}, clip, width=1.1\columnwidth]{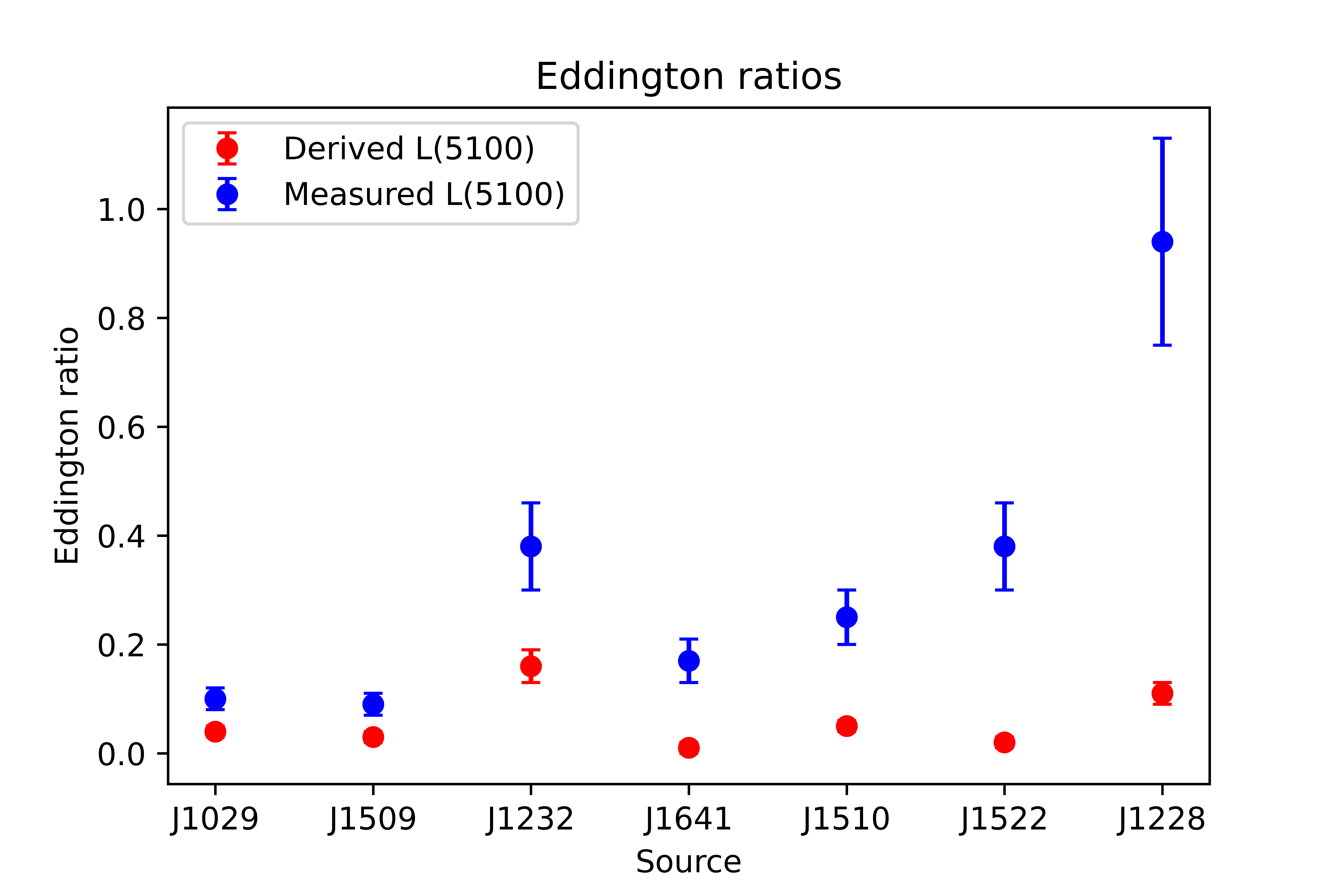}
\caption{Eddington ratios calculated using the measured continuum luminosity at 5100\r{A} (blue dots), and the derived continuum luminosity (red dots). The not visible error bars, due to their small values, are inside the size of the dots. \label{fig:Edd_ratios}}
\end{center}
\end{figure}

Focusing on the approaches we used to calculate the Eddington ratios, there is a clear difference in the results we got. The $\epsilon$ calculated using the derived continuum luminosity at 5100\r{A} are always smaller compared to the $\epsilon$ obtained with the measured continuum luminosity. As stated in Sec.~\ref{sec:Edd_ratio}, the continuum at 5100\r{A} can be contaminated mainly by the jet \citep{Foschini15} or by the host galaxy contribution, which is more likely in our case. Therefore, the difference in the Eddington ratio using the measured and derived continuum luminosity allows us to quantify them. We can analyze at this point the differences in the Eddington ratios. The smallest difference is 0.06 for J1029 and J1509, and the largest is 0.83 for the J1228. Considering the associated errors, and excluding J1228 and J1522, a weak decreasing trend related to the redshift seems to be present (Fig.~\ref{fig:Edd_ratios}). Indeed J1510 is the source with the largest difference for the Eddington ratios but it has the lowest redshift, among the first five sources in Fig.~\ref{fig:Edd_ratios}. On the other hand, J1029 has the lowest difference for the Eddington ratios, but it has the highest redshift of the whole sample. However, given the contaminants that we will discuss in a moment, we cannot test such a possible correlation with the redshift for the whole sample. Therefore with only few sources, a conclusion on this may not be reliable. This has led us to focus more on the difference of each source separately. Although not formally compatible, five sources out of seven show similar Eddington ratios regardless of the calculation method used. J1228 and J1522 are exceptions. For J1228 the resulting Eddington ratios are 0.11 and 0.94. Carefully looking at the spectrum, there are no signs of possible contamination by the host galaxy. Considering the redshift of 0.2627, the second highest in the sample, a very bright host galaxy would be necessary to produce such a difference in the Eddington ratio results. This would lead to very intense absorption lines which are totally missing. A more reasonable explanation may come from a relativistic jet contamination, which is less affected by the redshift due to its strength. Nevertheless, it might not produce recognizable features in the optical spectrum, increasing the difficulty of recognizing it. It is worth noting that J1228 is the source with the highest \textit{R}4570, proxy which suggests strong Fe~II multiplets emission. Such behaviour is associated with an intense ionizing continuum, which is produced by the accretion disk in a high accretion state \citep{Gaskell21}.
In J1522 we find the second largest Eddington ratios difference (0.36). This is the closest source in the sample (z=0.0769). Contrarily to the previous case, the optical spectrum of J1522 shows a slight increase in the continuum level toward red wavelengths and multiple absorption lines. This behavior suggests contamination of the continuum luminosity by the host galaxy emission, resulting in an increase of the Eddington ratio calculated with the measured continuum luminosity at 5100\r{A}. Behaviors that cannot be proven by a host galaxy modeling, since such modeling was not possible using the only available R1000R grism spectrum.

Two important conclusions can be drawn from such results. The former is that even without a strict host galaxy modeling, we can evaluate if the host galaxy contaminates the continuum emission by analysing the Eddington ratios calculated with the measured and derived continuum luminosity at 5100\r{A}. In the end, only J1522 showed a non-negligible host galaxy contamination. The latter is the demonstration that the Eddington ratio obtained using the derived continuum luminosity is more reliable compared to the measured continuum luminosity approach, being less contaminated by relativistic jets and host galaxy emission. However, it is worth noting that the relations we used have a rather large standard deviation. This means that while they work for large samples, they may be misleading for a single source.

Even considering different sources of uncertainty in Eq.~\ref{eq:L_bol} and Eq.~\ref{eq:k_bol}, we have to keep in mind that we are estimating a bolometric luminosity only using a luminosity in a small range of wavelength. This connection is based on the strong assumption that each value of $L_{\lambda}(5100\textrm{\r{A}})$ corresponds to a specific SED shape. To this purpose, \cite{Ferland20} found that similar spectral features are not always related to the same SED shape. Some authors also found a positive correlation between the bolometric correction in Eq.~\ref{eq:k_bol} and the Eddington ratio \citep{Vasudevan07,Jin12}. Correlation disproved by \cite{Cheng19}, who also showed that the bolometric correction factor changes greatly according to the disk model assumed. The different Eddington ratios that can be obtained depending on the assumption made for the bolometric luminosity derivation, i.e. the scaling laws used, is clearly demonstrated for instance by comparing \cite{Berton15a} and \cite{Tortosa23}. The former derived the Eddington ratio from optical spectra, using scaling laws similar to those used in this work, while the latter derived the Eddington ratio from X-ray data. Looking at sources analyzed in both papers, the $\epsilon$ found in \cite{Tortosa23} are three to four order of magnitude larger than those found in \cite{Berton15a}. Such huge difference cannot be entirely due to a variation of the accretion of the sources, especially in a timescale of few years, but more likely on the approach used by the two papers. It is clear how the bolometric correction, and then the Eddington ratio among all, are just an estimation without precise knowledge of the SED shape. Therefore, our results should be taken with a grain of salt and likely as a lower limit.

\subsection{\textit{R}4570-Eddington ratio discrepancy}
\label{sec:R4570-Edd_ratio}
An interesting parameter we can focus on is the \textit{R}4570. The sources we analyzed turned out to belong to the populations A3 and A4 of the quasar main sequence, therefore with a strong Fe~II emission. \cite{Gaskell07} with the Gaskell, Klimek \& Nazarova BLR model showed that the Fe~II emission comes from the outer part of the BLR, predominantly emitted at a radius twice that of H$\beta$. Moreover, \cite{Gaskell21} stated that the Fe~II emission is produced by photoionization, ruling out other possible hypotheses. They also found a positive correlation between the Fe~II strength and the Eddington ratio (see also \citealp{Boroson92,Wandel98,Sulentic00,Marziani01}). Our sources however seem to disagree with such a relation, except for J1228. As we described, despite all of them belonging to A3 and A4 populations, the resulting Eddington ratios are close to the lower boundary for the NLS1s class. Considering also the discussed biases which could affect the Eddington ratio calculation, we can think that the high \textit{R}4570 parameters might indicate higher Eddington ratios than what we estimated. A higher Eddington ratio means a soft X-ray excess, which is not present in low-Eddington ratio sources. Soft X-ray radiation breaks down the dust grains in the outer BLR, thus releasing the iron that then gets photoionized by the photons coming from the accretion disk \citep{Gaskell21}. \cite{Abramowicz13} described that for a typical Shakura–Sunyaev disk, the viscous heating is balanced by radiative cooling. In high-Eddington ratio sources instead, the accretion disk does not have enough time to cool down only by the radiation losses, therefore an advective cooling is established. This forms the so-called slim disk, in which the efficiency of transforming gravitational energy into radiative flux decreases with increasing accretion rate. A slim disk shows a moderate luminosity despite super-Eddington accretion. If this is the case, our sources could have higher Eddington ratios than what we measured. This shows that the Eddington ratio estimate obtained via optical spectroscopy may be, at least in some cases, misleading. The \textit{R}4570 parameter may thus be a very important, and more reliable, indicator of the real Eddington ratios of some AGN.

\begin{table}
    \caption{Pearson correlation coefficients.}
    \centering
    \begin{tabular}{l c c c c} \\
\hline\hline
Parameters & corr. coeff. & p-value \\
\hline
L(H$\alpha$)-L(5100\r{A}) & 0.89 & 0.04 \\
L(H$\beta$)-L(5100\r{A}) & 0.82 & 0.04 \\
L([O~III]$\lambda$5007)-L(5100\r{A}) & 0.88 & 0.02 \\
FWHM(H$\beta_b$)-R4570 & -0.75 & 0.15 \\
FWHM(H$\alpha$)-R4570 & -0.96 & 0.04 \\
R5007-R4570 & 0.90 & 0.02 \\
R5007-L(H$\beta$) & -0.70 & 0.12 \\
$\epsilon$ $L_m$(5100\r{A})-$\Delta$[O~III]$_c$ & 0.48 & 0.28 \\
$\epsilon$ $L_m$(5100\r{A})-$\Delta$[O~III]$_w$ & -0.26 & 0.57 \\
$\epsilon$ $L_d$(5100\r{A})-$\Delta$[O~III]$_c$ & 0.09 & 0.85 \\
$\epsilon$ $L_d$(5100\r{A})-$\Delta$[O~III]$_w$ & -0.2 & 0.96 \\
\hline
    \end{tabular}
    \label{tab:pearson}
\end{table}

\subsection{SDSS-GTC spectral continuum comparison}
Looking at the available SDSS spectra of the sample, we measured and compared the continuum flux densities only, and not the emission lines properties, due to the much lower S/N compared to the GTC spectra. Computing the ratio of the flux densities at 5100\r{A} in Tab~\ref{tab:spectra}, we found values between 0.90 for J1232 and 2.03 for J1641, the lowest and the highest respectively. For J1232 the SDSS continuum is lower than the GTC continuum. In the remaining cases, however, the SDSS flux density is higher. In Seyfert galaxies, a variability of the flux density at 5100\r{A}, sometimes even much larger than 2, has been found in several studies \citep{Zastrocky24,Shapovalova12,Grier12}. With a difference in the acquisition time of more than ten years between SDSS and GTC spectra, such values are in agreement with the well known long-term stochastic variability which characterize all the AGN.

\subsection{Comparison of the NLS1s properties}
In Tab.~\ref{tab:pearson} are reported the Pearson correlation coefficients we computed for several couples of parameters, setting 0.05 as threshold for the significance of the correlation analysis. We compared our results with the statistical analyses on large NLS1s samples available in the literature \citep{Zhang11,Rakshit17a,Paliya24}. The most significant correlations are those between H$\alpha$, H$\beta$ and [O~III]$\lambda$5007 luminosity and the continuum luminosity at 5100\r{A}, with Pearson coefficients around 0.8-0.9. This is in agreement to what was found by \cite{Rakshit17a} and \cite{Paliya24}. The same authors found, even though with a quite large scatter, a weak anti-correlation in the couples of parameters of FWHM(H$\beta_b$)-R4570, FWHM(H$\alpha$)-R4570 and R5007-R4570, and a moderately strong anti-correlation between R5007 and L(H$\beta$). We found instead three strong correlations and one strong anti-correlation for the same couples of parameters. Nevertheless, the FWHM(H$\beta_b$)-R4570 and R5007-L(H$\beta$) correlation coefficients are not statistically significant, due to the p-values of 0.15 and 0.12, respectively. It is important to note that we did not consider the source J1641 in the FWHM(H$\beta_b$)-R4570 correlation test, since the FWHM(H$\beta_b$) is much broader for intermediate Seyfert galaxies compared to NLS1s. Finally, we investigated the correlation between the Eddington ratios and the velocity shifts, with respect to the rest frame wavelength, of the core and the strongest wing components in the [O~III]$\lambda$5007, as found by \cite{Zhang11}. Nothing can be said in this case due to the large p-values. Even though all the correlation coefficients are on average in agreement with the most recent literature, the extremely small number of sources we analyzed, cannot represent a statistically valid sample. Nevertheless, the lacking of noticeable peculiarities in the correlation analysis prove, once again, that these sources do not show deviations from the NLS1s class properties.

\section{Conclusions}
\label{sec:conclusions}

In this study we derived the physical parameters of seven NLS1s, performing an analysis of the main emission lines in optical spectra observed with the GTC. The goal was to identify any common optical property of the sources in the sample, which could explain why they show similar extreme features in the radio band. Investigating the most significant parameters, we did not find strong similarities between the sources. All but one showed classical behavior for the NLS1s class. Only J1641 turned out to be an intermediate Seyfert, but with physical properties that are common for this population. There are instead some shared traits. On average the black holes are more massive than the median value for the class (M$_{BH}$ = 1×10$^7$, \citealp{Cracco16}), with Eddington ratios much closer to the lower boundary for NLS1s. On the other hand, according to the \textit{R}4570 parameter, the A3 and A4 nature of the sources \citep{Sulentic15} suggests an underestimation of the actual Eddington ratio, which could be much higher than what we measured. High Eddington ratios for NLS1s could mean sources in an early stage of evolution, with an ongoing intense accretion activity. Though only the early evolutionary stage is not enough to explain why these sources show extreme flares in radio, it is an environment where the phenomena described by the hypotheses in \cite{Jarvela23} can take place.

In particular AGN with a recently started activity, as NLS1s, can have regions with large amounts of gas, which is necessary to sustain a high Eddington ratio accretion \citep{Mathur00}. A gas-rich nuclear environment, coupled with the relatively massive black holes which might suggest the possible presence of relativistic jets, is in agreement with scenarios such as jet-cloud/star interaction, relativistic jet and free-free absorption with moving clouds and magnetic reconnection in the jet \citep{Jarvela23}.
Nevertheless with the results we obtained nothing more can be said to support or rule out some of the cited hypotheses.

To better strength the results we obtained in this study, optical spectra with wider wavelength ranges and higher S/N would be necessary. However, we would not expect much different results from those we obtained here since almost all the main emission lines were visible and we did not find substantial changes from the SDSS spectra. The only significant improvement we could get are spectra in which an host galaxy estimation can be done. In that case we could better constrain the Eddington ratio calculations. The similarities between the sources can be investigated more deeply with an analysis of the light curves, both in optical (Crepaldi et al., in prep.) and in radio frequencies. From the radio point of view, more observations at frequencies above 37~GHz would help us to better understand all the ongoing physical processes on these sources. Moreover trigger observations, for example at high radio frequencies with the upgraded detectors of the Sardinia Radio Telescope or with other facilities such as the Square Kilometer Array, can add more constraints to the viable hypotheses we already mentioned.

\begin{acknowledgements}
Based on observations made with the Gran Telescopio Canarias (GTC), installed at the Spanish Observatorio del Roque de los Muchachos of the Instituto de Astrofísica de Canarias, on the island of La Palma.
L.C. and M.B. acknowledge the ESO Science Support Discretionary Fund. G.L.M. is supported by the Italian Research Center on High Performance Computing Big Data and Quantum Computing (ICSC), a project funded by the European Union - NextGenerationEU - and National Recovery and Resilience Plan (NRRP) - Mission 4 Component 2 within the activities of Spoke 3 (Astrophysics and Cosmos Observations). The authors are grateful to Dr. L. Foschini for the helpful discussion on the topic of bolometric luminosity derivation.

\end{acknowledgements}

%
\bibliographystyle{aa} 
\bibliography{./biblio.bib} 
%



\begin{appendix}

\onecolumn
\section{Tables, spectra and lines profiles}
\label{appendix_tables}

\begin{table*}[h!]
    \caption{Observational and physical parameters derived from the optical spectra.}
    \centering
    \begin{tabular}{l c c c c c c c c c} \\
\hline \hline
Source & $L_{bol}^m$ & $L_{bol}^d$ & $L_{\rm{[O~III]}\lambda5007}$ & \textit{R}5007 & $R_{\rm{sub}}$ & $R_{\rm{out}}$ & $R_{\rm{NLR}}$ & $M_{\rm{BH}}^*$\\
\tiny{(1)} & \tiny{(2)} & \tiny{(3)} & \tiny{(4)} & \tiny{(5)} & \tiny{(6)} & \tiny{(7)} & \tiny{(8)} & \tiny{(9)} \\
\hline
J1029 & 8.39×10$^{44}$ & 3.81×10$^{44}$ & 1.23×10$^{41}$ & 0.12 & 0.25 & 7.50 & 906.05  & 9.32×10$^{7}$ \\
J1228 & 1.12×10$^{45}$ & 1.35×10$^{44}$ & 3.08×10$^{41}$ & 0.72 & 0.16 & 4.79 & 1459.28 & 2.23×10$^{7}$ \\
J1232 & 1.35×10$^{45}$ & 5.51×10$^{44}$ & 2.34×10$^{41}$ & 0.13 & 0.32 & 9.63 & 1263.67 & 5.99×10$^{7}$ \\
J1509 & 3.40×10$^{44}$ & 1.15×10$^{44}$ & 4.86×10$^{40}$ & 0.19 & 0.13 & 3.74 & 558.36  & 3.97×10$^{7}$ \\
J1510 & 2.64×10$^{44}$ & 5.45×10$^{43}$ & 8.61×10$^{40}$ & 0.71 & 0.09 & 2.55 & 751.76  & 1.22×10$^{7}$ \\
J1522 & 1.56×10$^{44}$ & 9.08×10$^{42}$ & 1.02×10$^{40}$ & 0.52 & 0.04 & 1.05 & 248.59  & 6.94×10$^{6}$ \\
J1641 & 3.81×10$^{44}$ & 2.57×10$^{43}$ & 7.33×10$^{40}$ & 1.12 & 0.06 & 1.89 & 691.75  & 1.80×10$^{7}$ \\
\hline
    \end{tabular}
    \tablefoot{Columns: (1) Source name; (2) bolometric luminosity calculated using the measured 5100\r{A} continuum luminosity [\ergs] (Eq.~\ref{eq:L_bol}); (3) bolometric luminosity calculated using the derived 5100\r{A} continuum luminosity [\ergs] (Eq.~\ref{eq:L_bol} and Eq.~\ref{eq:L5100_der}); (4) [O~III]$\lambda$5007 luminosity [\ergs]; (5) flux ratio between [O~III]$\lambda$5007 line and H$\beta$; (6) dust sublimation radius (i.e. boundary between the BLR and the molecular torus) [pc] \citep{Foschini19}; (7) outer radius of the molecular torus [pc] \citep{Foschini19}; (8) maximum extension of the NLR [pc] \citep{Foschini19}; (9) black hole mass [M$_{\odot}$] calculated using Eq.~2 in \cite{Shen24}.}
    \label{tab:app_results}
\end{table*}

\begin{table*}[h!]
    \caption{[O~III]$\lambda\lambda$4959,5007 models' parameters.}
    \centering
    \addtolength{\tabcolsep}{-0.3em}
    \begin{tabular}{l c c c c c c c} \\
\hline \hline
Param & J1029 & J1228 & J1232 & J1509 & J1510 & J1522 & J1641 \\
\hline
\tiny{(1)} Model & 4G & 6G & 6G & 6G & 4G & 4G & 4G \\
\tiny{(2)} f$_{c5007}$      & 1.25$\pm$0.30      & 13.23$\pm$0.66    & 2.35$\pm$0.42     & 2.84$\pm$0.53      & 16.80$\pm$0.34     & 4.14$\pm$0.87       & 6.61$\pm$0.22 \\
\tiny{(3)} f$_{c4959}$      & 0.42$\pm$0.10      & 4.41$\pm$0.22     & 0.78$\pm$0.14     & 0.54$\pm$0.09      & 5.60$\pm$0.11      & 1.38$\pm$0.29       & 2.20$\pm$0.07 \\
\tiny{(4)} $\Delta$c$_c$    & 31.14$\pm$34.73    & 71.25$\pm$4.49    & -37.72$\pm$30.54  & -60.48$\pm$29.94   & 1.20$\pm$2.99      & 98.20$\pm$77.84     & 38.32$\pm$4.19 \\
\tiny{(5)} w$_c$            & 145.21$\pm$44.91   & 110.69$\pm$14.37  & 117.09$\pm$35.93  & 144.50$\pm$34.73   & 137.64$\pm$6.59    & 239.41$\pm$116.16   & 239.30$\pm$8.98 \\
\tiny{(6)} f$_{w5007}$      & 0.72$\pm$0.30      & 3.21$\pm$0.44     & 3.02$\pm$0.38     & 0.83$\pm$0.55      & 2.23$\pm$0.35      & 2.12$\pm$0.80       & 1.22$\pm$0.22 \\
\tiny{(7)} f$_{w4959}$      & 0.24$\pm$0.10      & 1.07$\pm$0.15     & 1.01$\pm$0.13     & 0.28$\pm$0.18      & 0.74$\pm$0.12      & 0.71$\pm$0.27       & 0.41$\pm$0.08 \\
\tiny{(8)} $\Delta$c$_{w1}$ & -288.01$\pm$266.45 & -162.86$\pm$32.33 & -411.35$\pm$42.51 & -361.06$\pm$135.32 & -252.68$\pm$50.30  & -648.46$\pm$316.15  & -58.08$\pm$55.69 \\
\tiny{(9)} w$_{w1}$         & 343.85$\pm$247.29  & 149.58$\pm$44.31  & 221.68$\pm$37.72  & 153.96$\pm$116.76  & 606.28$\pm$95.20   & 326.89$\pm$283.22   & 770.93$\pm$83.83 \\
\tiny{(10)} f$_{w5007}$      &                    & 2.80$\pm$0.25     & 2.79$\pm$0.37     & 0.54$\pm$0.09      & & & \\
\tiny{(11)} f$_{w4959}$      &                    & 0.93$\pm$0.08     & 0.93$\pm$0.12     & 0.18$\pm$0.03      & & & \\
\tiny{(12)} $\Delta$c$_{w2}$ &                    & -371.24$\pm$35.33 & -595.77$\pm$10.78 & -192.80$\pm$264.06 & & & \\
\tiny{(13)} w$_{w2}$         &                    & 497.35$\pm$25.15  & 587.48$\pm$46.11  & 895.15$\pm$16.17   & & & \\

\hline
    \end{tabular}
    \tablefoot{Rows: (1) Fitting model; (2) peak flux density of the [O~III]$\lambda$5007 core component [×10$^{\textbf{-}17}$ \ergs~cm$^{-2}$~$\r{A}^{-1}$]; (3) peak flux density of the [O~III]$\lambda$4959 core component [×10$^{\textbf{-}17}$ \ergs~cm$^{-2}$~$\r{A}^{-1}$]; (4) shift of the core component [\kms]; (5) width of the core component [\kms]; (6) peak flux density of the first [O~III]$\lambda$5007 wing component [×10$^{\textbf{-}17}$ \ergs~cm$^{-2}$~$\r{A}^{-1}$]; (7) peak flux density of the first [O~III]$\lambda$4959 wing component [×10$^{\textbf{-}17}$ \ergs~cm$^{-2}$~$\r{A}^{-1}$]; (8) shift of the first wing component [\kms]; (9) width of the first wing component [\kms]; (10) peak flux density of the second [O~III]$\lambda$5007 wing component [×10$^{\textbf{-}17}$ \ergs~cm$^{-2}$~$\r{A}^{-1}$]; (11) peak flux density of the second [O~III]$\lambda$4959 wing component [×10$^{\textbf{-}17}$ \ergs~cm$^{-2}$~$\r{A}^{-1}$]; (12) shift of the second wing component [\kms]; (13) width of the second wing component [\kms].}
\end{table*}

\begin{table*}[h!]
    \caption{H$\beta$ models' parameters.}
    \centering
    \addtolength{\tabcolsep}{-0.3em}
    \begin{tabular}{l c c c c c c c} \\
\hline \hline
Param & J1029 & J1228 & J1232 & J1509 & J1510 & J1522 & J1641 \\
\hline
\tiny{(1)} Model & 3G & 3G & 3G & 3G & 3G & LG & 2G \\
\tiny{(2)} f$_{b1}$         & 1.42$\pm$0.07      & 4.23$\pm$0.15      & 13.23$\pm$0.27    & 3.85$\pm$0.08      & 2.64$\pm$0.19      & 3.35$\pm$0.42      & 1.11$\pm$0.05 \\
\tiny{(3)} $\Delta$c$_{b1}$ & 15.23$\pm$25.28    & -117.36$\pm$15.42  & -79.74$\pm$7.40   & 41.75$\pm$10.48    & -397.95$\pm$49.34  & -285.09$\pm$70.92  & 23.25$\pm$76.47 \\
\tiny{(4)} w$_{b1}$         & 773.06$\pm$48.72   & 541.14$\pm$28.98   & 601.43$\pm$18.50  & 737.35$\pm$20.35   & 395.13$\pm$48.72   & 685.14$\pm$109.15  & 1767.11$\pm$117.17 \\
\tiny{(5)} f$_{b2}$         & 0.68$\pm$0.06      & 0.96$\pm$0.14      & 5.74$\pm$0.32     & 0.82$\pm$0.07      & 1.37$\pm$0.12      & & \\
\tiny{(6)} $\Delta$c$_{b2}$ & -187.66$\pm$131.35 & -596.52$\pm$160.34 & -216.64$\pm$31.45 & 440.74$\pm$148.62  & 9.68$\pm$156.64    & & \\
\tiny{(7)} w$_{b2}$         & 3426.33$\pm$317.59 & 2465.03$\pm$297.24 & 2055.15$\pm$74.00 & 3091.26$\pm$254.69 & 2609.44$\pm$289.84 & & \\
\tiny{(8)} f$_{n}$          & 0.18$\pm$0.10      & 2.02$\pm$0.26      & 2.65$\pm$0.38     & 0.45$\pm$0.13      & 2.12$\pm$0.48      & 2.13$\pm$0.31      & 1.08$\pm$0.10 \\
\tiny{(9)} w$_{n}$          & 145.21             & 110.69             & 117.09            & 144.50             & 137.64             & 239.41             & 239.30\\

\hline
    \end{tabular}
    \tablefoot{Rows: (1) fitting model; (2) peak flux density of the first broad component [×10$^{\textbf{-}17}$ \ergs~cm$^{-2}$~$\r{A}^{-1}$]; (3) shift of the first broad component [\kms]; (4) width of the first broad component [\kms]; (5) peak flux density of the second broad component) [×10$^{\textbf{-}17}$ \ergs~cm$^{-2}$~$\r{A}^{-1}$]; (6) shift of the second broad component [\kms]; (7) width of the second broad component [\kms]; (8) peak flux density of the narrow component [×10$^{\textbf{-}17}$ \ergs~cm$^{-2}$~$\r{A}^{-1}$]; (9) width of the narrow component constrained with the width of the [O~III]$\lambda$5007 core component [\kms].}
\end{table*}

\begin{table*}[h!]
    \caption{H$\alpha$+[N II]$\lambda\lambda$6548,6583 and [S II]$\lambda\lambda$6716,6731 models' parameters.}
    \centering
    \addtolength{\tabcolsep}{-0.3em}
    \begin{tabular}{l c c c c c c c} \\
\hline \hline
Param & J1029 & J1228 & J1232 & J1509 & J1510 & J1522 & J1641 \\
\hline
\tiny{(1)} f$_{b1}$           & & 16.19$\pm$0.11  & 41.33$\pm$0.16   & 9.61$\pm$0.06    & 6.11$\pm$0.15    & 5.61$\pm$0.57  & \\
\tiny{(2)} f$_{b2}$           & & 1.35$\pm$0.05   & 7.82$\pm$0.07    & 0.90$\pm$0.03    & 2.40$\pm$0.06    &                & \\
\tiny{(3)} f$_{n}$            & & 14.42$\pm$0.19  & 10.03$\pm$0.26   & 2.50$\pm$0.10    & 14.83$\pm$0.23   & 13.56$\pm$0.70  & \\
\tiny{(4)} f$_{6583}$         & & 8.17$\pm$0.13   & 2.42$\pm$0.16    & 1.50$\pm$0.08    & 8.19$\pm$0.17    & 4.65$\pm$0.23  & \\
\tiny{(5)} f$_{6716}$         & & 1.93$\pm$0.13   & 1.28$\pm$0.17    & 0.51$\pm$0.08    & 2.54$\pm$0.18    &                & 1.69$\pm$0.09 \\
\tiny{(6)} f$_{6731}$         & & 1.77$\pm$0.13   & 1.02$\pm$0.17    & 0.49$\pm$0.08    & 2.33$\pm$0.17    &                & 1.67$\pm$0.10 \\
\tiny{(7)} $\Delta$c$_{6731}$ & & 22.49$\pm$12.03 & -30.51$\pm$35.63 & -24.72$\pm$21.83 & 22.94$\pm$11.14  &                & -51.89$\pm$17.82 \\
\tiny{(8)} w$_{6731}$         & & 152.59$\pm$13.36& 180.35$\pm$31.62 & 97.20$\pm$41.86  & 135.08$\pm$15.59 &                & 217.67$\pm$12.47 \\

\hline
    \end{tabular}
    \tablefoot{Rows: (1) Peak flux density of the first H$\alpha$ broad component [×10$^{\textbf{-}17}$ \ergs~cm$^{-2}$~$\r{A}^{-1}$]; (2) peak flux density of the second H$\alpha$ broad component [×10$^{\textbf{-}17}$ \ergs~cm$^{-2}$~$\r{A}^{-1}$]; (3) peak flux density of the first H$\alpha$ narrow component [×10$^{\textbf{-}17}$ \ergs~cm$^{-2}$~$\r{A}^{-1}$]; (4) peak flux density of the [N II]$\lambda$6583 [×10$^{\textbf{-}17}$ \ergs~cm$^{-2}$~$\r{A}^{-1}$]; (5) peak flux density of the [S II]$\lambda$6716 [×10$^{\textbf{-}17}$ \ergs~cm$^{-2}$~$\r{A}^{-1}$]; (6) peak flux density of the [S II]$\lambda$6731 [×10$^{\textbf{-}17}$ \ergs~cm$^{-2}$~$\r{A}^{-1}$]; (7) shift of the [S II]$\lambda$6731 [\kms]; (8) width of the [S II]$\lambda$6731 [\kms].}
\end{table*}

\twocolumn


\begin{figure*}[htbp]
    \centering
    \includegraphics[trim={0 0 0 0.7cm}, clip, width=\textwidth]{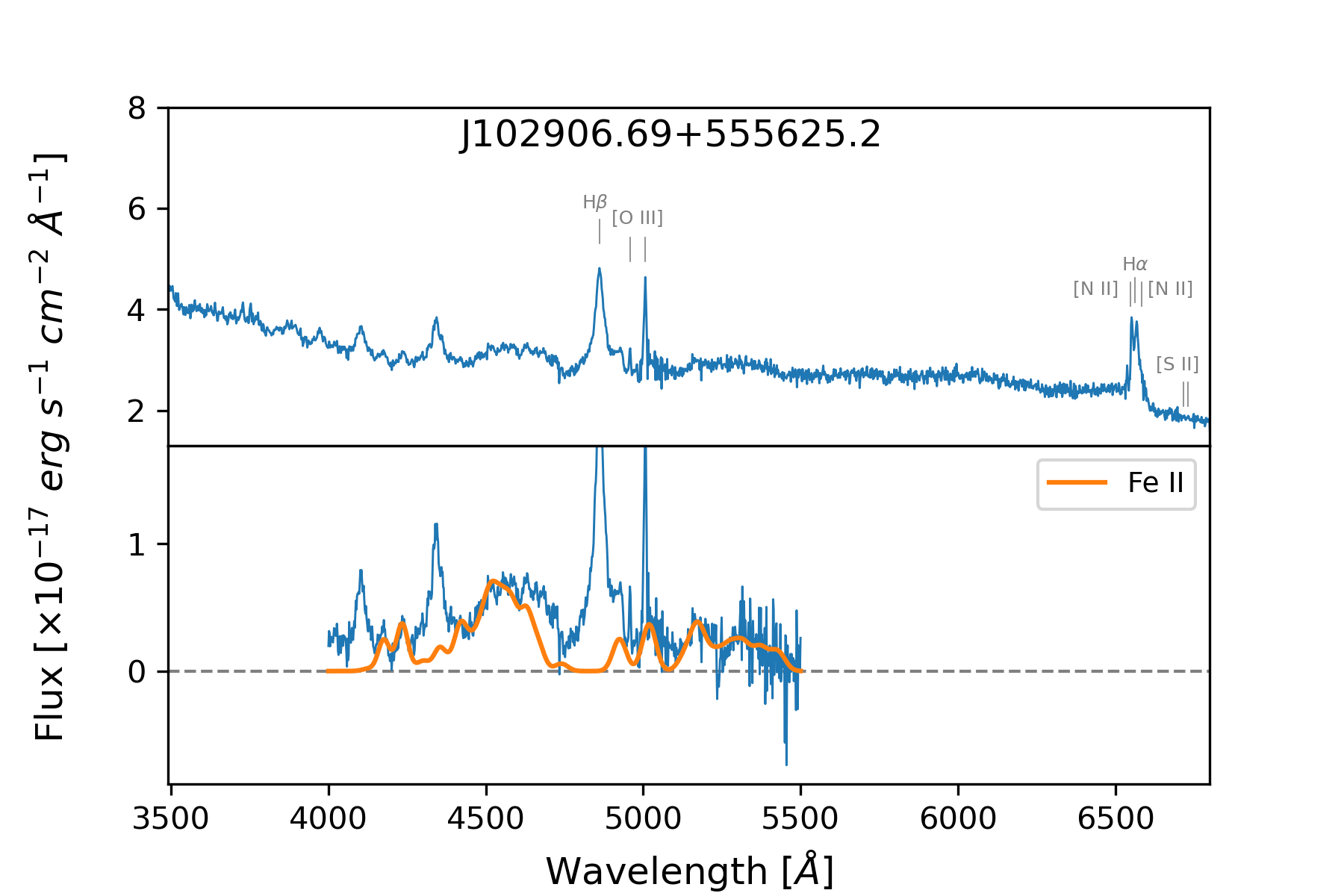}
    \caption{Spectrum of J1029.}

    \vspace{5em}

    \begin{minipage}[b]{0.49\linewidth}
        \centering
        \includegraphics[width=\textwidth]{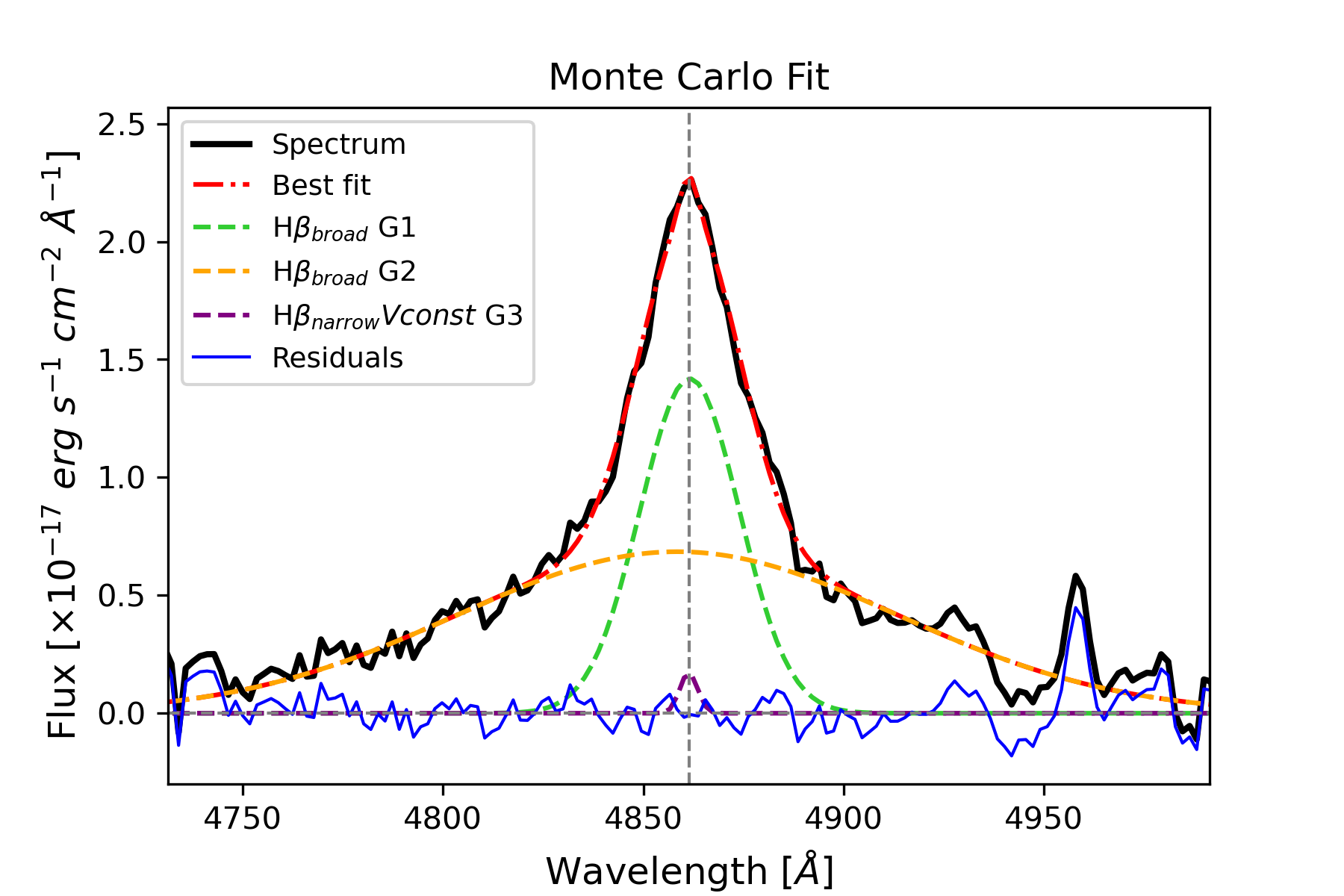}
        \caption{H$\beta$ line profile of J1029.}
    \end{minipage}
    \hfill
    \begin{minipage}[b]{0.49\linewidth}
        \centering
        \includegraphics[width=\textwidth]{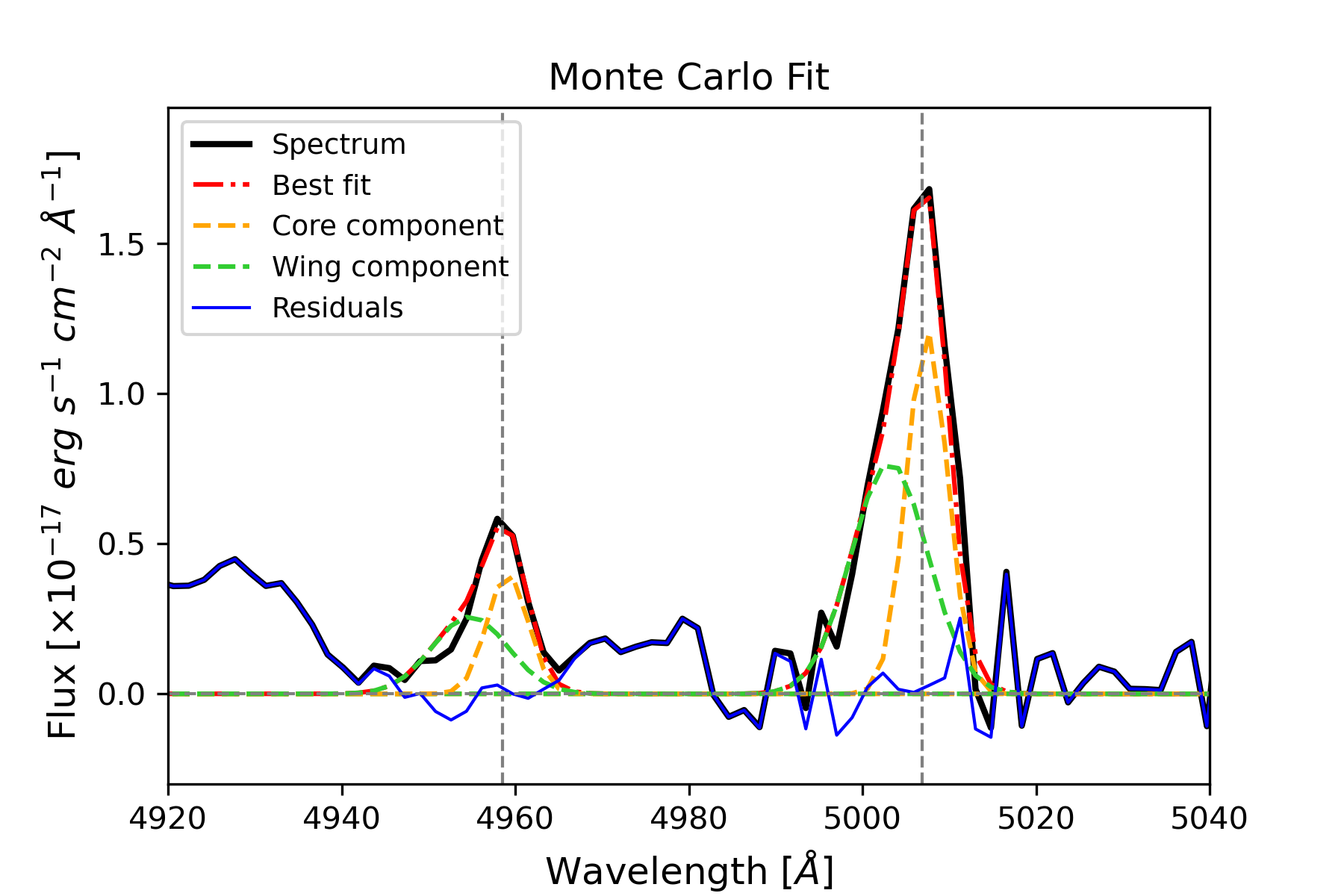}
        \caption{[O~III]$\lambda\lambda$4959,5007 lines profile of J1029.}
    \end{minipage}
\end{figure*}

\clearpage

\begin{figure*}[htbp]
    \centering
    \includegraphics[trim={0 0 0 0.7cm}, clip, width=\textwidth]{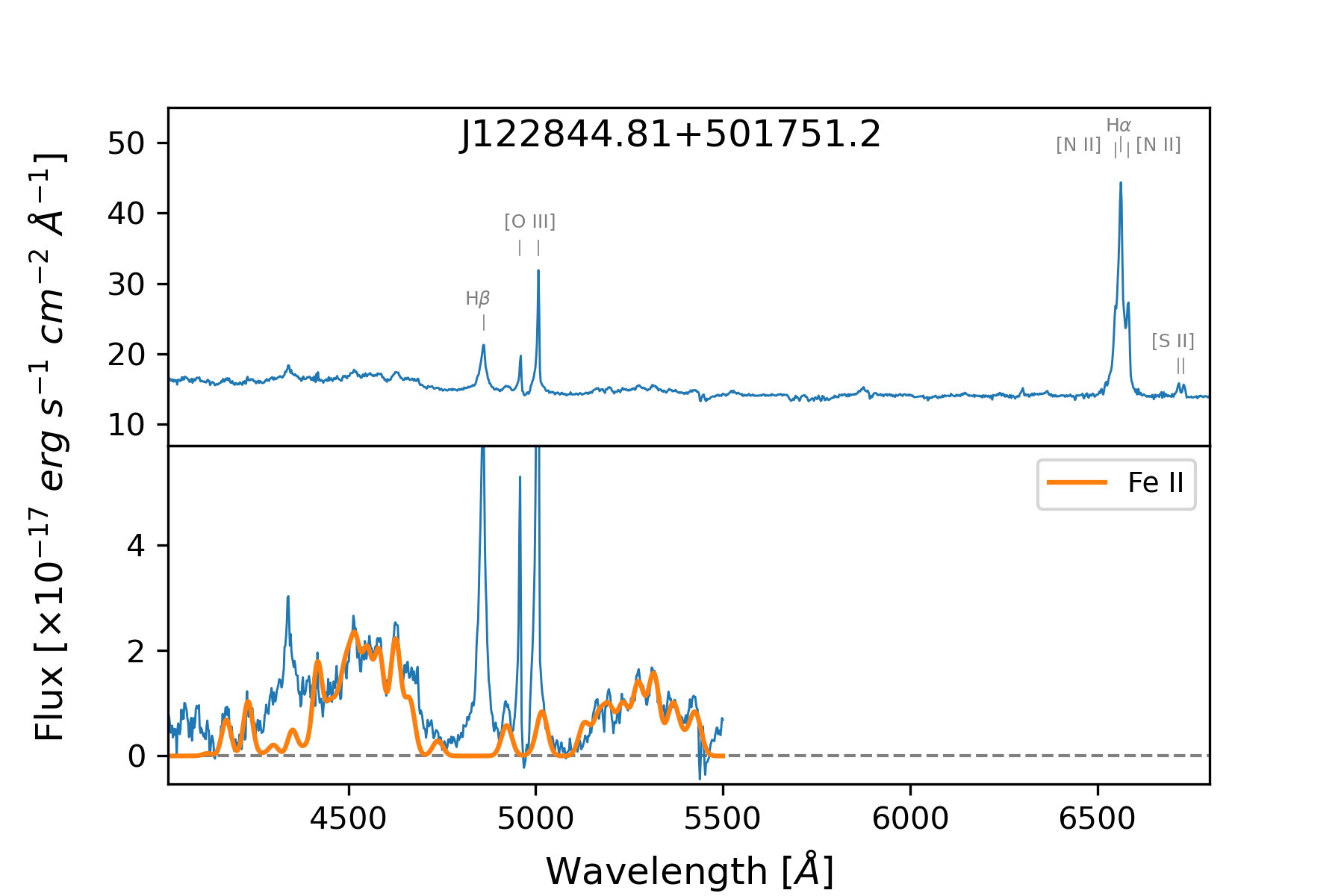}
    \caption{Spectrum of J1228.}

    \vspace{1em}

    \begin{minipage}[b]{0.49\linewidth}
        \centering
        \includegraphics[trim={0 0 0 0.6cm}, clip, width=\textwidth]{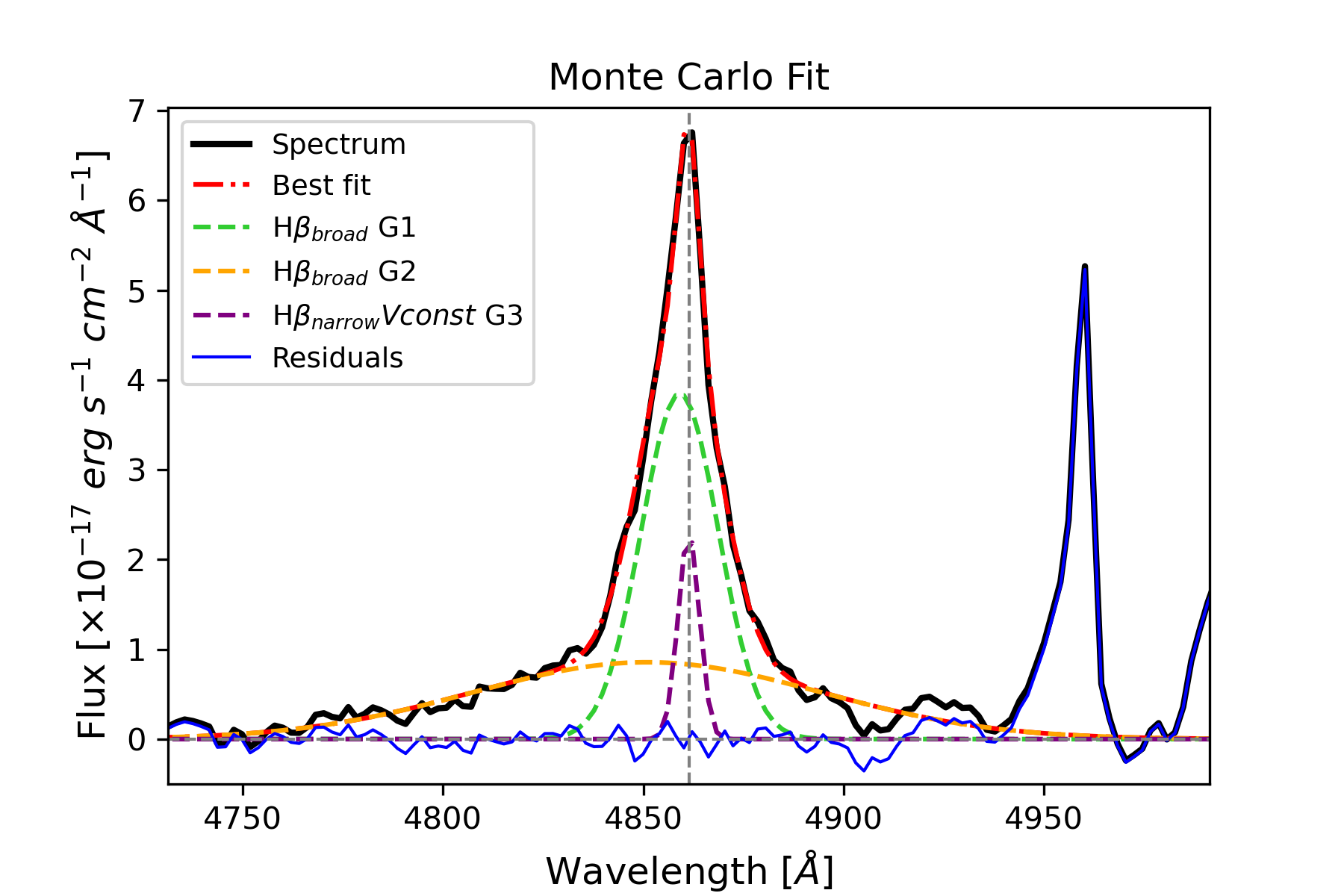}
        \caption{H$\beta$ line profile of J1228.}

    \end{minipage}
    \hfill
    \begin{minipage}[b]{0.49\linewidth}
        \centering
        \includegraphics[trim={0 0 0 0.6cm}, clip, width=\textwidth]{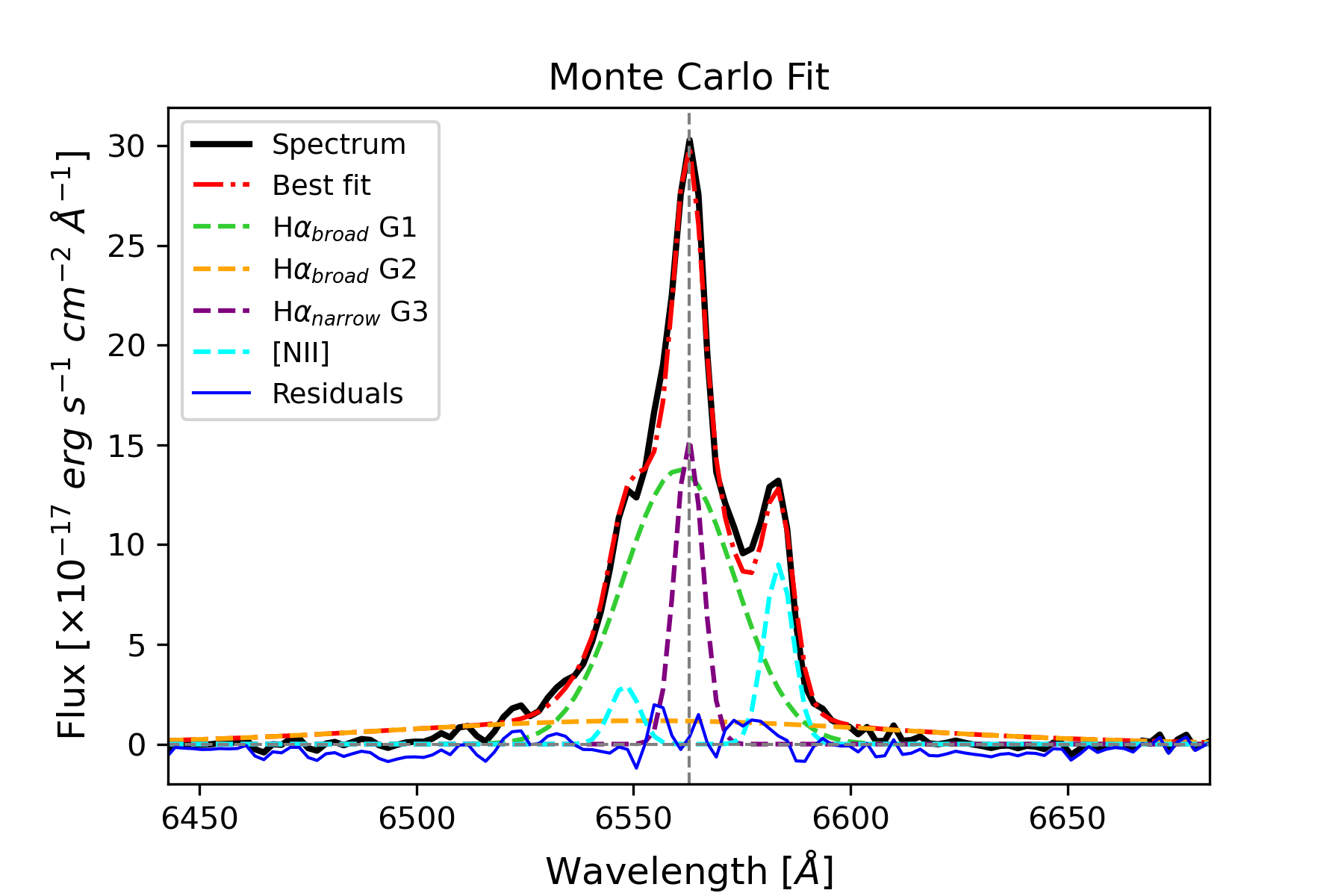}
        \caption{H$\alpha$+[N~II]$\lambda\lambda$6548,6583 lines profile of J1228.}

    \end{minipage}

    \vspace{1em} 
    \begin{minipage}[b]{0.49\linewidth}
        \centering
        \includegraphics[trim={0 0 0 0.7cm}, clip, width=\textwidth]{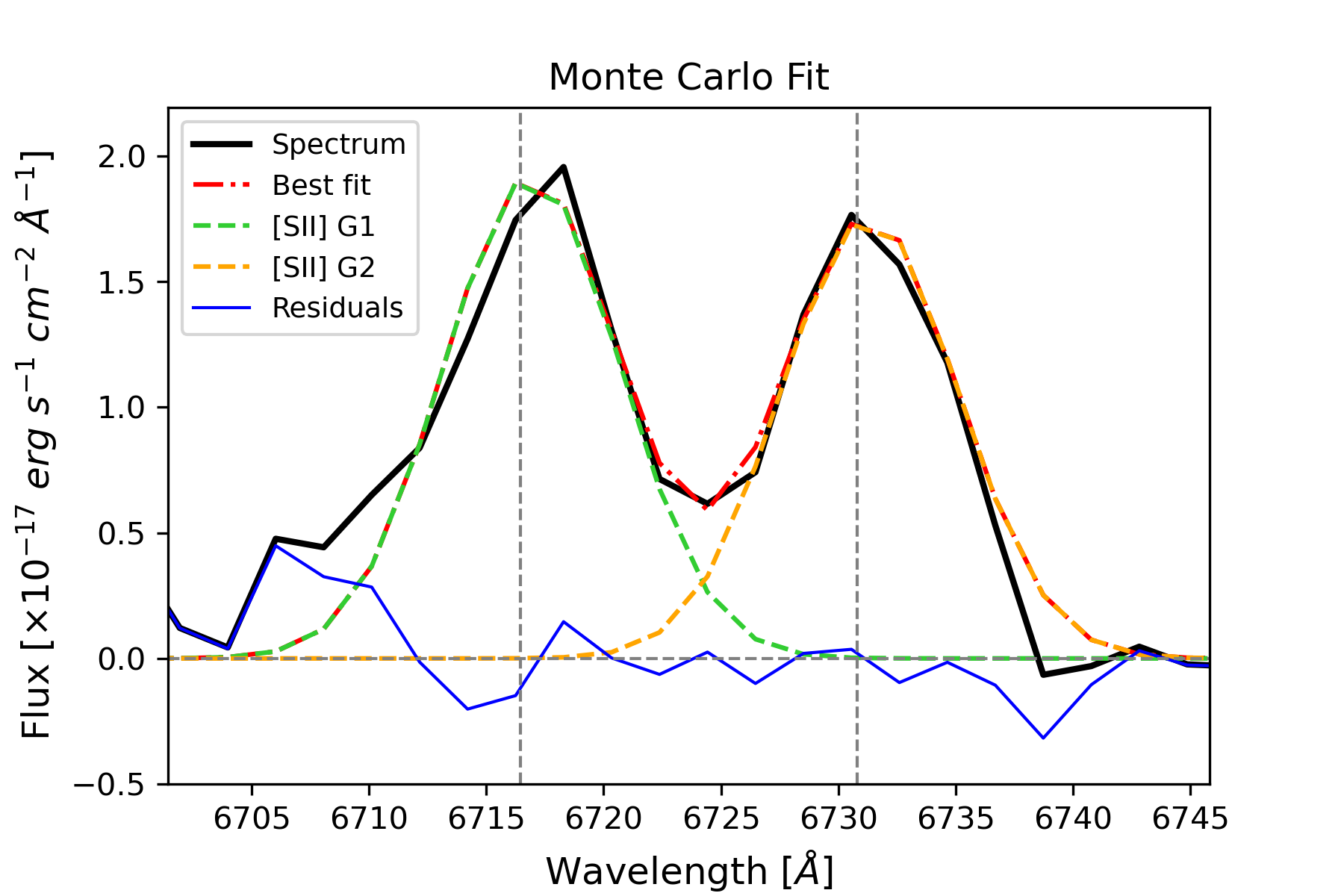}
        \caption{[S~II]$\lambda\lambda$6716,6731 lines profile of J1228.}
    \end{minipage}
\end{figure*}

\clearpage

\begin{figure*}[htbp]
    \centering
    \includegraphics[trim={0 0 0 0.7cm}, clip, width=\textwidth]{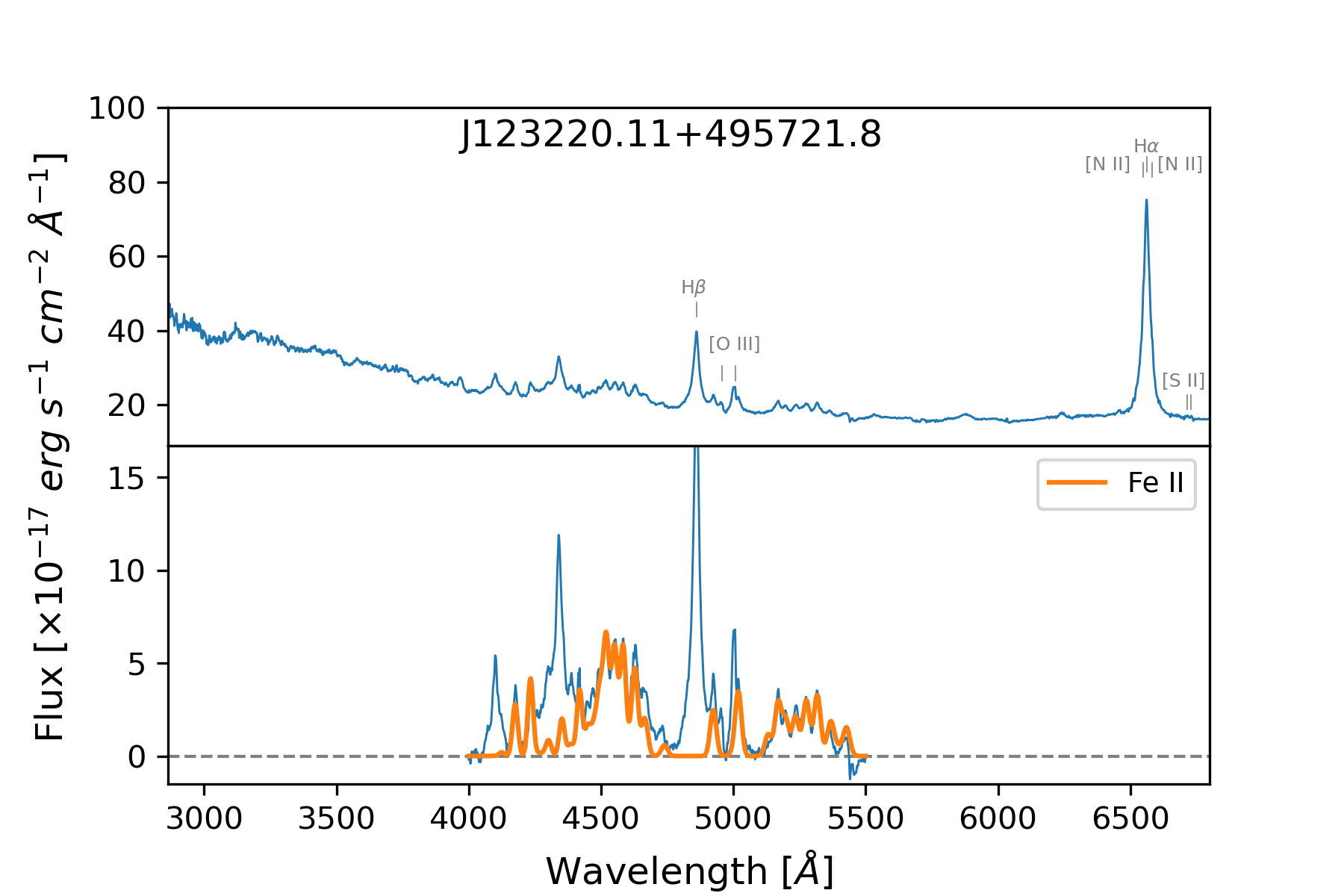}
    \caption{Spectrum of J1232.}

    \vspace{5em}

    \begin{minipage}[b]{0.49\linewidth}
        \centering
        \includegraphics[width=\textwidth]{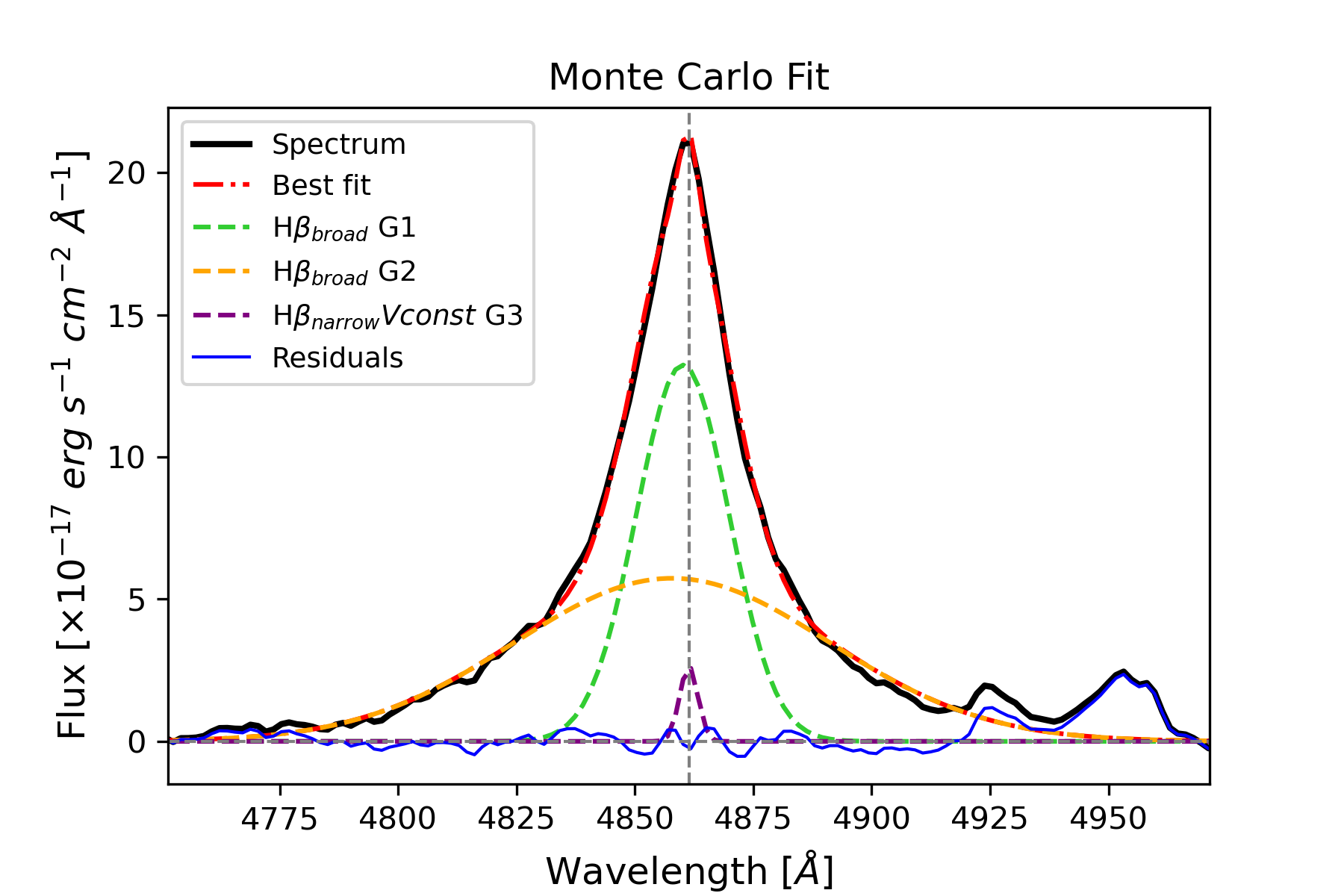}
        \caption{H$\beta$ line profile of J1232.}

    \end{minipage}
    \hfill
    \begin{minipage}[b]{0.49\linewidth}
        \centering
        \includegraphics[width=\textwidth]{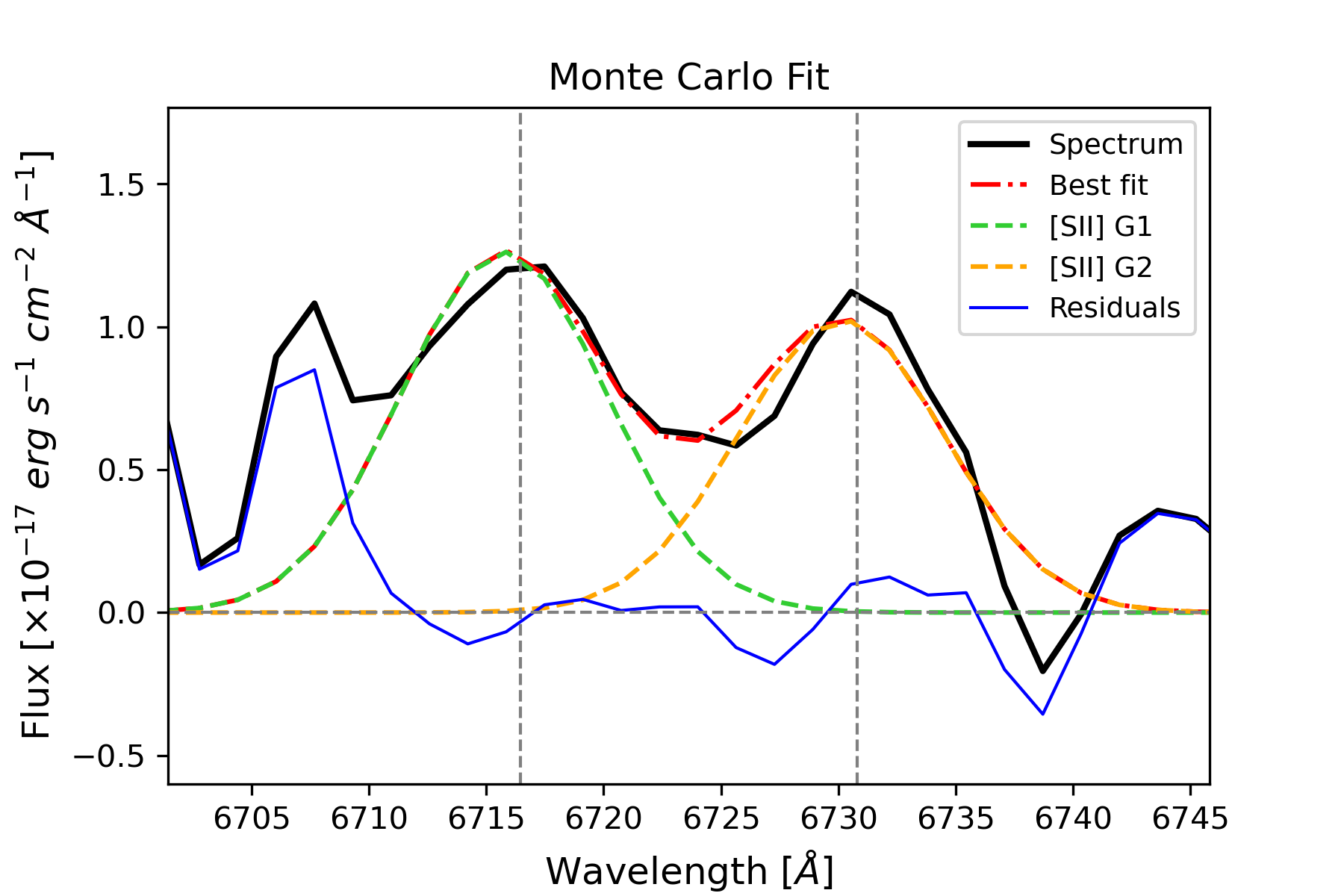}
        \caption{[S~II]$\lambda\lambda$6716,6731 lines profile of J1232.}
    \end{minipage}
\end{figure*}

\clearpage

\begin{figure*}[htbp]
    \centering
    \includegraphics[trim={0 0 0 0.7cm}, clip, width=\textwidth]{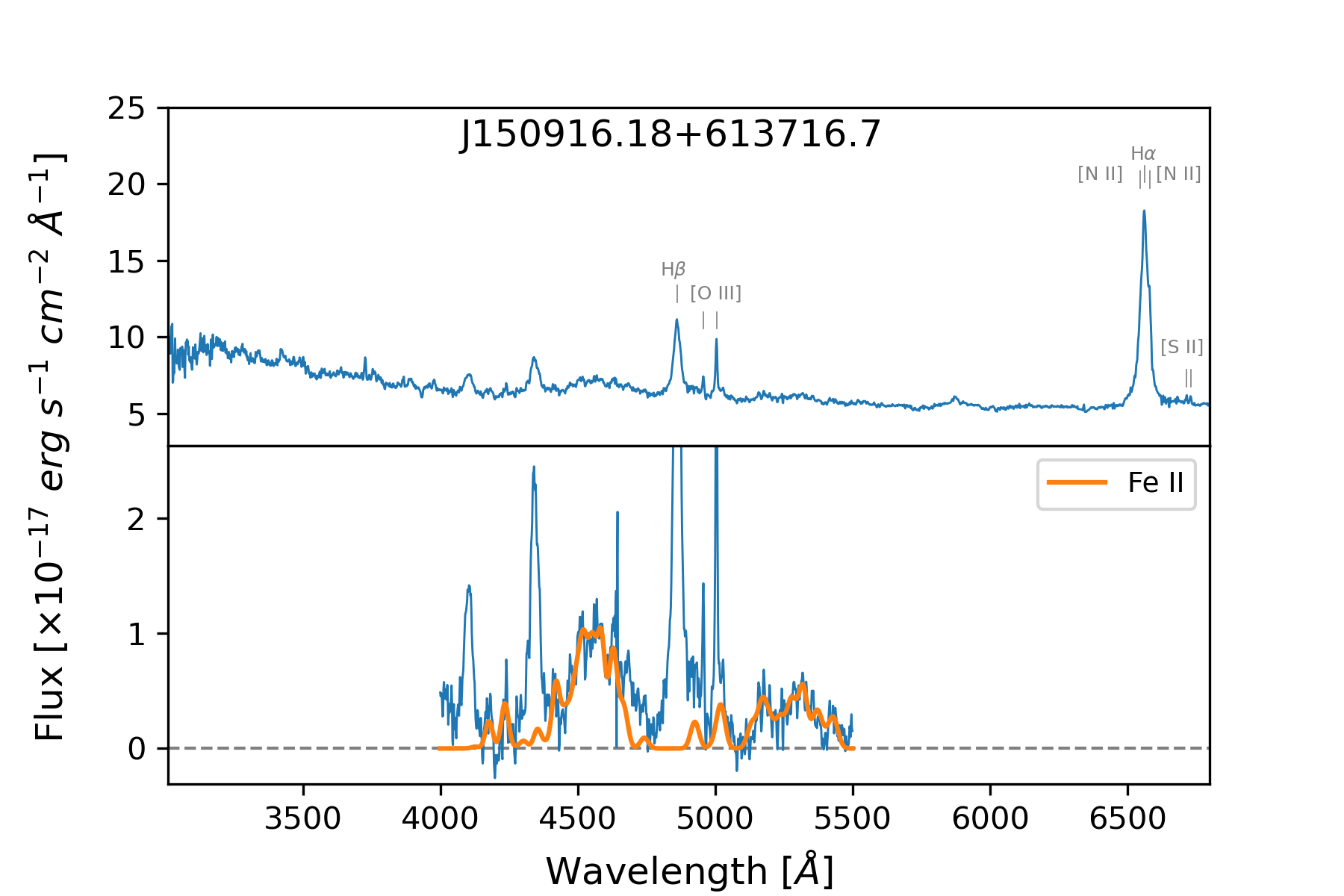}
    \caption{Spectrum of J1509.}

    \begin{minipage}[b]{0.49\linewidth}
        \centering
        \includegraphics[trim={0 0 0 0.6cm}, clip, width=\textwidth]{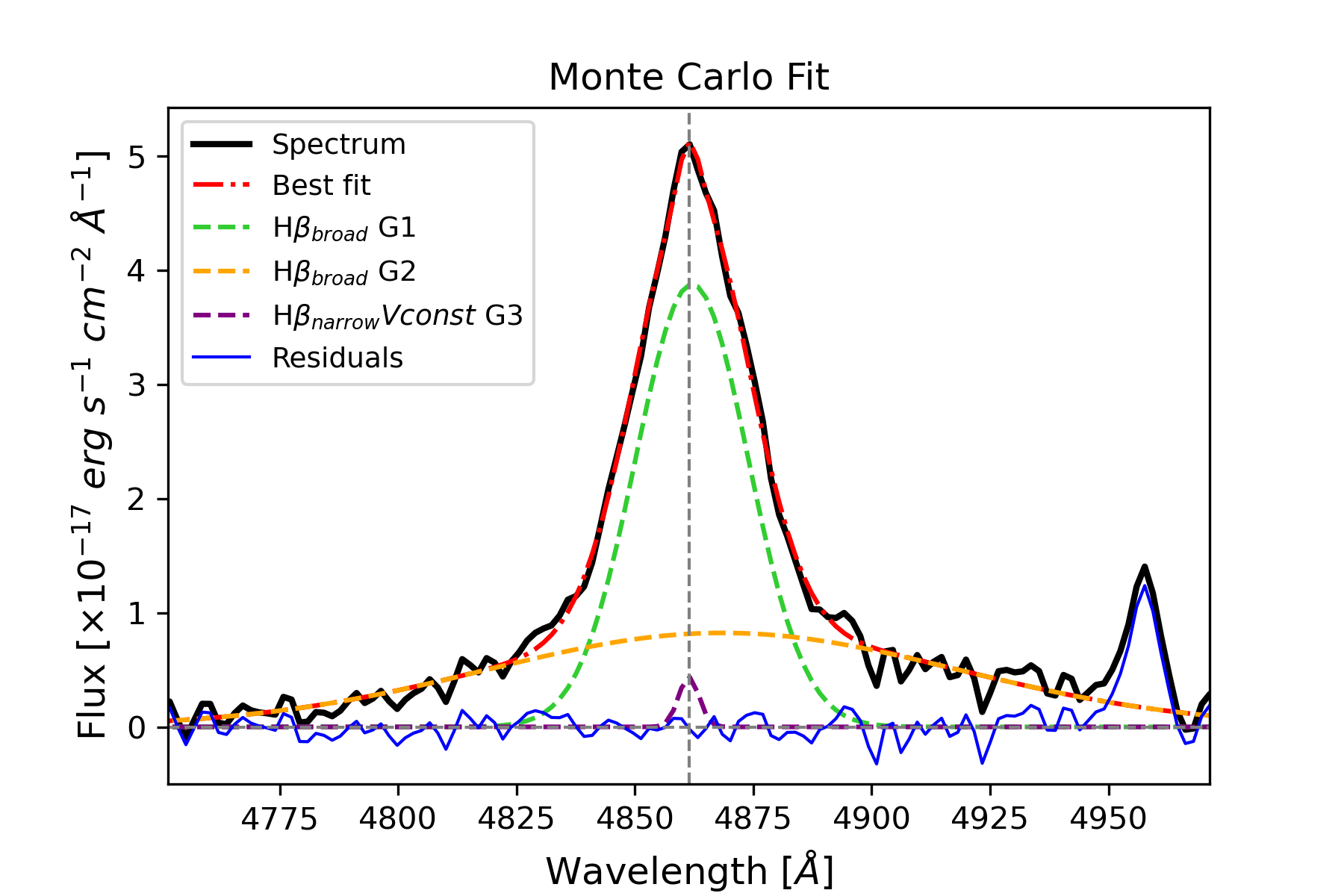}
        \caption{H$\beta$ line profile of J1509.}

    \end{minipage}
    \hfill
    \begin{minipage}[b]{0.49\linewidth}
        \centering
        \includegraphics[trim={0 0 0 0.6cm}, clip, width=\textwidth]{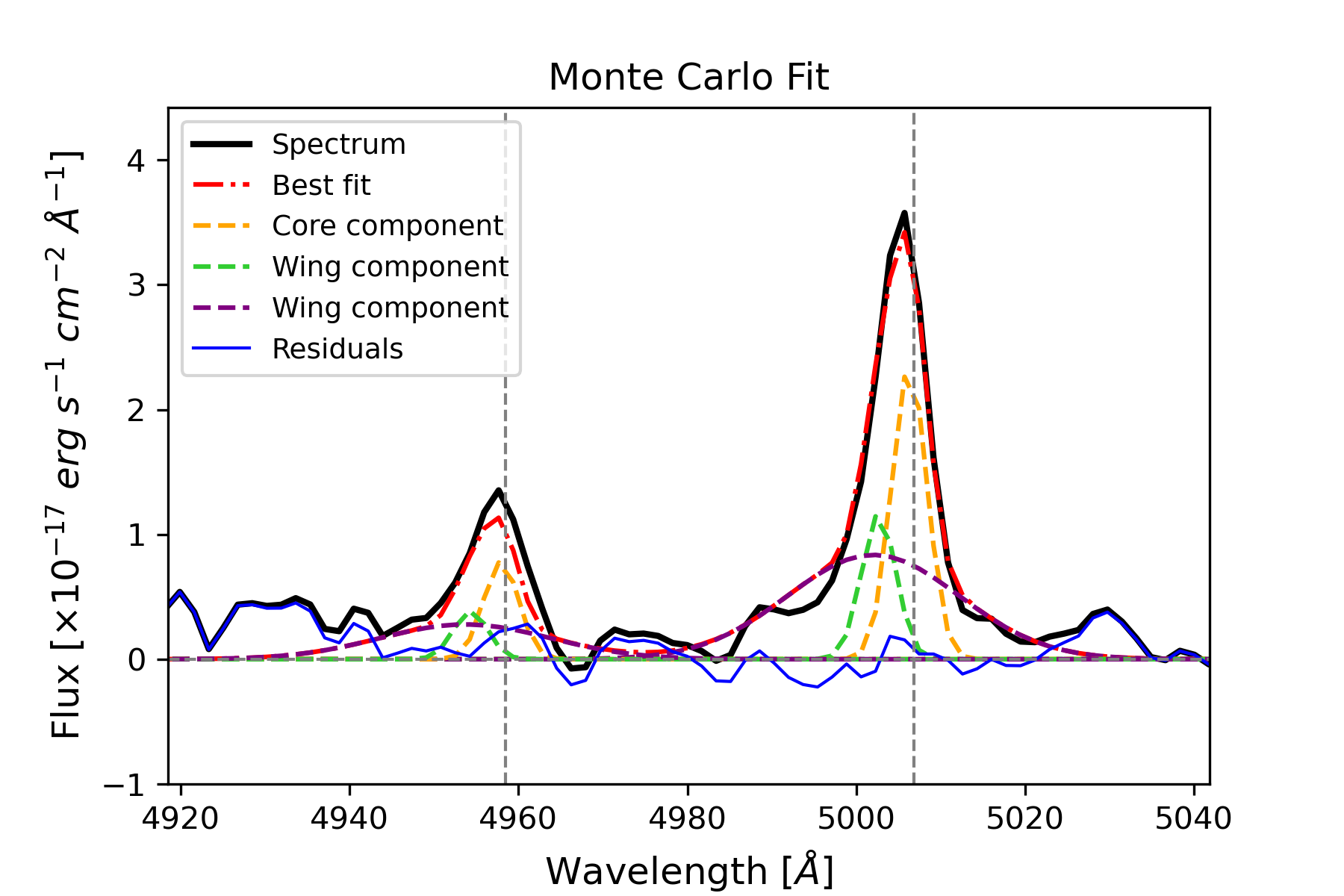}
        \caption{[O~III]$\lambda\lambda$4959,5007 lines profile of J1509.}

    \end{minipage}

    \vspace{1em}
    
    \begin{minipage}[b]{0.49\linewidth}
        \centering
        \includegraphics[trim={0 0 0 0.6cm}, clip, width=\textwidth]{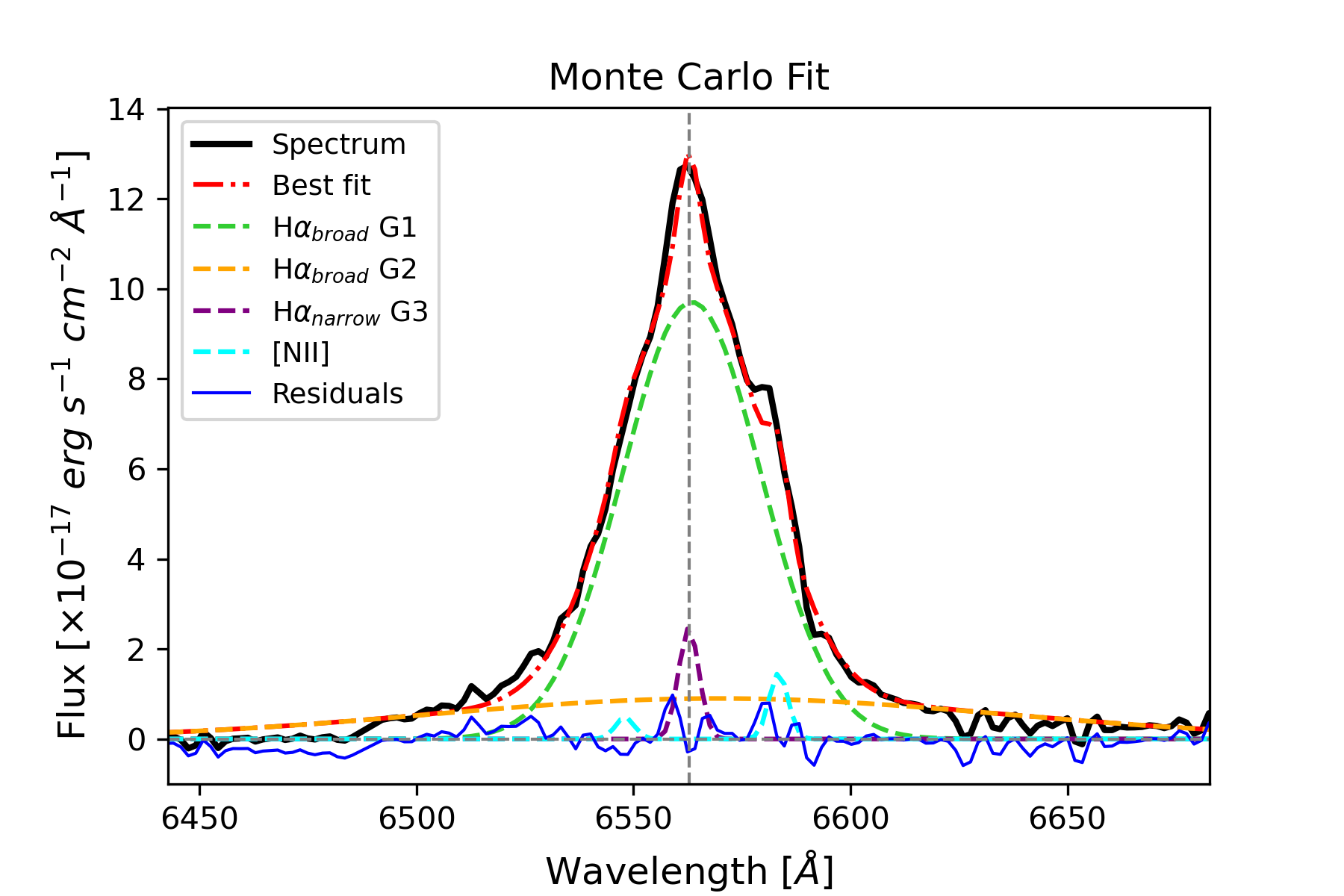}
        \caption{H$\alpha$+[N~II]$\lambda\lambda$6548,6583 lines profile of J1509.}

    \end{minipage}
    \hfill
    \begin{minipage}[b]{0.49\linewidth}
        \centering
        \includegraphics[trim={0 0 0 0.6cm}, clip, width=\textwidth]{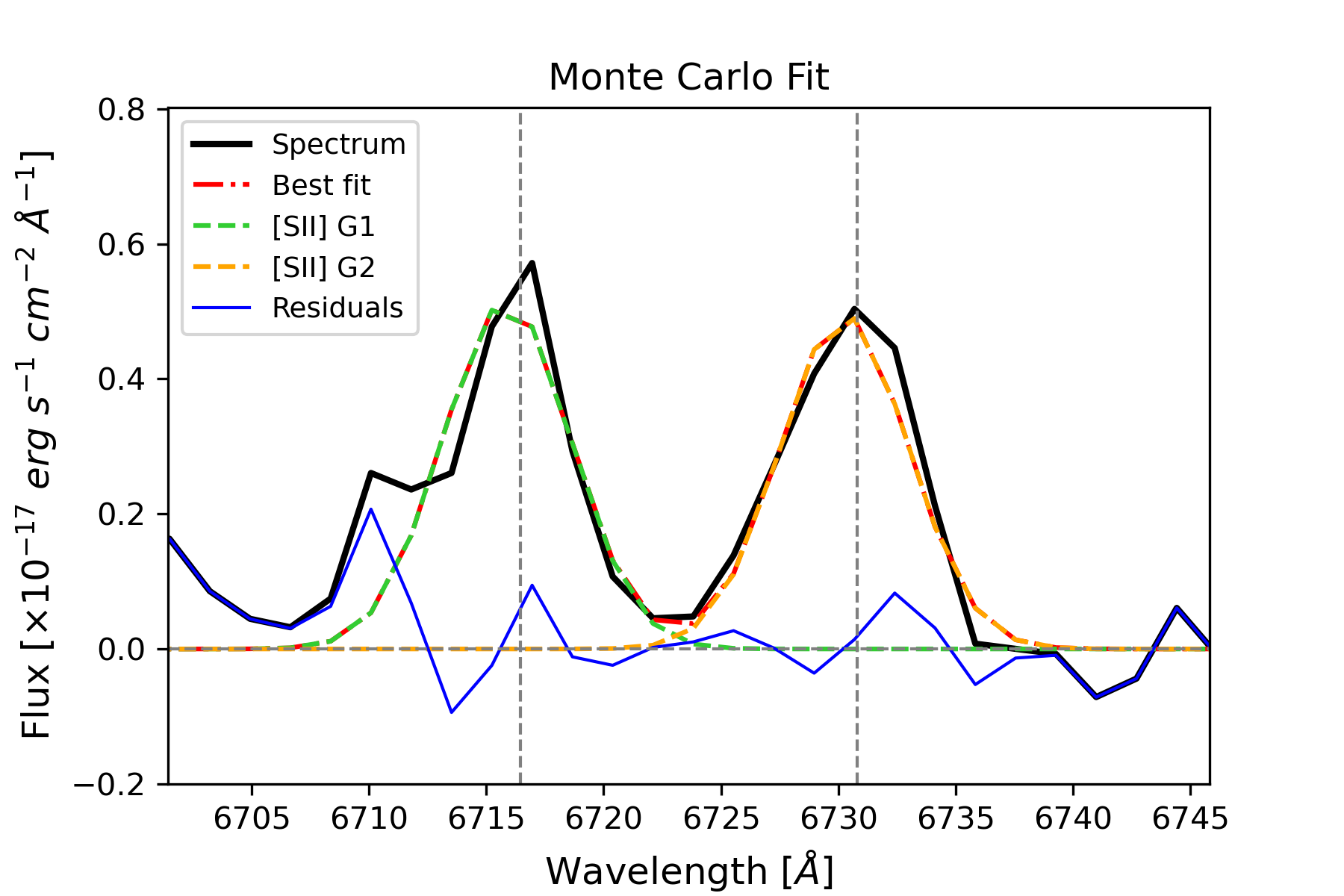}
        \caption{[S~II]$\lambda\lambda$6716,6731 lines profile of J1509.}

    \end{minipage}
\end{figure*}

\clearpage

\begin{figure*}[htbp]
    \centering
    \includegraphics[trim={0 0 0 0.7cm}, clip, width=\textwidth]{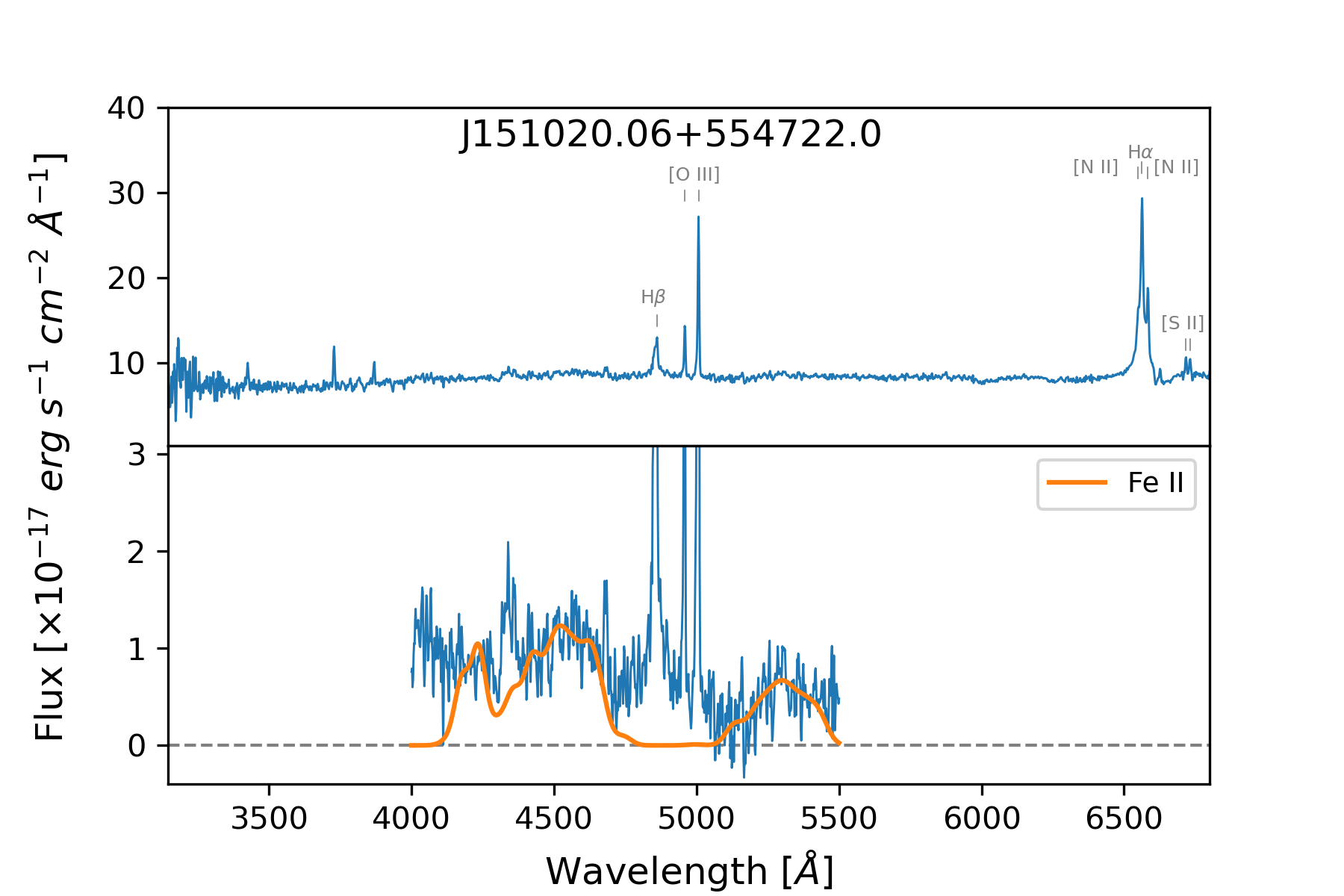}
    \caption{Spectrum of J1510.}

    \begin{minipage}[b]{0.49\linewidth}
        \centering
        \includegraphics[trim={0 0 0 0.6cm}, clip, width=\textwidth]{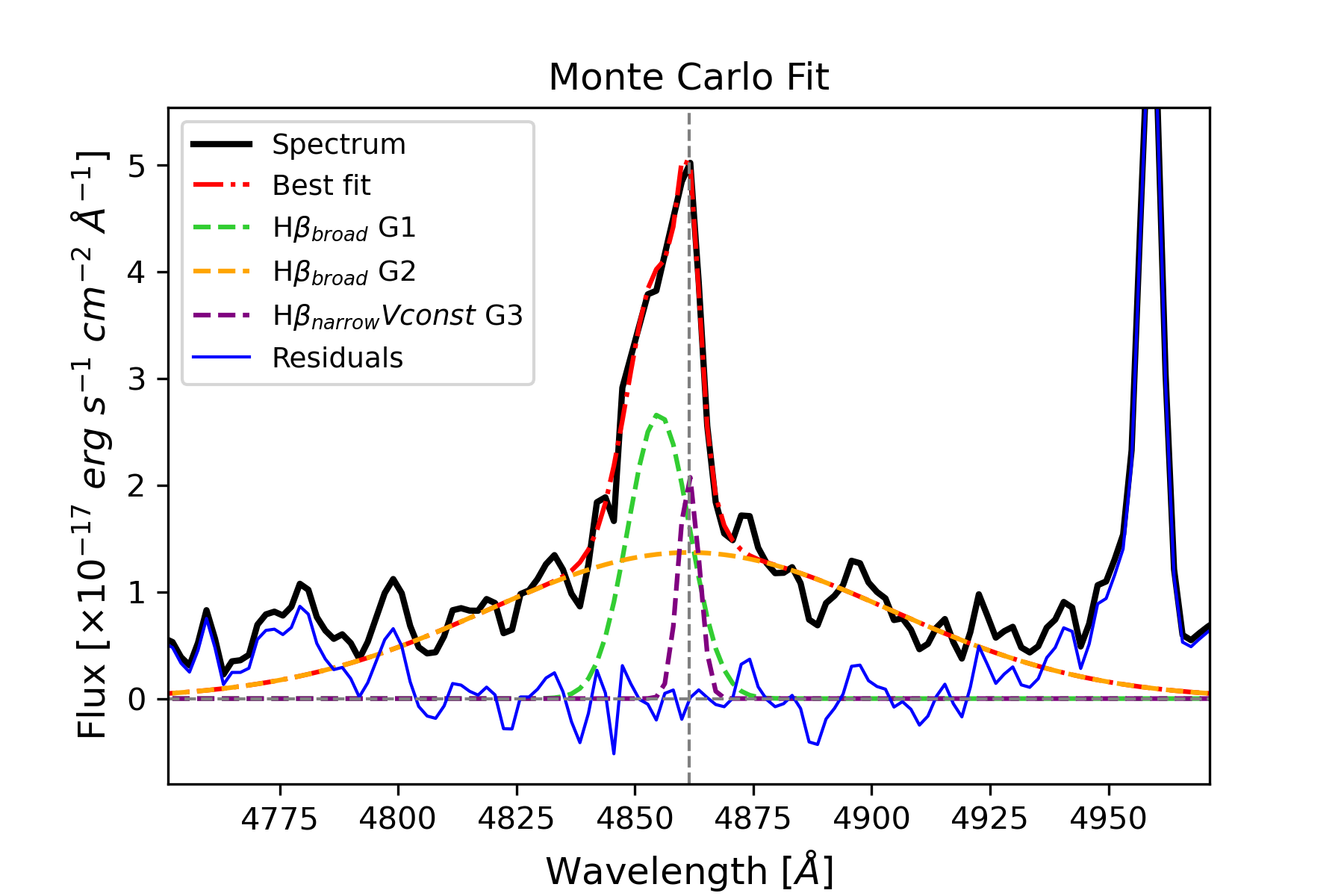}
        \caption{H$\beta$ line profile of J1510.}
    \end{minipage}
    \hfill
    \begin{minipage}[b]{0.49\linewidth}
        \centering
        \includegraphics[trim={0 0 0 0.6cm}, clip, width=\textwidth]{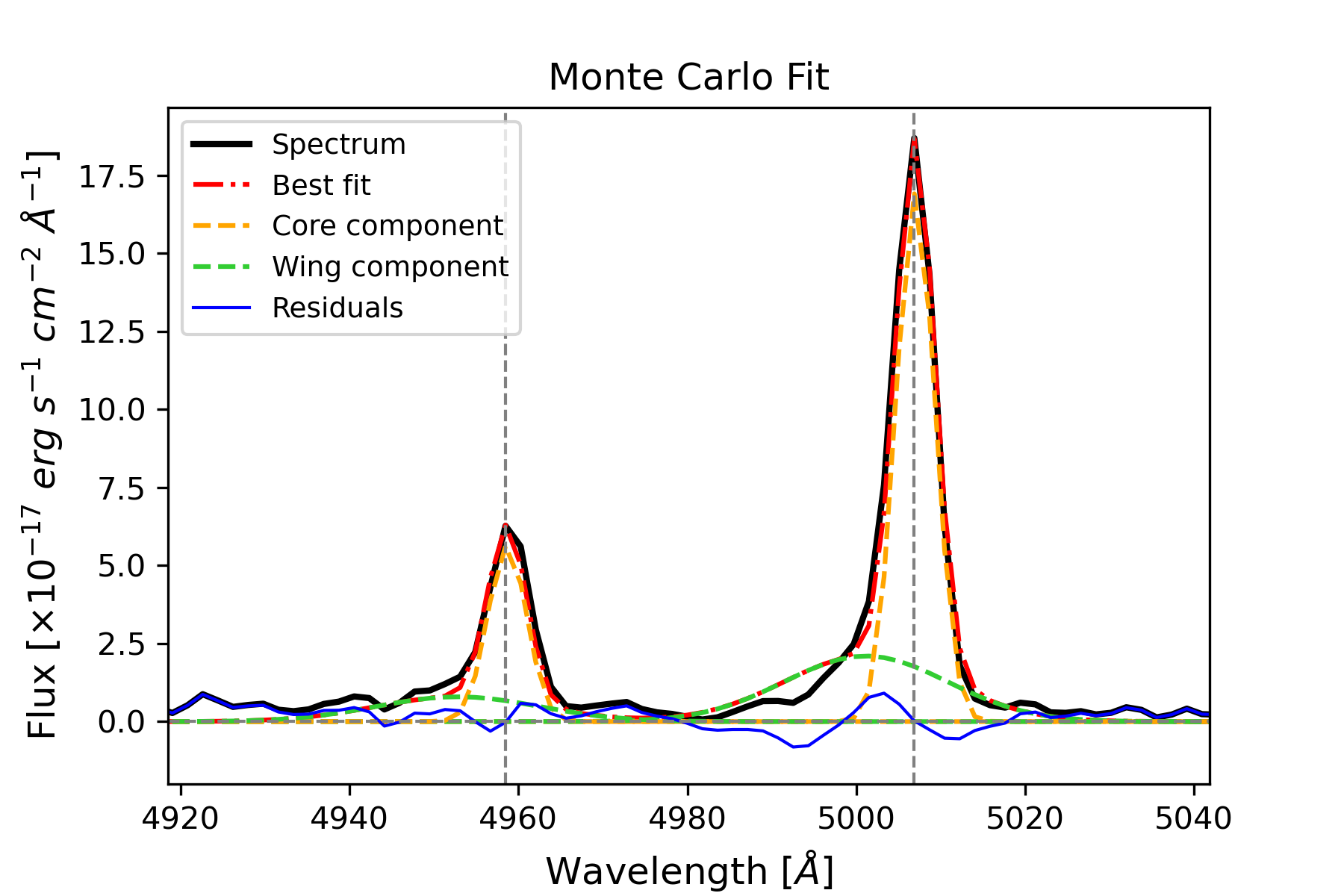}
        \caption{[O~III]$\lambda\lambda$4959,5007 lines profile of J1510.}
    \end{minipage}

    \vspace{1em}

    \begin{minipage}[b]{0.49\linewidth}
        \centering
        \includegraphics[trim={0 0 0 0.6cm}, clip, width=\textwidth]{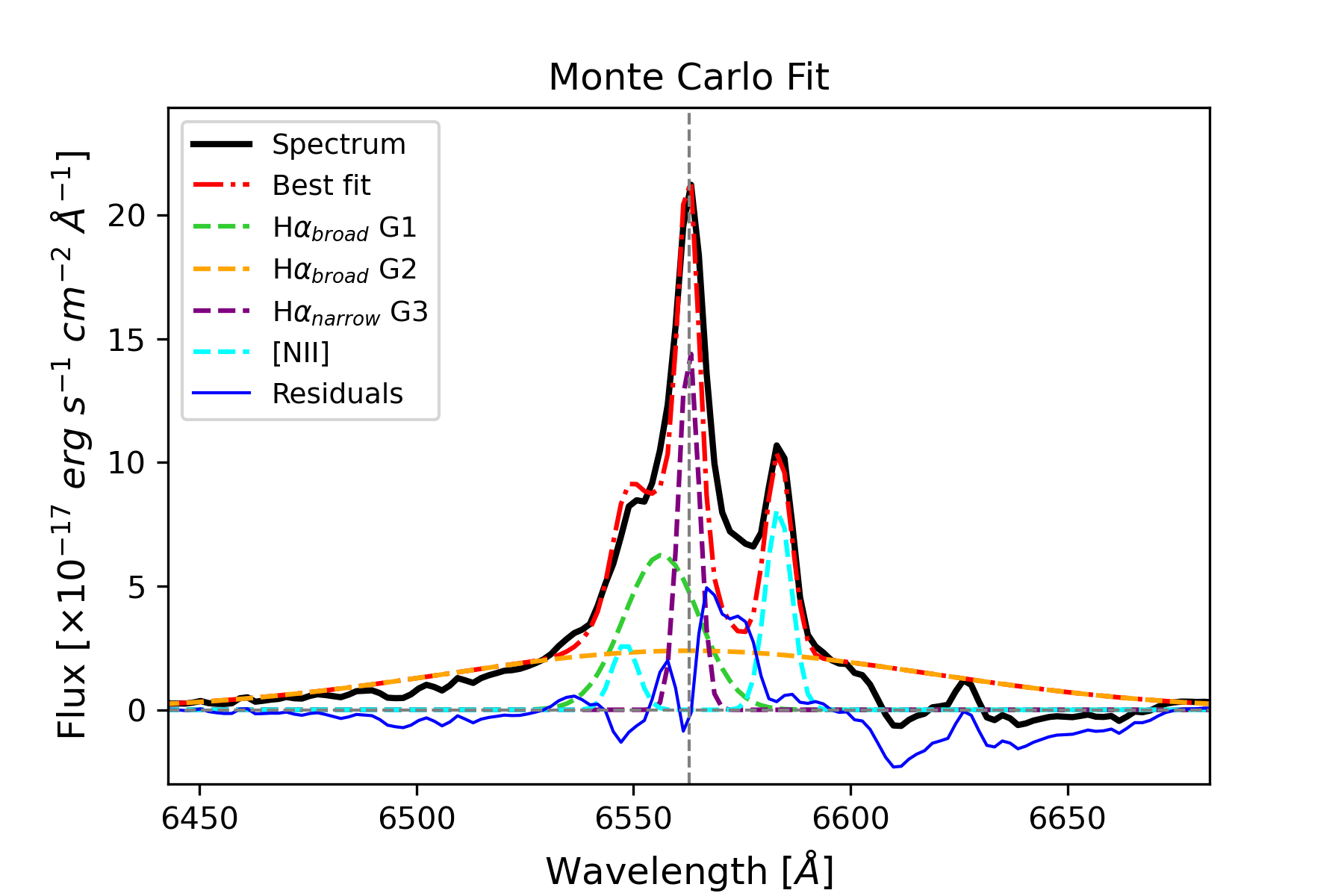}
        \caption{H$\alpha$+[N~II]$\lambda\lambda$6548,6583 lines profile of J1510.}

    \end{minipage}
    \hfill
    \begin{minipage}[b]{0.49\linewidth}
        \centering
        \includegraphics[trim={0 0 0 0.6cm}, clip, width=\textwidth]{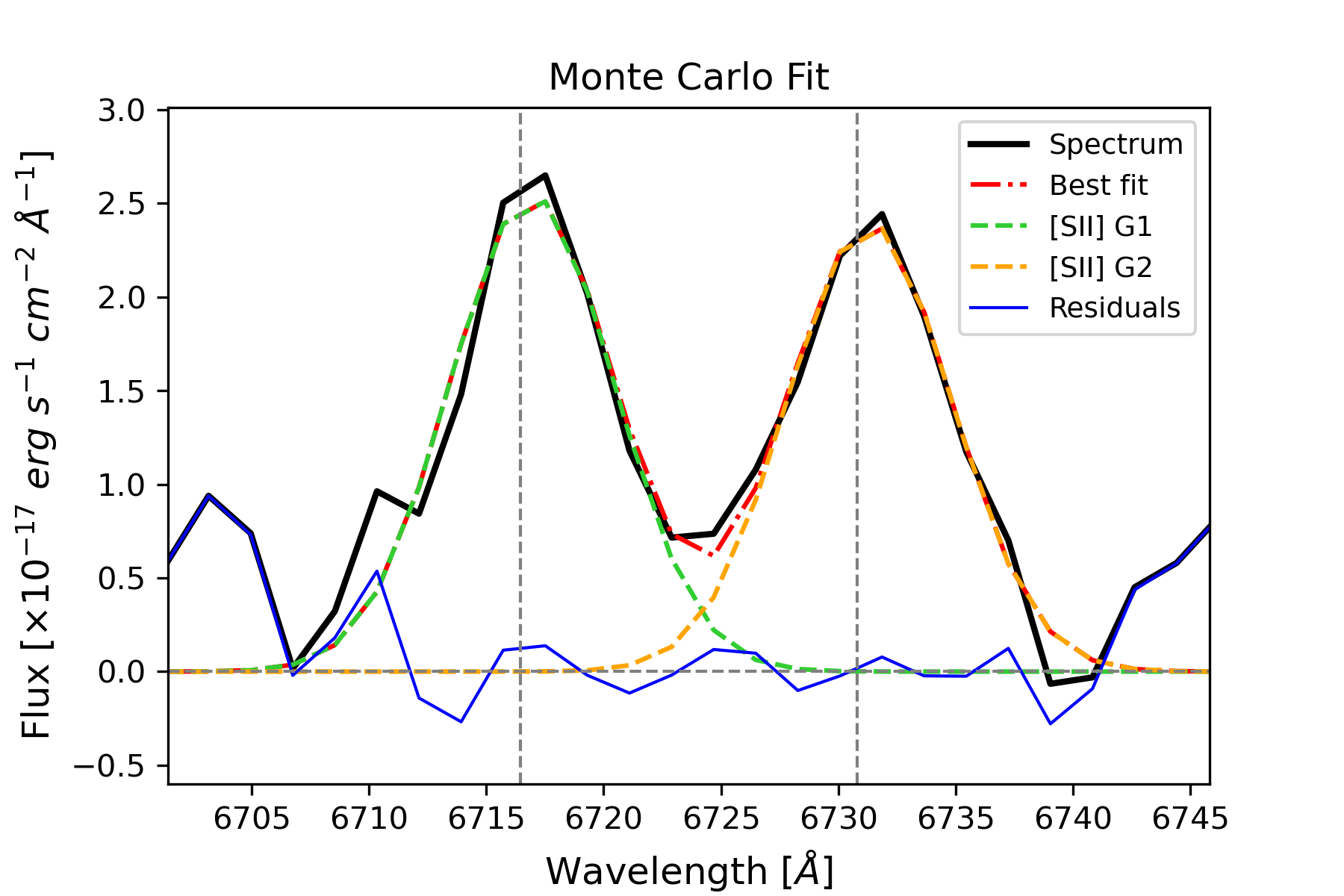}
        \caption{[S~II]$\lambda\lambda$6716,6731 lines profile of J1510.}

    \end{minipage}
\end{figure*}

\clearpage

\begin{figure*}[htbp]
    \centering

    \includegraphics[trim={0 0 0 0.7cm}, clip, width=\textwidth]{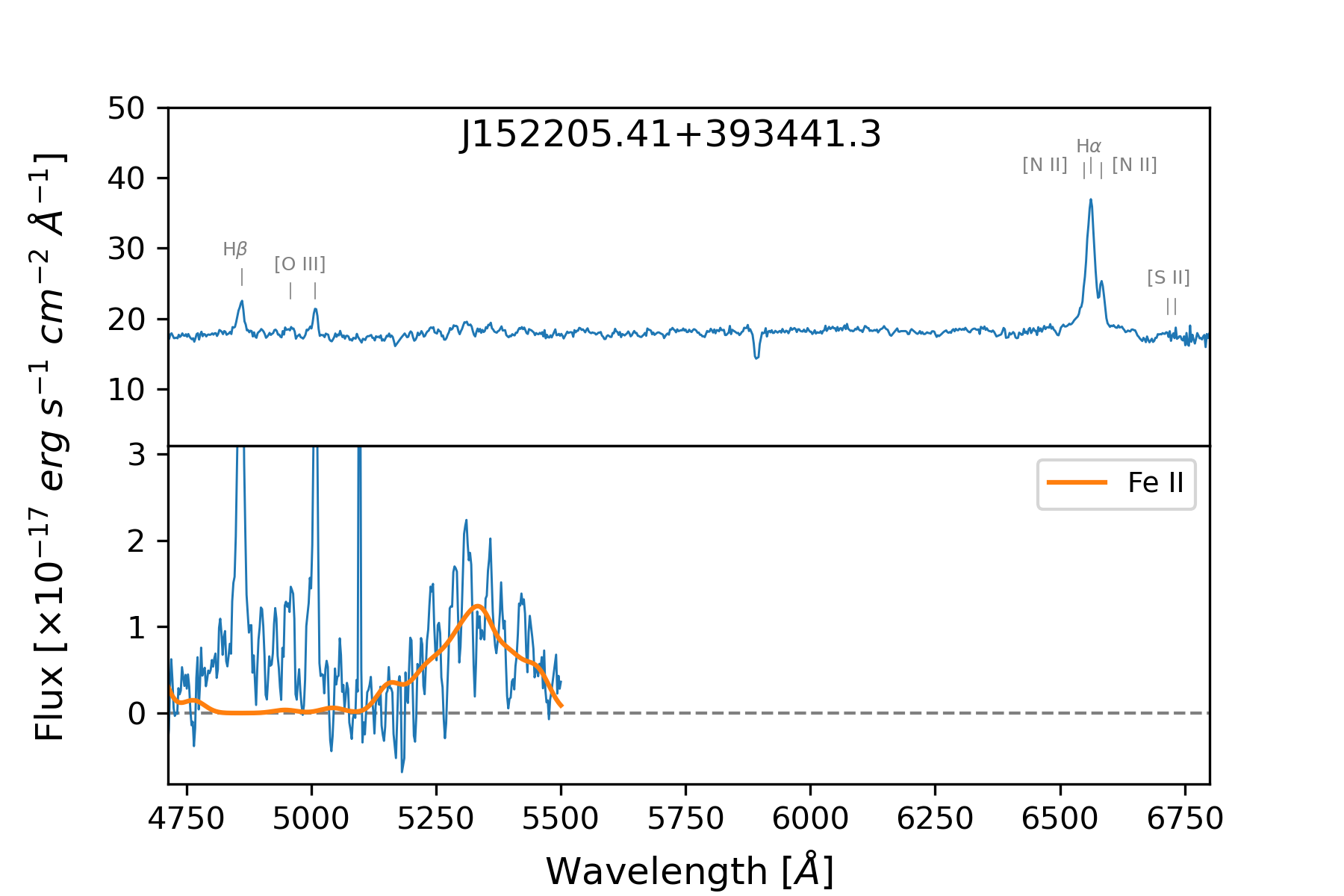}
    \caption{Spectrum of J1522.}

    \vspace{1em}

    \begin{minipage}[b]{0.49\linewidth}
        \centering
        \includegraphics[trim={0 0 0 0.6cm}, clip, width=\textwidth]{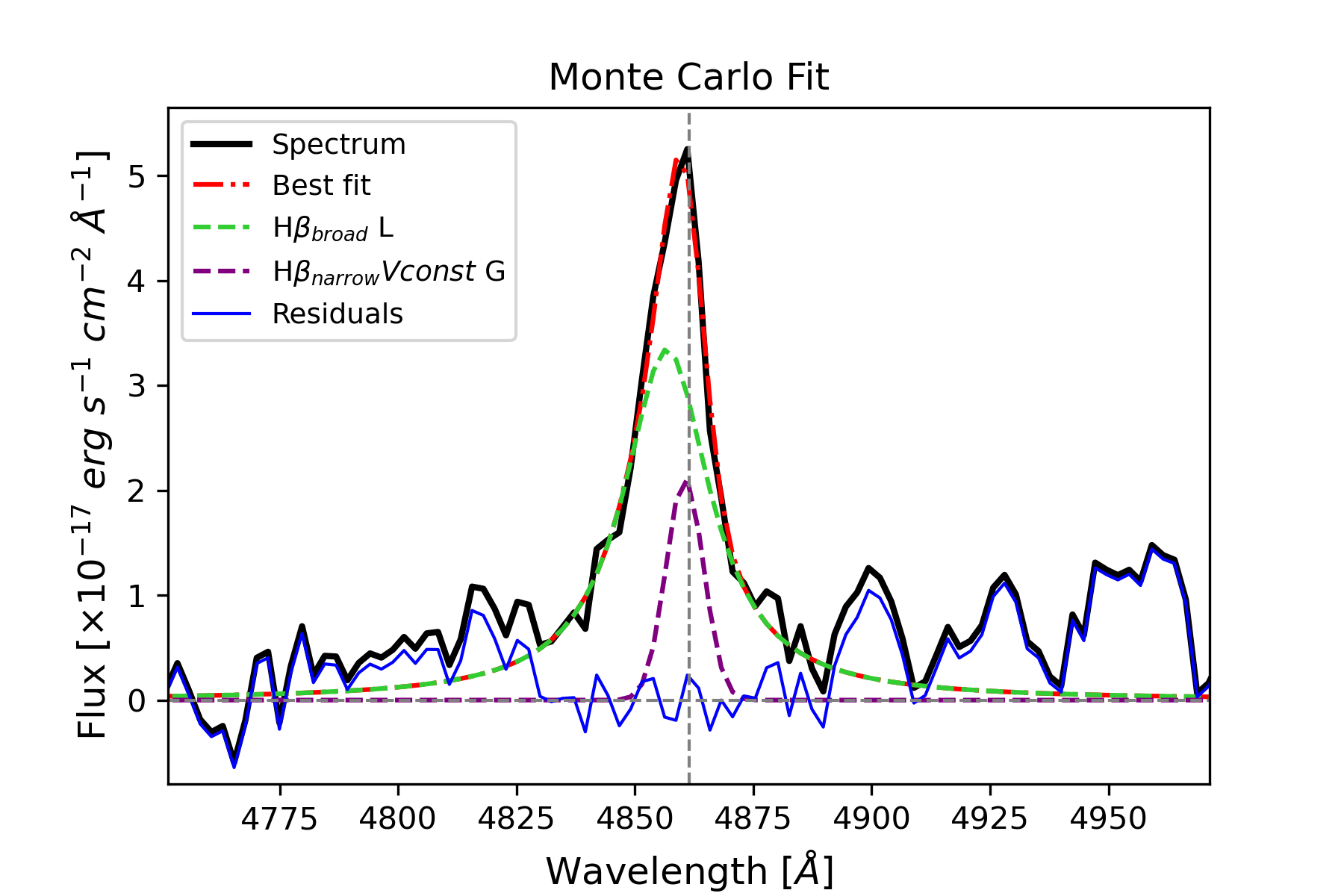}
        \caption{H$\beta$ line profile of J1522.}

    \end{minipage}
    \hfill
    \begin{minipage}[b]{0.49\linewidth}
        \centering
        \includegraphics[trim={0 0 0 0.6cm}, clip, width=\textwidth]{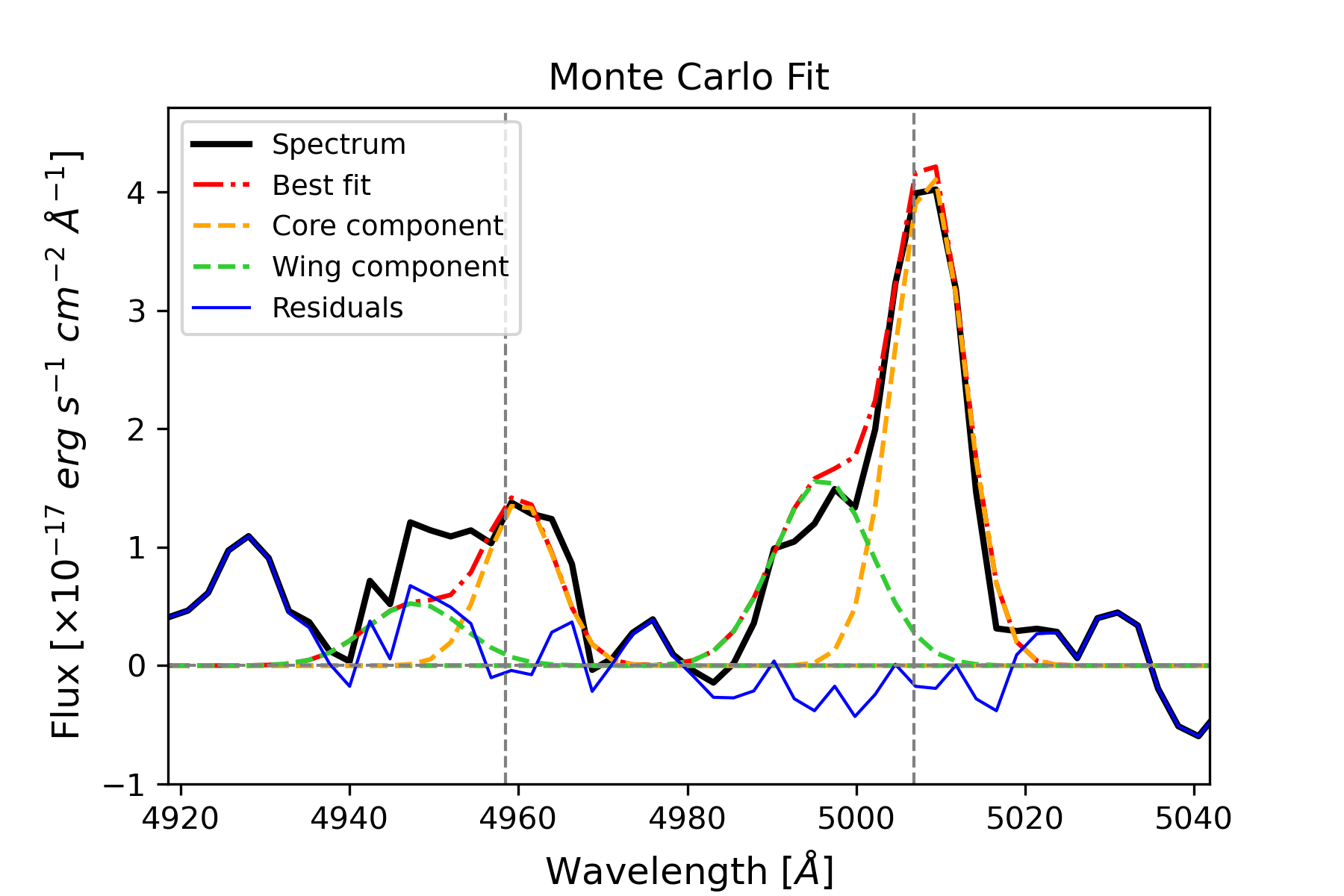}
        \caption{[O~III]$\lambda\lambda$4959,5007 lines profile of J1522.}

    \end{minipage}

    \vspace{1em}
    \begin{minipage}[b]{0.49\linewidth}
        \centering
        \includegraphics[trim={0 0 0 0.7cm}, clip, width=\textwidth]{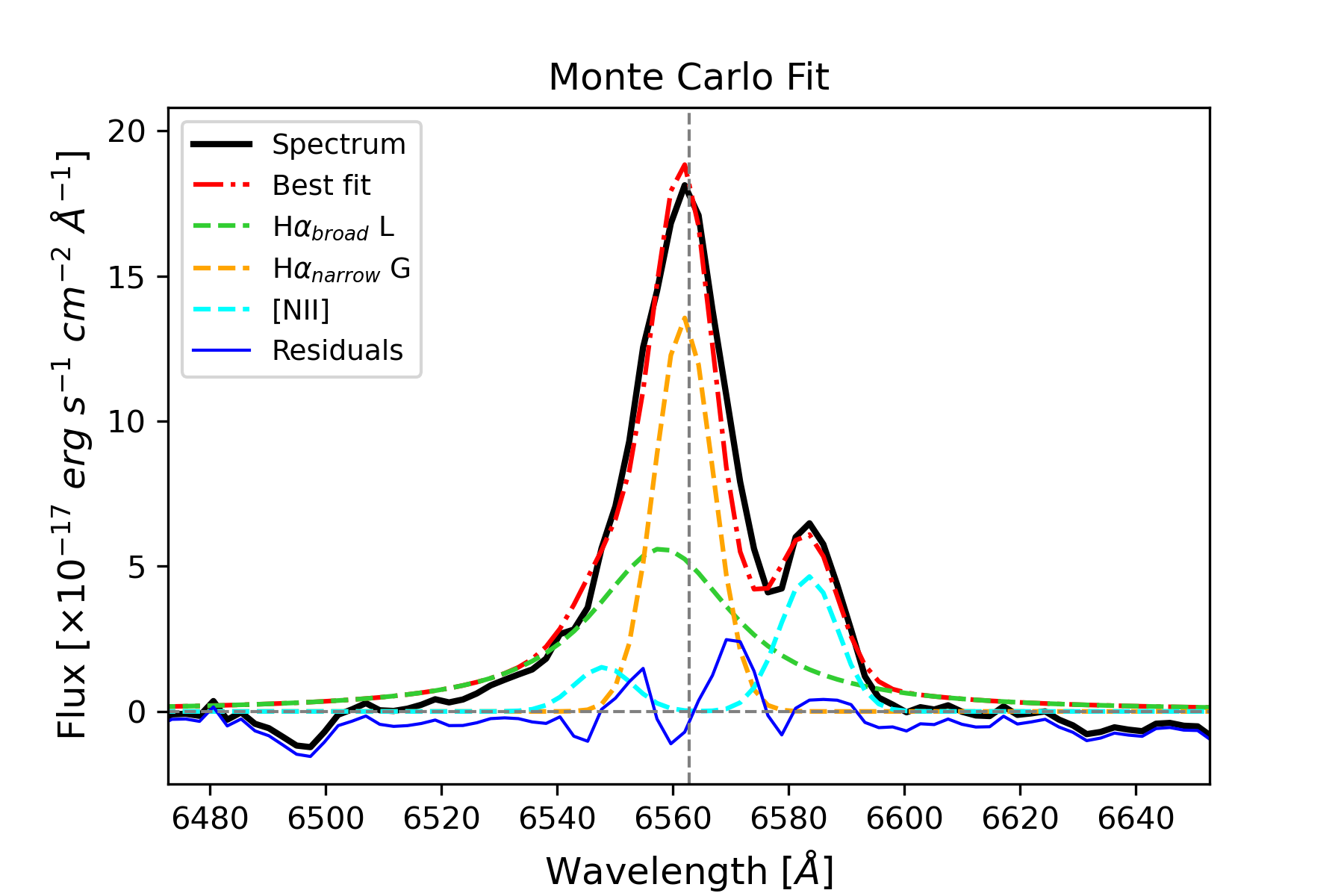}
        \caption{H$\alpha$+[N II]$\lambda\lambda$6548,6583 lines profile of J1522.}

    \end{minipage}
\end{figure*}

\clearpage

\begin{figure*}[htbp]
    \centering

    \includegraphics[trim={0 0 0 0.7cm}, clip, width=\textwidth]{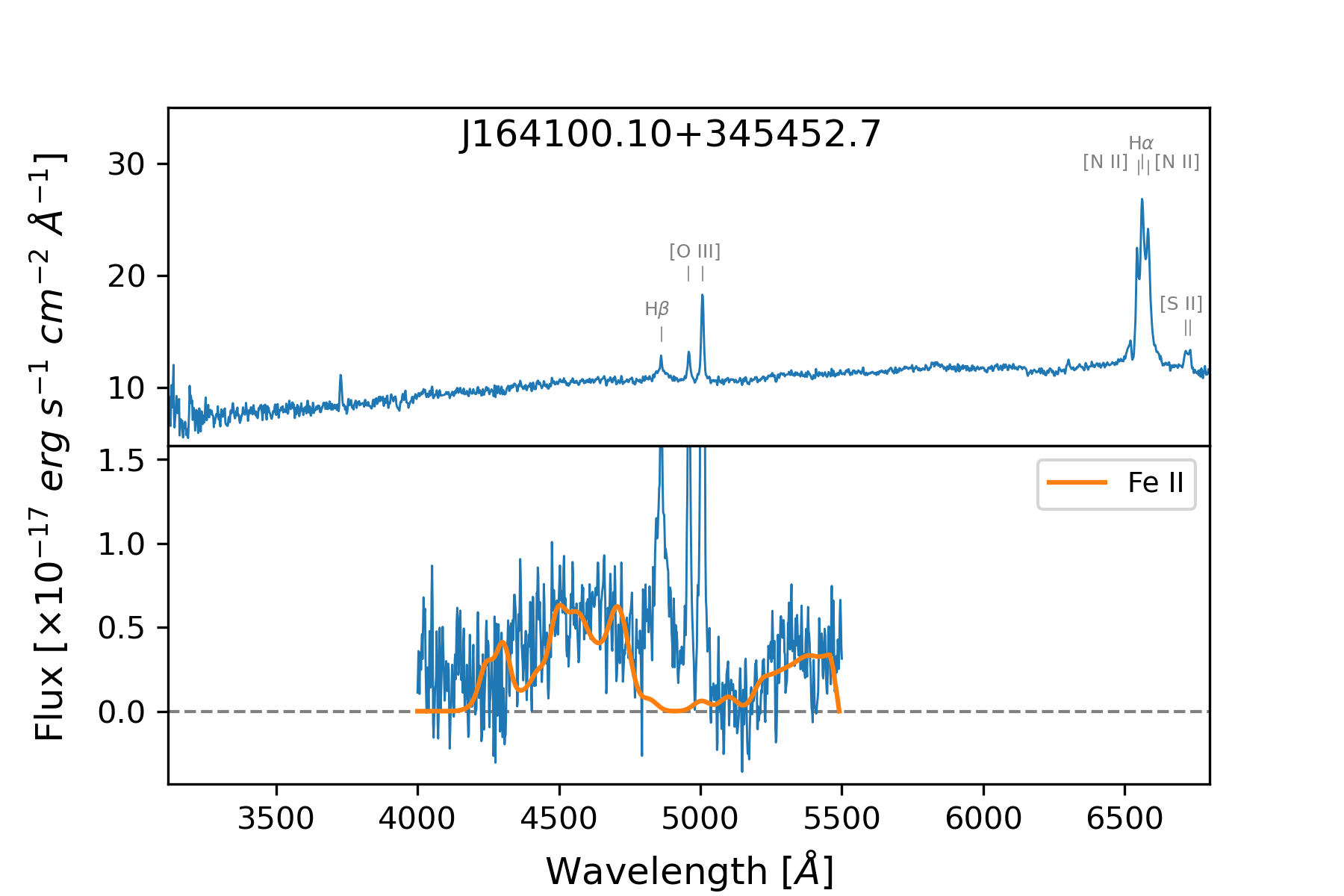}
    \caption{Spectrum of J1641.}

    \vspace{5em}

    \begin{minipage}[b]{0.49\linewidth}
        \centering
        \includegraphics[width=\textwidth]{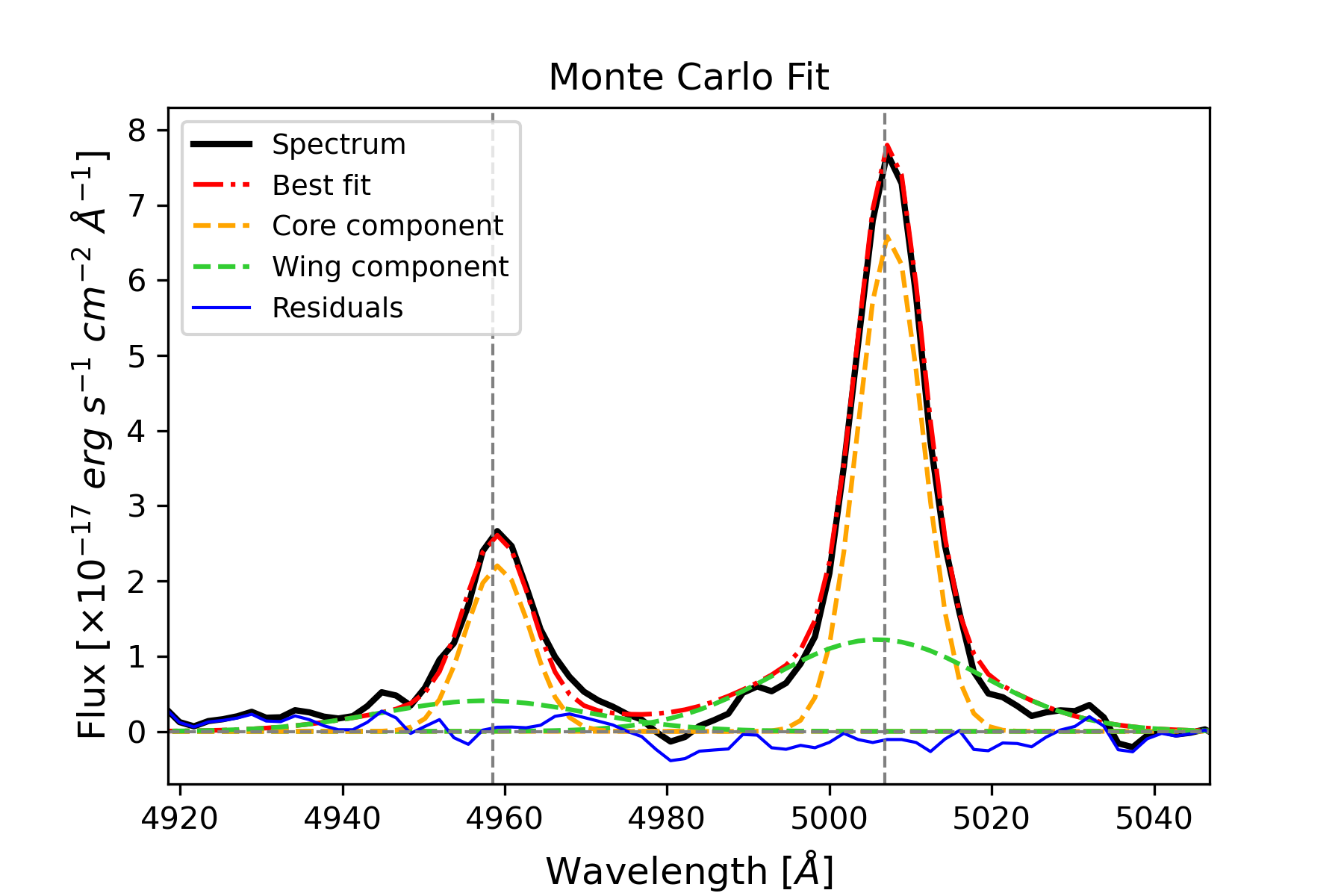}
        \caption{[O~III]$\lambda\lambda$4959,5007 lines profile of J1641.}

    \end{minipage}
    \hfill
    \begin{minipage}[b]{0.49\linewidth}
        \centering
        \includegraphics[width=\textwidth]{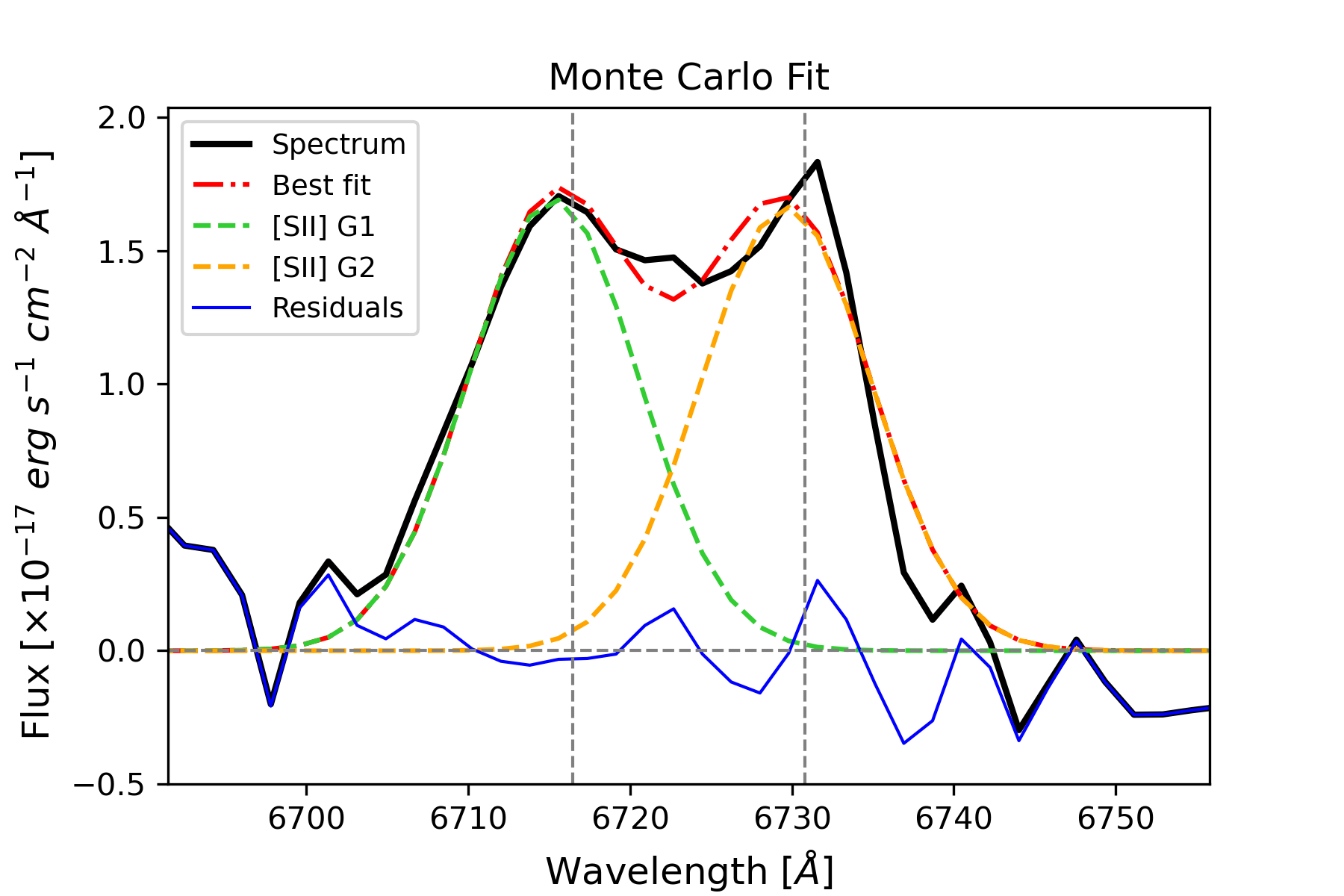}
        \caption{[S~II]$\lambda\lambda$6716,6731 lines profile of J1641.}
    \end{minipage}
\end{figure*}

\end{appendix}

\end{document}